\documentclass[12pt]{article}
\usepackage[onehalfspacing]{setspace}
\usepackage{amsthm}
\usepackage[hidelinks]{hyperref}
\usepackage{graphics,graphicx,amssymb,mathtools}
\usepackage{caption}
 \usepackage{booktabs}
 \usepackage{url}
 \usepackage{orcidlink}
\usepackage{dsfont}
\usepackage{xcolor}
\usepackage{bm}
\usepackage{siunitx}
\usepackage{booktabs}
\usepackage{multirow}
\usepackage{enumitem}
 \usepackage{xr} 
\onecolumn 
\usepackage{subcaption}
 \usepackage{natbib}
\usepackage{footmisc}
\usepackage{dsfont}
\usepackage{xr}
\usepackage{dsfont}
\usepackage{amssymb}
\usepackage{algorithm}
\usepackage{algorithmic}
\usepackage{array}
\usepackage{threeparttable}
\usepackage[section]{placeins}  

\newcommand{\blind}{1}

\newcommand{\qq}{{\mathbf q}}
\newcommand{\hh}{{\mathbf h}}
\newcommand{\ee}{{\mathbf e}}
\newcommand{\uu}{{\mathbf u}}
\newcommand{\vv}{{\mathbf v}}
\newcommand{\yy}{{\mathbf y}}
\newcommand{\zz}{{\mathbf z}}
\newcommand{\YY}{{\mathit{Y} }}
\newcommand{\AAA}{{\mathit{A}}}
\newcommand{\ThTh}{{\mathit \Theta}}
\newcommand{\PhPh}{\mathit \Phi}
\newcommand{\PsPs}{\mathit \Psi}
\newcommand{\XiXi}{{\mathbf{\Xi} }}

\newcommand{\xx}{\mathbf{x}}

\newcommand{\HH}{\mathcal{H}}
\newcommand{\HHH}{\mathbb{H}}

\let\hat\widehat
\let\tilde\widetilde

\newcommand{\EE}{\mathit{E}}

\newcommand{\ZZ}{{\mathit Z}}

\newtheorem{assump}{Assumption}
\newtheorem{remark}{Remark}

\newtheorem{lemma}{Lemma}
\newtheorem{theorem}{Theorem}

\def\aa{\mathbf a }
\newcommand{\bbb}{{\boldsymbol \beta}}
\newcommand{\thth}{\boldsymbol \theta}
\newcommand{\rhrh}{\boldsymbol \rho}
\def\ggg {\boldsymbol \gamma}

\newcommand{\RR}{{\mathbf R}}
\newcommand{\rr}{{\mathbf r}}

\newcommand{\ttt}{{\boldsymbol \theta}}

\newcommand{\LL}{{\mathit L}}

\newcommand{\argmax}{\operatornamewithlimits{arg\,max}}
\newcommand{\argmin}{\operatornamewithlimits{arg\,min}}
\newcommand{\MM}{{\mathbf M}}
\newcommand{\UU}{{\mathit U}}
\newcommand{\VV}{{\mathbf V}}
\newcommand{\DD}{{\mathbf D}}
\newcommand{\vvv}{{\mathbf v}}

\newcommand{\SigSig}{\mathit{\Sigma}}
\newcommand{\zeze}{\boldsymbol \zeta}
\newcommand{\mumu}{\boldsymbol \mu}
\newcommand{\omom}{\boldsymbol \omega}
\def\ggg {\boldsymbol \gamma}
\newcommand{\sss}{{\mathbf s}}
\newcommand{\chichi}{\mathbf \chi}

\newcommand{\XX}{{\mathit{X}}}

\newcommand{\BB}{{\mathit{B}}}

\newcommand{\GG}{{\mathbf \Gamma}}

\def\aa{\mathbf a }

\date{} 

\begin{document}

\def\spacingset#1{\renewcommand{\baselinestretch}%
{#1}\small\normalsize} \spacingset{1}


  \title{\bf A Latent Variable Approach to Learning
High-dimensional Multivariate longitudinal Data}
\if1\blind
{
  \author{Sze Ming Lee$^a$\orcidlink{0009-0008-1487-2367}, Yunxiao Chen$^{a,*}$\orcidlink{0000-0002-7215-2324} and Tony Sit$^b$\orcidlink{0000-0001-5006-1838}~\\~\\$^a$\footnotesize\textit{Department of Statistics, London School of Economics and Political Science}~\\$^b$\footnotesize\textit{Department of Statistics, The Chinese University of Hong Kong}~\\~\\
  }
  \maketitle
} \fi

\if0\blind
{
  \bigskip
  \bigskip
  \bigskip
  \begin{center}
    {\LARGE\bf Title}
\end{center}
  \medskip
} \fi

\bigskip
\begin{abstract}
High-dimensional multivariate longitudinal data, which arise when many outcome variables are measured repeatedly over time,
are becoming increasingly common in social, behavioral and health sciences.  
We propose a latent variable model for 
drawing statistical inferences on covariate effects and predicting future outcomes based on high-dimensional multivariate longitudinal data. 
This model introduces unobserved factors to account for the between-variable and across-time dependence and assist the prediction. Statistical inference and prediction tools are developed under a general setting that allows outcome variables to be of mixed types and possibly unobserved for certain time points, for example, due to right censoring. A central limit theorem is established for drawing statistical inferences on regression coefficients. Additionally, an information criterion is introduced to choose the number of factors. The proposed model is applied to customer grocery shopping records to predict and understand shopping behavior.
\end{abstract}

\noindent%
\vfill

\newpage
\spacingset{1.9} 

\section{Introduction}
High-dimensional multivariate longitudinal data are becoming increasingly common, especially in social, behavioral and health sciences, where many outcomes are measured repeatedly within individuals. Examples include ecological momentary assessment data collected by smartphones or wearable devices for understanding within-subject social, psychological, and physiological processes in everyday contexts \citep{bolger2013intensive,wang2014studentlife}, electronic health record data for predicting and understanding health-related conditions \citep{lian2015multitask, zhang2020adverse}, computer logfile data for understanding human-computer interactions from solving complex computer-simulated tasks \citep{chen2019statistical,chen2020continuous}, and grocery shopping data for market basket analysis \citep{wan2017modeling,wan2018representing}.  These data may involve outcome variables of different types. For example, for ecological momentary assessment, physiological processes are typically measured by continuous variables, such as blood pressure, while psychological processes are recorded by participants' responses to survey items that involve binary or categorical variables. In addition, many multivariate longitudinal data may be derived from multitype recurrent event history  \citep[Chapter 2.5,][]{cook_etal-2007-recurbook}, for which an outcome variable records whether a specific type of event occurs (e.g., purchasing a merchandise item) or the count of its occurrences within a time interval (e.g., the number of purchases). 

In this paper, we study high-dimensional multivariate longitudinal data, aiming to (1) infer the effect of covariates on each outcome variable and (2) predict future outcomes based on covariates and historical data. These tasks involve three challenges. First, due to the nature of the data, there is a complex within-individual dependence structure which exists between outcome variables and across time. Valid statistical inference and accurate prediction become a challenge if one fails to account for the dependence properly. Second, the presence of many outcome variables implies a substantial number of item-specific parameters, bringing challenges to the statistical inference. The classical theory for  M- or Z-estimators no longer applies, and new asymptotic results concerning the consistency and asymptotic normality under a high-dimensional regime are needed. Third, some observation units may be lost to follow-up or observed only intermittently, resulting in incomplete data. For example, grocery shopping records based on membership may be incomplete if customers occasionally shop without using their membership card. 

To tackle these challenges, we propose a high-dimensional generalized latent factor model. In this model, low-dimensional factors are introduced within each observation unit to capture the between-item and across-time dependence that is not attributable to the covariates. The model is very flexible, allowing for many types of outcome variables, including binary, count, and continuous variables. 
In addition, a computationally efficient joint likelihood estimator is proposed that estimates the unobserved factors, loading parameters, and regression coefficients simultaneously, which treats the factors as fixed parameters. Asymptotic properties of this estimator are established, including a central limit theorem for drawing statistical inferences on regression coefficients and an information criterion for choosing the number of factors. Moreover, we introduce a missing indicator approach \citep[see Chapter 26,][]{Molenberghs_Verbeke-2005-Springer} to account for data missingness. 
Under a Missing at Random (MAR) assumption, this approach can handle many missingness patterns, including right-censoring that is common to recurrent event data.

Various statistical methods have been proposed for analyzing multivariate longitudinal data. Generalized estimating equation (GEE) methods \citep[e.g.,][]{liang1986longitudinal,prentice1988correlated,carey1993modelling,gray2000multidimensional} are widely used for drawing statistical inferences on regression parameters relating the means of outcome variables to a set of covariates and parameters characterizing the marginal association between outcome variables. These methods often provide valid statistical inferences on parameters of interest without a need to specify a full joint distribution for the outcome variables. On the other hand, many joint models have been proposed for multivariate longitudinal data that are better at making predictions while still capable of drawing statistical inferences on parameters of interest, though the latter may be jeopardized by model misspecification. Joint models for multivariate longitudinal data include transition models \citep{liang1989class,zeng2007transition} that are specified through a sequence of conditional probabilities of outcome variables given historical outcome variables and covariates, copula-based models \citep{lambert2002copula,smith2010modeling,panagiotelis2012pair} that specify a joint distribution via copulas, and latent variable models  \citep{ten1999mixed,oort2001three,liu2006mixed,Hsieh_etal-2010-MBR,proust2013analysis,wang2016second,ounajim2023mixture,sorensen2023longitudinal} that capture the complex dependence structure by introducing latent variables. Latent variable models are very popular, thanks to their flexibility and interpretability. However, the statistical inference for these traditional latent variable models is carried out based on a marginal likelihood, where the latent variables are treated as random variables and marginalized out. This approach can hardly be extended to the high-dimensional setting with many latent variables due to the high computational cost of optimizing the marginal likelihood. Our method extends the traditional latent variable models to the high-dimensional setting and further overcomes their computational challenge using the proposed joint likelihood estimator. 

The proposed method is also related to high-dimensional factor models for multivariate cross-sectional data or panel data that do not directly apply to the current problem. These models are estimated by minimizing a loss function of both unobserved factors and loading parameters. In other words, although unobserved factors may be regarded as random variables in the model specification, they are conditioned upon and treated as unknown parameters at the estimation stage.  In this direction,  \cite{stock_Watson-2002-JASA} and \cite{Bai_Li-2012-Aos} considered linear factor analysis and proposed estimation methods based on quadratic and likelihood-based loss functions, respectively. \cite{Chen_Li_Zhang-2020-JASA} and \cite{liu2023generalized} considered generalized latent factor models that allow for various data types 
and proposed likelihood-based estimation procedures. Moreover, \cite{Chen_etal-2021-Econometrica}
introduced a quantile factor model for multivariate data and proposed estimators based on the check loss function for quantile estimation. Although \cite{liu2023generalized} and \cite{Chen_etal-2021-Econometrica} established some asymptotic normality results, they focused on factor models without covariates, and their results are not directly applicable to the current setting.

 To summarize, our main contribution is three-fold: First, this study introduces a novel flexible latent variable modeling framework designed to address the analytical challenges inherent in high-dimensional multivariate longitudinal datasets characterized by heterogeneous outcome variables and incomplete observations—a methodological gap that conventional statistical approaches fail to adequately address. This framework accommodates a range of correlation structures by allowing both time-invariant and time-varying regression parameters and factor loadings, as well as structured forms of time-dependent intercepts. Notably, this modeling flexibility is novel compared to the existing high-dimensional factor models that are not tailored to longitudinal settings.
Second, we advance a principled approach for statistical inference concerning covariate effects, accompanied by a thorough exposition of the underlying statistical theory. We specifically provide identifiability conditions for latent variables that ensure the parametric estimability of covariate effects, while establishing a central limit theorem that demonstrates the asymptotic properties—namely, consistency, validity, and efficiency—of our proposed inference methodology. Third, our theoretical framework facilitates statistical inference procedures for generalized latent factor models with integrated covariate effects, representing a previously unexamined methodological paradigm within the statistical literature. This contribution thus possesses intrinsic theoretical merit and may stimulate further research in the development of covariate-adjusted latent variable modeling approaches.

The rest of the paper is organized as follows.  Section \ref{sect: Proposed Method} introduces a factor model for 
high-dimensional longitudinal data and proposes a likelihood-based estimator, along with several extensions and variants. Section \ref{sect: Theoretical results} establishes the theoretical properties of the proposed estimator. Specifically, a central limit theorem is established for statistical inference on regression coefficients, and
an information criterion is introduced for choosing the number of factors. 
The proposed method is evaluated by simulation studies in Section \ref{sect: Simulation Study} regarding its finite sample performance and is further applied to a grocery shopping dataset in Section \ref{sect: Application to The Complete Journey dataset } for understanding and predicting customers’ shopping behavior. 
The paper is concluded with discussions in Section \ref{sect: Discussions}. A software implementation
for R is available at https://github.com/Arthurlee51/LVHML. Further details about the computation and proofs of the theoretical results are given in the online supplementary material.

\section{Proposed Method} \label{sect: Proposed Method}

\subsection{Setting and Proposed Model}

Consider multivariate longitudinal data with $N$ individuals and $J$ outcome variables observed on discrete time points $t = 1, ..., T$. Let $\YY_i=(y_{ijt})_{j = 1, ..., J, t = 1, ..., T}$ be a $J \times T$ data matrix for each individual $i$, where $y_{ijt}$ is a random variable indicating the measurement of the $j$th outcome at time $t$. We further use the vector $\yy_{it} = (y_{i1t}, \dots, y_{iJt})^{\top}$ to denote all of the individual's outcomes at time $t$. Besides the measured outcomes, a set of $p$ covariates are collected for each individual $i$, denoted by $\xx_i = (x_{i1}, ..., x_{ip})^\top$. To facilitate clarity of presentation and methodological exposition, we restrict our initial analysis to the setting in which covariates are static; the theoretical extension to accommodate time-varying covariate structures will be comprehensively addressed in Section \ref{subsect: ext}. Furthermore, to account for missing observations, we let $r_{it}$ be a missing indicator for individual $i$ at time $t$, where $r_{it} = 1$ if $\yy_{it}$ is observed and $r_{it} = 0$ otherwise. We then partition $\YY_i$ into $\YY_i^o$ and $\YY_i^m$, where $\YY_i^o$ contains those $\yy_{it}$ for which $r_{it}=1$ and $\YY_{i}^m$ contains the remaining components. Let $\rr_i = (r_{i1}, \dots, r_{iT})^{\top}$ denote the vector of the individual's missing indicators. We observe independent and identically distributed (i.i.d.) copies of the triplet $\YY_i^o$, $\rr_i$ and $\xx_i$, $i = 1, \dots, N$. 

Within this analytical framework, we develop a high-dimensional factor model designed to achieve two primary methodological objectives. Our first objective involves conducting rigorous statistical inference regarding the  covariates' effects on  individual outcome variables based on the estimated model parameters derived from the training process. The second objective encompasses the development of predictive capabilities, wherein we leverage the up-to-date information to construct a trained model framework for forecasting future realizations of outcome variables $y_{ijt}$ at time point $(T+1)$ across all individuals $i = 1, ..., N$ and outcome dimensions $j=1, ..., J$.

To address the complex dependency structure inherent in longitudinal data, we introduce individual-specific latent variables $\ttt_i = (\theta_{i1}, ..., \theta_{iK})^\top$, commonly referred to as latent factors. These latent constructs serve to model the within-individual correlational patterns and systematic variations that remain unexplained by the observable covariate effects, where $K$ represents the predetermined dimensionality of the factor space. The latent dimension $K$ is assumed to satisfy the condition where $K$ is small relative to $N$ and $J$,  but can still be large in absolute value, ensuring computational tractability.

Suppose that each of $y_{ijt}, i = 1, \dots, N, j = 1, \dots, J, t = 1, \dots, T$, follows an exponential family distribution with natural parameter $\gamma_{jt} + \aa_j^{\top}\thth_i + \bbb_j^{\top}\xx_i$, and possibly a scale dispersion parameter $\phi_j$. Here $\gamma_{jt}$, $\aa_j = (a_{j1}, \ldots, a_{jK})^{\top}$ and $\bbb_j = (\beta_{j1}, \ldots, \beta_{jp})^\top$ are item-specific parameters. Specifically, $\gamma_{jt}$ is a variable- and time-specific intercept capturing the baseline intensity,  
$\aa_j$ is a vector of the loading parameters, and $\bbb_j$ contains the regression coefficients. More precisely, the probability density/mass function for $y_{ijt}$ takes the form 
\begin{align}\label{eq: dist asssmption of data}
    &f(y_{ijt} \mid \gamma_{jt}, \aa_j, \thth_i, \bbb_j, \xx_i, \phi_j)\nonumber \\
    = &\exp\left( \frac{y_{ijt}(\gamma_{jt} + \aa_j^{\top}\thth_i + \bbb_j^{\top}\xx_i )  - b_j(\gamma_{jt} + \aa_j^{\top}\thth_i + \bbb_j^{\top}\xx_i)}{\phi_j}  + c_j(y_{ijt}, \phi_j) \right),
\end{align}
where $b_j(\cdot)$ and $c_j(\cdot)$ are pre-specified variable-specific functions that are determined by the choice of 
the exponential family distribution. This model assumption 
allows us to model outcome variables of mixed types, including binary, count and continuous data. For example, for a binary variable, \eqref{eq: dist asssmption of data} leads to a logistic model where 
\begin{equation}\label{eq:model}
 P(y_{ijt}=1 \mid \gamma_{jt}, \aa_j, \thth_i, \bbb_j, \xx_i) = \frac{\exp( \gamma_{jt} + \aa_j^{\top}\thth_i + \bbb_j^{\top}\xx_i )}{1+\exp(\gamma_{jt} + \aa_j^{\top}\thth_i + \bbb_j^{\top}\xx_i)},
\end{equation}
for which $b_j(\cdot ) = \log( 1 + \exp(\cdot)), \phi_j=1$ and $c_j(\cdot, \cdot) =0.$ For count data, \eqref{eq: dist asssmption of data} gives
\begin{align}\label{eq:possion model}
P(y_{ijt}=y \mid \gamma_{jt}, \aa_j, \thth_i, \bbb_j, \xx_i) = \frac{ \exp( y( \gamma_{jt} + \aa_j^{\top}\thth_i + \bbb_j^{\top}\xx_i ) - \exp(\gamma_{jt} + \aa_j^{\top}\thth_i + \bbb_j^{\top}\xx_i ))  }{y!},
\end{align}
a Poisson model for which $b_j(\cdot ) = \exp(\cdot) , \phi_j=1$ and $c_j(y, \phi_j) = - \log(y !) .$  

For each individual $i$, we assume that $y_{ijt}$s are conditionally independent given the latent variables $\thth_i$ and covariates $\xx_i$. Furthermore, we assume the missing outcome variables to be MAR, such that the missing indicator $\rr_i$ is conditionally independent of the unobserved data $\YY_i^m$ given the observed data $Y_i^o$.  
We denote $\AAA = (a_{jk})_{J\times K}$ as the loading matrix, $\ThTh = (\theta_{ik})_{N\times K}$ as the matrix for factor scores, and $\XX = (x_{il})_{N\times p}$ as the covariate matrix. 

To ensure the identifiability of the regression coefficients $\bbb_j$, we impose the restriction 
\begin{align}\label{eq: normalization for beta}
\ThTh^{\top} \XX = \mathit{0}_{K \times p},
\end{align}
where  $\mathit{0}_{K \times p}$ is a $K \times p$ matrix with all the entries being zero. This constraint requires the latent factors to be uncorrelated with the observed covariates,  coinciding the assumption in traditional random effects models that random effects and regressors are orthogonal.

We present several theoretical observations regarding the proposed modeling framework. First, it is assumed that the within-individual dependence, both between items and across time, is completely captured by covariates $\xx_i$ and low-dimensional factors $\ttt_i$. Conditioning on these variables, the outcome variables are independent, and the right-hand side of \eqref{eq: dist asssmption of data} does not depend on the outcomes at other time points.  Second, one can regard the latent variables $\ttt_i$ as random effects capturing unobserved within-individual heterogeneity. Conventional latent variable modeling frameworks for multivariate longitudinal data characteristically require the specification of restrictive parametric distributional assumptions—most commonly normality—concerning the underlying latent variable structure. Statistical inference procedures in these traditional approaches subsequently rely upon marginal likelihood computations wherein the latent variables are analytically or numerically integrated out from the joint distribution, thereby yielding parameter estimates based solely on the observed data likelihood. While this approach works well for low-dimensional latent variable models, it becomes computationally challenging when the latent dimension becomes moderately large (Chapter 6, \citealp{skrondal2004generalized}) and unsuitable for the current setting where $K$ can be large. In this work, we adopt an approach commonly applied in high-dimensional factor models \citep{Chen_Li_Zhang-2020-JASA,Chen_etal-2021-Econometrica,liu2023generalized}, optimizing an objective function involving both the fixed parameters such as $\gamma_{jt}$, $\aa_j$ and $\bbb_j$ and the latent variables $\ttt_i$, without requiring distributional assumption on the latent variables. Third, except for $\bbb_j$, the remaining unknown parameters in \eqref{eq: dist asssmption of data} are not identifiable without additional constraints. For example, one can add a constant to each entry of $\Theta$ and compensate it by adjusting the intercepts $\gamma_{jt}$, without changing the density/probability \eqref{eq: dist asssmption of data}. It is important to acknowledge that analogous indeterminacy issues constitute a fundamental characteristic of factor analytic methodologies, and the identifiability of these parameters can be rigorously established through the implementation of appropriate normalization constraints analogous to those proposed in, for instance, \cite{Bai_Li-2012-Aos}. This indeterminacy does not affect making predictions but affects interpretation of the factors and inference of the corresponding loading parameters. As we are mainly interested in drawing statistical inferences on the regression coefficients, we do not impose constraints to fix the rotational indeterminacies. As will be rigorously demonstrated in Section \ref{subsect: Estimation and Prediction}, the proposed estimator for $\bbb_j$ exhibits both consistency and asymptotic normality properties, irrespective of the identifiability status of the remaining model parameters.

Finally, although the covariates, latent variables, and most of the parameters are assumed to be time-independent in \eqref{eq: dist asssmption of data}, we can extend our model to allow them to be time-dependent. Some of such extensions are discussed in Sections~\ref{subsect: model extensions} and \ref{sect: Discussions}, respectively. However, we should note that these extensions also introduce more model parameters, which may lead to a higher variance in prediction and additional challenges with interpretations.

\subsection{Estimation}\label{subsect: Estimation and Prediction}

We consider the estimation of the proposed model 
based on the joint log-likelihood function
\begin{equation}\label{eq:loglik}
 l(\XiXi) = \sum_{i=1}^N\sum_{j=1}^J\sum_{t=1}^T r_{it} \left\{y_{ijt}(\gamma_{jt} + \aa_j^{\top}\thth_i + \bbb_j^{\top}\xx_i )  - b_j(\gamma_{jt} + \aa_j^{\top}\thth_i + \bbb_j^{\top}\xx_i)\right\},   
\end{equation} 
where $\XiXi$ is a vector containing unknown quantities including $\gamma_{jt}, \aa_{j}, \bbb_{j}$ and $\ttt_i$, for $i=1, \dots, N, j=1, \dots, J, t = 1, \dots, T$. 
To estimate the regression coefficients $\bbb_j$, we maximize $l(\XiXi)$  with respect to $\XiXi$ under certain compactness constraints on the model parameters. 
More specifically, let $\|\cdot\|$ denote the Euclidean norm, we solve the optimization problem 
\begin{equation}\label{eq: estimator}
\begin{aligned}
\hat{\XiXi} &= \argmax_{\XiXi} l(\XiXi)  \text{ such that} \\
     & 
     ~  \Vert\ttt_i\Vert  \leq c_1 \sqrt{K}, \mbox{and~} \Vert(\aa_j^\top, \ggg_j^\top,\bbb_j^\top)^\top\Vert \leq c_2 \sqrt{T+p+K}, i = 1,\dots, N, j=1, \dots , J,
\end{aligned}   
\end{equation} 
where $c_1$ and $c_2$ are two constraint parameters, and $\ggg_j = (\gamma_{j1}, ..., \gamma_{jT})^\top$. Numerically, the compactness constraints prevent parameters from taking extreme values, which may happen when observed variables are discrete and certain categories are rarely observed. Theoretically, this constraint plays a crucial role in establishing the estimation consistency;  see Section \ref{sect: Theoretical results} for the details.  
This optimization problem is solved by a projected gradient descent algorithm \citep[Chapter 4,][]{bertsekas1999nonlinear} that is guaranteed to converge to a critical point. See Section \ref{app: Estimation Method} in the online supplementary material for the computational details.  

We designate the expression in \eqref{eq:loglik} as a joint log-likelihood function to establish a clear  distinction from the marginal log-likelihood formulations adopted in conventional latent variable modeling frameworks. This nomenclature reflects the fundamental characteristic that our proposed log-likelihood function explicitly incorporates both the structural model parameters and the latent factor realizations as estimable quantities, contrasting with traditional approaches that marginalize out the latent variables.

We shall acknowledge that, from a theoretical standpoint, the complete joint log-likelihood function assumes a formulation that differs subtly from the simplified representation $l(\XiXi)$ presented in \eqref{eq:loglik}. Specifically, the complete joint log-likelihood takes the form $\sum_{i=1}^N\sum_{j=1}^J\sum_{t=1}^T r_{it}  \log f(y_{ijt} \mid \gamma_{jt}, \aa_j, \thth_i, \bbb_j, \xx_i, \phi_j)$. The discrepancy arises because our formulation $l(\XiXi)$ omits the scale parameters $\{\phi_j\}_{j=1}^J$ from the likelihood specification. Notably, these two formulations are asymptotically equivalent (differing only by an additive constant) when the response variables are exclusively binary or count data that adhere to the Bernoulli or Poisson distributional specifications outlined in \eqref{eq:model} and \eqref{eq:possion model}, respectively.

The proposed estimator is suitable when all the scale parameters are close to each other. In scenarios characterized by substantial scale parameter heterogeneity, we recommend using the complete joint log-likelihood function to facilitate simultaneous estimation of both $\XiXi$ and the entire collection of scale parameters $\{\phi_j\}_{j=1}^J$. The asymptotic theoretical properties established in Section~\ref{sect: Theoretical results} can be appropriately extended to accommodate this more comprehensive estimation framework through straightforward analytical adaptations.

\subsection{Model Extensions}\label{subsect: model extensions}
\subsubsection{Extension to Incorporating Time-dependent Covariates}\label{subsect: ext}
In scenarios where each individual $i$ is associated with a time-dependent covariate vector $\zz_{it} = (z_{i1t}, z_{i2t}, \ldots, z_{ip_{z}t})^{\top}$ at each time point $t$, with corresponding regression parameters $\vv_j = (v_{j1}, \ldots, v_{jp_{z}})^{\top}$, our model adapts accordingly. The natural parameter of the exponential family distribution can be modeled as $\gamma_{jt} + \aa_j^{\top}\thth_i + \bbb_j^{\top}\xx_i + \vvv_j^{\top}\zz_{it}$. The conditional probability density/mass function $f(y_{ijt} \mid \gamma_{jt}, \aa_j, \thth_i, \bbb_j, \xx_i, \vvv_j, \zz_{it}, \phi_j)$ then becomes 
$
  \exp\left( \phi_j^{-1} \{y_{ijt}(\gamma_{jt} + \aa_j^{\top}\thth_i + \bbb_j^{\top}\xx_i  + \vvv_j^{\top}\zz_{it} )  - b_j(\gamma_{jt} + \aa_j^{\top}\thth_i + \bbb_j^{\top}\xx_i+\vvv_j^{\top}\zz_{it})\}+ c_j(y_{ijt}, \phi_j) \right).
$

This modification maintains the structure of the likelihood as in equation \eqref{eq:loglik}, and, thus, the estimation algorithm described in Section \ref{subsect: Estimation and Prediction} can still be applied.  By incorporating additional assumptions for time-dependent covariates, we can derive theorems akin to
those established under the current model in Section~\ref{sect: Theoretical results}. The specific assumptions and proofs of these theorems are elaborated in Sections~\ref{app: Assumptions and Additional Theoretical Results}~and~\ref{app: Proof} in the online supplementary material. 

\subsubsection{Extension for Time-dependent Loadings and Coefficients}\label{subsect: ext 2}
To better accommodate the effects of time in complex datasets, the loadings and the coefficients of the covariates may also be made time-dependent by modelling the natural parameter as $\gamma_{jt} + \aa_{jt}^{\top}\thth_i + \bbb_{jt}^{\top}\xx_i$. Similar to the extension discussed in Section \ref{subsect: ext}, we could also include time-dependent covariate $\ZZ_{t}$, and the estimation procedure outlined in Section~\ref{subsect: Estimation and Prediction} can be adapted to incorporate this extension. Theoretical results analogous to those in Section~\ref{sect: Theoretical results} can be established. The required modifications, assumptions and proofs are detailed in Sections~\ref{app: Estimation Method}, \ref{app: Assumptions and Additional Theoretical Results} and~\ref{app: Proof} of the online supplementary material, respectively.

\subsubsection{Imposing Dependence Structure on Intercepts}\label{subsect: gamma_structure}
In practical data analysis, it is often desirable to impose structured dependence on \(\ggg_j\) to enhance estimation efficiency and predictive accuracy, or reflect prior knowledge. An important example is \(\gamma_{jt} = t\gamma_j\), which fits naturally under the framework developed in Section~\ref{subsect: ext}, treating \(\gamma_j\) as the coefficient for the time-dependent covariate \(t\). Consequently, the asymptotic results established for that framework continue to hold.

This structure can also be incorporated into the extension in Section~\ref{subsect: ext 2}, with additional assumptions required for valid asymptotic theory. These conditions and related proof adjustments are provided in Sections~\ref{app: subsect ext 2} and~\ref{app: Proof} of the online supplementary material.

\section{Theoretical Results}\label{sect: Theoretical results}
\subsection{Consistency and Asymptotic Normality}
We now establish the asymptotic properties for the estimated regression coefficients $\hat{\bbb}_j$. Let $\uu_j^* = \left({\ggg^*_j}^{\top},{\bbb^*_j}^{\top},{\aa^*_j}^{\top} \right)^{\top}$ denote the vector of true values of item-specific parameters. Additionally, define $\DD_{it} = (D_{it1}, \dots, D_{itT})^{\top}$ as a vector of dummy variables indicating the time periods, where $D_{itt}=1$ and $D_{itt^{'}}=0$ for $t \neq t^{'}, i = 1 \dots N$. We further define $\ee^*_{it} = (\DD_{it}^{\top}, \xx_i^{\top} ,{\thth^*_i}^{\top} )^{\top}$ as the vector of true and observed individual-specific quantities. Let $K^*$ denote the true dimension of the latent variables $\thth^*_i$, and $P = T+p+K^*$ denote the dimension of $\uu_j^*$. $P$ is assumed to be fixed that does not vary with $N$ and $J$. Let $\XiXi^* = \left({\uu^*_1}^{\top}, \dots, {\uu^*_J}^{\top}, {\thth^*_1}^{\top}, \dots, {\thth^*_{N}}^{\top}\right)^{\top}$ denote the vector of true parameters.
 Let $\mathcal{U} \subset \mathbb{R}^{P}$, $\mathbf{\Theta} \subset \mathbb{R}^{K^*} $ and define the space of possible parameters
            $\mathcal{H}^{K^*} =
            \left\{\XiXi \in \mathbb{R}^{NK^* + PJ}: \uu_j \in \mathcal{U}, \thth_i \in  \mathbf{\Theta} \text{ for all } i,j,\text{ }
            \ThTh^{\top} \XX = \mathit{0}_{K^* \times p}\right\}.$
For positive sequences ${a_n}$ and ${b_n}$, we write $a_n \lesssim b_n$ if $a_n \leq C b_n$ for some $C > 0$, and $a_n \asymp b_n$ if both $a_n \lesssim b_n$ and $b_n \lesssim a_n$. The following regularity conditions ensure consistency of $\hat{\bbb}_j$.

\begin{assump}\label{assp:1}
 $\mathcal{U}$ and $\mathbf{\Theta} $ are compact sets and $\XiXi^* \in \mathcal{H}^{K^*}$. Moreover, $\xx_i \in \mathcal{X}$ for all $i$, where $\mathcal{X} \subset \mathbb{R}^{p}$ is a compact set.
\end{assump}

\begin{assump}\label{assp: 1.5}
   For any compact set $\mathcal{C} \subset \mathbb{R}$, there exists $\bar{b}>\underline{b} >0$ (depending on $\mathcal{C}$) such that $\bar{b} \geq b_j^{''}(s) \geq \underline{b}$ and $|b^{'''}_j(s)|\leq \bar{b}$ for all $s \in \mathcal{C}, j = 1, \dots, J$. Moreover, $\{\phi_j\} \lesssim 1$.
\end{assump}
\begin{assump}\label{assp:2}
    $J^{-1}{\AAA^*}^{\top}\AAA^*$ converges to a positive definite matrix as $J$ tends to infinity. Also, $N^{-1}{\ThTh^*}^{\top}\ThTh^*$ converge to a positive definite matrix as $N$ tends to infinity.
\end{assump}
\begin{assump}\label{assp:3}
    There exists $\kappa_1>0$ such that $\inf_{i=1,\dots, N,t=1, \dots, T}P(r_{it} =1) \geq \kappa_1$. 
\end{assump}

\begin{assump}\label{assp:4}
   There exists $\kappa_2>0$ such that 
        $\liminf_{N \to \infty} \pi_{\min}\left(\left(\XX, \mathbf{1_{N}} \right)^{\top}\left(\XX, \mathbf{1_{N}} \right)\right)/N \geq \kappa_2,$
    where  $\pi_{\min}(\cdot)$ is the minimum eigenvalue of a matrix, $\mathbf{1_{N}}$ is  length-$N$ vector of ones. 
\end{assump}

The following theorem establishes the consistency of $\hat{\bbb}_j$:
\begin{theorem}\label{thm: consistency}
  Under Assumptions \ref{assp:1} to \ref{assp:4},
     $ \Vert \hat \bbb_j - \bbb_j^* \Vert = o_P\left(1\right)$, as $N$ and $J$ grow to infinity.
\end{theorem}

Define $\BB^* = (\beta^*_{jl})_{J \times p}$ and $\GG^*_{t} = (\gamma^*_{1t}, \dots, \gamma^*_{Jt})^{\top}$  for $t = 1, \dots, T$. Furthermore, let $\hat{\ThTh} = (\hat{\theta}_{ik})_{N \times K^*},
 \hat{\AAA} = (\hat{a}_{jk})_{J \times K^*},$
 $ \hat{\BB} = (\hat{\beta}_{jl})_{J \times p}$ and $\hat{\GG}_{t} = (\hat{\gamma}_{1t}, \dots, \hat{\gamma}_{Jt})^{\top}, t = 1, \dots, T$  be the estimated parameters from $\hat{\XiXi}$. The following theorem provides the average rate of convergence of $\hat{\XiXi}$ and $\hat{\BB}$:
\begin{theorem}\label{Thm: convergence of Betas } 
Under Assumptions \ref{assp:1} to \ref{assp:4}, we have 
\begin{align} 
\max_{t = 1, \dots, T} \frac{\left\| \hat{\Theta}\hat{A}^{\top} - \Theta^* {A^*}^{\top} + X(\hat{B} -B^*) ^{\top} + \mathbf{1_{N}}(\hat{\GG}_{t} -\GG^*_{t}) ^{\top} \right\|_{F}}{\sqrt{NJ}}  &= O_P(min\{\sqrt{N}, \sqrt{J}\}^{-1})\label{eq:Xi loss},\\
 \frac{1}{\sqrt{J}} \left\| \hat{B}^{\top} - {B^*}^{\top} \right\|_{F} &= O_P(min\{\sqrt{N}, \sqrt{J}\}^{-1}).\label{eq: Bloss}
\end{align}
\end{theorem}
We comment on the rate of convergence for $ J^{-1/2} \| \hat{B}^{\top} - {B^*}^{\top}\|_{F}$. One might expect a $N^{-1/2}$ rate given that the parameter matrix $B$ represents the regression coefficients associated with directly observable covariates. However, our analytical framework necessitates the simultaneous estimation of latent variable components, thereby introducing an additional source of statistical uncertainty that can be seen as measurement error. Specifically, the estimated latent component $\hat{\Theta}\hat{A}^{\top}$ has an estimation error rate of $(NJ)^{-1/2}\left\| \hat{\ThTh}\hat{A}^{\top}  - \ThTh^* {A^*}^{\top}\right\|_{F} = O_P(min\{\sqrt{N}, \sqrt{J}\}^{-1}) $. This measurement error dominates the estimation error of the regression component, resulting in the convergence rate stated in Theorem \ref{Thm: convergence of Betas }.
To establish the asymptotic normality for each $\hat \bbb_j$, we need two additional assumptions. 
\begin{assump} \label{assp:5}
The limits 
    $\PhPh_{j} =\lim_{N \to \infty} \frac{1}{N}\sum_{i=1}^{N}\sum_{t=1}^{T} -\phi_j^{-1}E\left(r_{it}\right) b_j^{''}({\uu^*_{jt}}^{\top}\ee^*_{it}) \ee^*_{it}{\ee^*_{it}}^{\top}$ and 
    $\PsPs_{i} =\lim_{J \to \infty} \frac{1}{J}\sum_{j=1}^{J}\sum_{t=1}^{T} -\phi_j^{-1}E\left(r_{it}\right) b_j^{''}({\uu^*_{jt}}^{\top}\ee^*_{it})\aa^*_{j}{\aa_{j}^*}^{\top}$
exist for $i= 1, \dots, N$ and $j = 1, \dots, J$. Moreover, there exists $\kappa_3>0$ such that $\pi_{\min}(\PhPh_{j}^{\top}\PhPh_{j}) \geq \kappa_3 $ and $\pi_{\min}(\PsPs_{i}^{\top}\PsPs_{i}) \geq \kappa_3 .$

\end{assump}
\begin{assump} \label{assp:6}
    As $N, J \to \infty$, $N \asymp J$.  
\end{assump}
     
\begin{theorem}\label{thm: Normality}
  Under Assumptions \ref{assp:1} to \ref{assp:6} , we have 
            $\sqrt{N}\left(\hat{\bbb}_j - \bbb_j^* \right) \overset{d}{\to} \mathcal{N}\left(0, \SigSig_{\EE,j} \right),$
  where the asymptotic variance $\SigSig_{\EE,j} = (-\PhPh_j^{-1})_{(T+1):(T+p),(T+1):(T+p)}$ is a submatrix of $-\PhPh_j^{-1}$ that corresponds to its $T+1$ to $(T+p)$th rows and columns.  $\SigSig_{\EE,j}$ is uniquely determined by the true model without being affected by the indeterminacy of $\gamma_{jt}$, $\aa_j$, and $\ttt_i$. 
\end{theorem}
Theorem \ref{thm: Normality} establishes the asymptotic normality of \(\hat{\bbb}_j\), and specifies the form of the asymptotic variance. This shows that $\hat{\bbb}_j$ is efficient, as its asymptotic variance matches the maximum likelihood estimator in generalized linear model regression, where the latent factors $\thth_i^*$ are directly observable.
Assumptions \ref{assp:1}, \ref{assp:2} and \ref{assp:5} are standard in the literature of factor analysis (see e.g., \citealp{Bai_Ng-2002-Econometrica} and \citealp{bai-2003-Econometrica}). Assumption \ref{assp: 1.5} concerns the regularity conditions of the exponential family. The condition regarding the derivatives of $b_j(\cdot)$ is straightforward to verify and applies to a wide range of commonly used models under the exponential family, including the logistic model for binary data and Poisson model for count data. The condition for the scale parameter is a mild assumption ensuring the variance of \( y_{ijt} \) does not explode. Assumptions \ref{assp:3} and \ref{assp:4} are mild conditions that provide lower bounds for the probability of observing the data and guarantee a degree of variability in the values of \(\XX\), respectively. Assumption \ref{assp:6} ensures that \(N\) and \(T\) grow at the same rate. 
Similar assumptions are made when deriving asymptotic normality for high-dimensional factor models; see e.g., \cite{Bai_Li-2012-Aos},
\cite{Galvao_Kato-2016-JE} and \cite{Chen_etal-2021-Econometrica}.  

\begin{remark}\label{rmk: Asymptotic variance estimation}
    In practice, asymptotic variance is unknown and needs to be estimated. Define $\hat{\ee}_{it} = (\DD_{it}^{\top}, \xx_i^{\top} ,{\hat{\thth}_i}^{\,\top} )^{\top}$. We can estimate $\PhPh_j$ by 
    $ \hat{\PhPh}_{j} = N^{-1}\sum_{i=1}^{N}\sum_{t=1}^{T} -\hat{\phi}_j^{-1}r_{it} b_j^{''}(\hat{\uu}_j^{\top}\hat{\ee}_{it} ) \hat{\ee}_{it}{\hat{\ee}_{it}}^{\top},$
    and estimate $\hat{\SigSig}_{\EE,j}$ by the corresponding submatrix. We show in Section \ref{subsect: Proof of consistency results for asymptotic variance } in the online supplementary material that $\hat{\SigSig}_{\EE,j}$ is a consistent estimator for the true asymptotic variance $\SigSig_{\EE,j}$, where $\hat{\phi}_j$ is any consistent estimator of the scale parameter $\phi_j$.  
\end{remark}

\begin{remark}\label{rmk: uniform convergence rate}
    Under the conditions of Theorem \ref{thm: Normality}, the established asymptotic normality for each \( j \in \{1, \dots, J\} \) implies a uniform convergence rate of $\max_{1\leq j\leq J}\|\hat{\bbb}_j - \bbb_j^*\| = O_P\left(N^{-1/2}\sqrt{\log J}\right).$ This result follows by applying a union bound over $j = 1, \dots, J $ to the sub-Gaussian tail probabilities of $\hat{\beta}_{jl} - \beta^*_{jl}$ for $l = 1, \dots, p$, noting that $p$ is assumed fixed.
\end{remark}

\subsection{Determining the Number of Factors}\label{subsect: Determining the Number of Factors}
In real application, the true number of factors $K^*$ is unknown and, thus, needs to be estimated. To do so, we consider a finite set $\mathcal K$, containing the candidate numbers of factors. For each value of $K \in \mathcal K$, we estimate the proposed model and obtain the estimate $\hat{\XiXi}_K$ and the corresponding log-likelihood function value, $l(\hat{\XiXi}_K)$. We then construct an information criterion taking the form 
     $\text{IC}(K) = -2l(\hat{\XiXi}_K ) + K \Lambda_{NJ},$
where $\Lambda_{NJ}$ is a penalty term to be discussed in the sequel. We then set 
\begin{equation}\label{eq:IC}
    \hat K = \argmin_{K \in \mathcal K} \text{IC}(K).
\end{equation}

\begin{theorem}\label{Thm: selection of factors}
  Suppose that Assumptions \ref{assp:1} to \ref{assp:4} hold and $K^* \in \mathcal K$. If the penalty term $\Lambda_{NJ}$ satisfies  $\max{\left\{N, J\right\}} \lesssim \Lambda_{NJ} \lesssim NJ $, then  
  $\lim_{N, J \rightarrow \infty} P(\hat{K} = K^*) = 1.$  
\end{theorem}

This result is an extension of an information criterion for a generalized latent factor model proposed by \cite{chen_Li-2022-Biometrika} to the current model. Following the choice in \cite{chen_Li-2022-Biometrika}. we set
     $\Lambda_{NJ} = {\max\{N,J\}} \times \log\left(\max\{N,J\}^{-1}J\sum_{i=1}^{N}\sum_{t=1}^{T} r_{it} \right) $
in implementation, where $J\sum_{i=1}^{N}\sum_{t=1}^{T} r_{it}$ records the total number of data points being observed. It is easy to see that the requirement on $\Lambda_{NJ}$ is satisfied with this choice. 

\section{Simulation Study}\label{sect: Simulation Study}

\subsection{Simulation Setting}
We assess the finite sample performance of the proposed method via Monte Carlo simulations under a variety of settings. Specifically, we consider $J = 100, 200, 300, 400$ with $N = 5J$ or $10J$, yielding eight combinations. For each setting, we generate 100 replications with the number of time points $T = 4$ and true latent dimensions $K^* = 3$ and $8$. Model selection is performed over the candidate set $\mathcal{K} = \{1, 2, \dots, 10\}$.

We simulate data in a binary response setting to mimic the real data example. The simulated data follows the logistic model given in \eqref{eq:model}:
\begin{equation}\label{eq: sim logistic}
 P(y_{ijt}=1 \mid \gamma_{jt}, \aa_j, \thth_i, \bbb_j, \xx_i) = \frac{\exp( \gamma_{jt} + \sum_{k=1}^{K^*}a_{jk}\theta_{ik} + \sum_{l=1}^{5} \beta_{jl}x_{il}  )}{1+\exp(\gamma_{jt} + \sum_{k=1}^{K^*}a_{jk}\theta_{ik} + \sum_{l=1}^{5} \beta_{jl}x_{il} )}.
\end{equation}
The variables are generated according to the following procedure, with a minor notational inconsistency arising from the use of identical symbols both prior to and following normalization. Intercepts $\gamma_{jt}$ are sampled from a uniform distribution $U[-1, 1]$, and regression coefficients $\beta_{jl}$ from $U[0.5, 1]$. The latent variables $\theta_{ik}$ and $a_{jk}$ are sampled from truncated standard normal distributions on $[-3, 3]$. The covariates ($x_{i1}$, $x_{i2}$) and ($x_{i3}$,$x_{i4}$) are two pairs of dummy variables, each derived independently from a binomial distribution $\text{Bin}(2,0.5)$. The last covariate $x_{i5}$ is sampled from $U[-1,1]$. The normalization procedures described in Section~\ref{app: Normalization algorithm} of the online supplementary material is applied to ensure identifiability of regression coefficients. We then independently set half of each normalized coefficients pair $(\beta_{j1}, \beta_{j2})$ and $(\beta_{j3}, \beta_{j4})$ to zero, as well as half of the $\beta_{j5}$ coefficients. The missingness indicator $\rr_{i}$ is sampled from all possible binary combinations of $0$ and $1$ with equal probability, excluding the all-zero case, resulting in approximately 47\% of the values in $r_{it}$ being $0$. 

\subsection{Evaluation Criteria}\label{subsect: Evaluation Criteria}
The performance of the proposed estimator is assessed based on several performance metrics, as given in \hyperref[tab: result under proposed]{Table \ref{tab: result under proposed}}. Specifically, in each replication, at the true number of factors $K^*$, we compute the ``Loss'' metric defined on the left-hand side of \eqref{eq:Xi loss} to evaluate the convergence of $\hat{\XiXi}$ in finite sample. Additionally, we compute ``Bloss'' as defined on  the left-hand side of \eqref{eq: Bloss} to quantify the convergence of $\hat{B}$. The mean ``Loss'' and ``Bloss'' across 100 simulations are reported in \hyperref[tab: result under proposed]{Table \ref{tab: result under proposed}}. 

To further assess the estimator's performance on individual parameters, the mean squared error (MSE) for each $\beta_{jl}$, where $j= 1, \dots, J$ and $l = 1, \dots, 4$, is computed across all trials. The maximum of these MSE values is reported as ``MMSE''. Additionally, the proportion of instances where the correct number of factors is accurately identified is denoted by $P(\hat{K}= K^*)$. The asymptotic variance for each simulation is estimated following the methodology proposed in Remark \ref{rmk: Asymptotic variance estimation}, based on which 95\% confidence intervals for $\beta_{jl}$s are constructed. The empirical coverage probability (ECP) is then determined by aggregating the coverage probabilities across all parameters and simulation repetitions.

In addition, the recovery of the coefficients $\bbb_j$s is evaluated against a baseline approach that assumes no factors, that is, $K=0$. The corresponding likelihood is given by 
     $\prod_{i=1}^N \prod_{j=1}^{J} \prod_{t=1}^T   \left\{\exp((\gamma_{jt} + \boldsymbol{\beta}_j^{\top}\mathbf{x}_i)y_{ijt} )(1+\exp(\gamma_{jt} + \boldsymbol{\beta}_j^{\top}\mathbf{x}_i))^{-1}\right\}^{r_{it}}.$
The optimization is carried out using the \texttt{glm} function in R, leveraging a logistic regression (LR) approach. Additionally, we compare the proposed method to a logistic regression model incorporating a random intercept $\alpha_{ij}$, where for each $j$, the $\alpha_{ij}$ are assumed to be identically and independently distributed as normal random variables. The likelihood for this model is
     $ \prod_{i=1}^N \prod_{j=1}^{J} \prod_{t=1}^T \left\{\exp((\gamma_{jt} + \alpha_{ij} + \boldsymbol{\beta}_j^{\top}\mathbf{x}_i)y_{ijt})(1+\exp(\gamma_{jt} + \alpha_{ij} + \boldsymbol{\beta}_j^{\top}\mathbf{x}_i) )^{-1}\right\}^{r_{it}}.$
This model is optimized using the \texttt{glmer} function from the \texttt{lme4} package in R. The results, including the metrics ``Bloss'' and ``MMSE'' from both approaches, are reported in \hyperref[tab: result under proposed]{Table \ref{tab: result under proposed}}.

Furthermore, when drawing statistical inferences on a large set of regression coefficients, it is imperative to account for multiple testing. We present the average false discovery rate (FDR) across 100 simulation runs, computed using the Benjamini–Yekutieli (BY) procedure \citep{Benjamini_Yejutieli-2001-Aos}, which maintains validity under arbitrary dependence structures among the hypotheses. Since $x_{i1}$, $x_{i2}$ are dummy variables for a single covariate, we test the hypotheses $H_{0j}: \beta_{j1} = \beta_{j2} = 0$ for $j = 1, \dots, J$. The Wald test is applied using the estimator for asymptotic variance derived in Theorem~\ref{thm: Normality}, as detailed in Remark~\ref{rmk: Asymptotic variance estimation}. We reject hypotheses at a significance level of $0.05$ based on the BY-adjusted $p$-values. The same procedure is applied to the coefficients associated with $x_{i3}$ and $x_{i4}$. For the continuous covariate $x_{i5}$, we test $H_{0j}: \beta_{j5} = 0$ for each $j = 1, \dots, J$. For each covariate, we compute the mean FDR (MFDR) across the 100 replications and report the maximum as ``MMFDR'' in Table~\ref{tab: result under proposed}. In addition, we report the maximum of the mean false non-discovery rates (MMFNR), as the proportion of true alternative hypotheses that are incorrectly not rejected, among all hypotheses not rejected. 
The results in Table~\ref{tab: result under proposed} are based on setting the constraint parameters in Equation~\eqref{eq: estimator} to $c_1 = c_2 = 5$. To assess sensitivity, we repeat the simulations with $c_1 = c_2 = c$ for $c \in \{3, 4, 5, 6, 7\}$. Results are shown in Table~\ref{tab: tuning_sensitivity_combined} of the online supplementary material.
\subsection{Results}
The simulation results align with the theory for the proposed method. As $N$ and $J$ increase, the metrics ``Loss'', ``Bloss'', and ``MMSE'' under the proposed method show a decreasing trend, which occurs regardless of the number of factors and the ratio between $N$ and $J$. Moreover, the information criterion for determining the true number of factors $K^*$, introduced in Section \ref{subsect: Determining the Number of Factors}, proves to be effective. This is evidenced by $P(\hat{K} = K^*)$ achieving 1 in every scenario, which means our method always identifies the correct number of factors.
Additionally, as $N$ and $J$ increase, the empirical coverage probability (ECP) approaches the nominal 95\% confidence interval level. This validates our asymptotic normality results in Theorem~\ref{thm: Normality} and the asymptotic variance estimator in Remark \ref{rmk: Asymptotic variance estimation}. Furthermore, the metric “MMFDR” consistently remains below the 0.05 significance threshold across all experimental conditions, thereby confirming the efficacy of the Benjamini–Yekutieli (BY) procedure in controlling the false discovery rate. Additionally, the “MMFNR” metric exhibits a decreasing trend with increasing values of $N$ and $J$, indicating that the proposed methodology achieves asymptotically high statistical power. 

In comparison, the ``Bloss'' and ``MMSE'' metrics under the logistic regression (LR) method are consistently higher than those of the proposed method. In addition, 
they do not further improve as $N$ and $J$ increase when they are sufficiently large, suggesting that this simplified model suffers from a large bias. In contrast, the logistic regression model with a random intercept (LRRI) outperforms the basic LR model by accounting for the effects of unobserved random intercepts. However, our proposed method continues to demonstrate superior performance as $J$ and $N$ increase, highlighting the importance of considering correlations among different outcomes to achieve optimal results. 

Finally, as demonstrated in Table~\ref{tab: tuning_sensitivity_combined}, the proposed estimator exhibits consistent performance across varying selections of the constraint parameter $c$. This stability indicates that the method is robust to the specification of constraint values. Accordingly, we adopt
$c_1 = c_2 = 5$ in Section \ref{sect: Application to The Complete Journey dataset }.

\begin{table}[!t]
\centering
\tiny
\caption{Summary statistics for the simulation study. The results for the proposed method, logistic regression (LR), and logistic regression with a random intercept (LRRI) across different combinations of $N$, $K^*$ and $J$ are reported.}
\begin{tabular}{llrrrrrrrrrrrrrr}
\toprule
\multicolumn{2}{l}{} & \multicolumn{7}{c}{\textbf{Proposed}} && \multicolumn{2}{c}{\textbf{LR}} && \multicolumn{2}{c}{\textbf{LRRI}} \\
\cmidrule(lr){3-9} \cmidrule(lr){11-12} \cmidrule(lr){14-15}
$N$ & $J$ & Loss & $P(\hat{K}=K^*)$ & ECP & MMFDR & MMFNR & Bloss & MMSE && Bloss & MMSE && Bloss & MMSE \\
\midrule
\multicolumn{15}{l}{\textbf{$K^* = 3$}} \\
5J  & 100 & 0.55 & 1 & 0.94 & 0.01 & 0.20 & 0.49 & 0.14 && 0.545 & 0.512 && 0.485 & 0.137 \\
5J  & 200 & 0.36 & 1 & 0.95 & 0.01 & 0.05 & 0.32 & 0.05 && 0.453 & 0.426 && 0.338 & 0.099 \\
5J  & 300 & 0.29 & 1 & 0.95 & 0.00 & 0.00 & 0.25 & 0.03 && 0.427 & 0.407 && 0.268 & 0.104 \\
5J  & 400 & 0.24 & 1 & 0.95 & 0.00 & 0.02 & 0.22 & 0.03 && 0.398 & 0.328 && 0.237 & 0.056 \\
10J & 100 & 0.48 & 1 & 0.94 & 0.01 & 0.01 & 0.33 & 0.06 && 0.469 & 0.400 && 0.333 & 0.063 \\
10J & 200 & 0.32 & 1 & 0.95 & 0.01 & 0.00 & 0.22 & 0.03 && 0.409 & 0.380 && 0.233 & 0.047 \\
10J & 300 & 0.26 & 1 & 0.95 & 0.00 & 0.00 & 0.18 & 0.01 && 0.399 & 0.388 && 0.194 & 0.030 \\
10J & 400 & 0.22 & 1 & 0.95 & 0.00 & 0.00 & 0.16 & 0.01 && 0.418 & 0.379 && 0.185 & 0.059 \\
\midrule
\multicolumn{15}{l}{\textbf{$K^* = 8$}} \\
5J  & 100 & 1.26 & 1 & 0.91 & 0.02 & 0.24 & 0.64 & 0.27 && 0.700 & 0.869 && 0.614 & 0.371 \\
5J  & 200 & 0.68 & 1 & 0.94 & 0.01 & 0.10 & 0.39 & 0.09 && 0.671 & 0.787 && 0.432 & 0.245 \\
5J  & 300 & 0.52 & 1 & 0.94 & 0.01 & 0.07 & 0.31 & 0.06 && 0.642 & 1.059 && 0.366 & 0.403 \\
5J  & 400 & 0.44 & 1 & 0.94 & 0.01 & 0.02 & 0.26 & 0.04 && 0.629 & 0.788 && 0.317 & 0.434 \\
10J & 100 & 1.12 & 1 & 0.91 & 0.02 & 0.11 & 0.45 & 0.15 && 0.672 & 0.955 && 0.482 & 1.119 \\
10J & 200 & 0.63 & 1 & 0.94 & 0.01 & 0.03 & 0.28 & 0.05 && 0.636 & 0.653 && 0.319 & 0.168 \\
10J & 300 & 0.48 & 1 & 0.94 & 0.01 & 0.00 & 0.22 & 0.03 && 0.613 & 0.582 && 0.260 & 0.113 \\
10J & 400 & 0.41 & 1 & 0.94 & 0.01 & 0.00 & 0.18 & 0.02 && 0.592 & 0.492 && 0.238 & 0.123\\
\bottomrule
\end{tabular}
\begin{tablenotes}
    \tiny
     \item \textbf{Loss:} Frobenius loss measuring the convergence of $\hat{\Xi}$. 
     \item \textbf{P($\hat{K} = K^*$):} Proportion of instances where the correct number of factors is identified.
     \item \textbf{ECP:} Empirical coverage probability of the confidence intervals.
      \item \textbf{MMFDR:} Maximum mean false discovery rate across all covariates.
     \item \textbf{MMFNR:} Maximum mean false non-discovery rate across all covariates.
     \item \textbf{Bloss:} Frobenius loss measuring convergence of $\hat{B}$.
     \item \textbf{MMSE:} Maximum mean squared error across all estimated $\beta_{jl}$s.
\end{tablenotes}
\label{tab: result under proposed}
\end{table}

\section{Application to Grocery Shopping Data}\label{sect: Application to The Complete Journey dataset }

\subsection{Background}

We illustrate the proposed method via an application to a grocery shopping dataset. This dataset encompasses household-level transactions over a span of two years from approximately $2,000$ frequent shoppers. It includes purchases made by each household, recorded daily, alongside demographic information such as age groups, household sizes, and income levels for around $800$ households. We focus on the subset of customers with demographic information to understand how customers' shopping behavior is associated with their demographic variables and evaluate prediction performance based on latent factors and demographic variables. 

In this analysis, daily transaction data are aggregated into 25 four-week periods, using the first $T=24$ intervals for statistical analysis and model training. The 25th interval is reserved for assessing the predictive performance of our proposed model. We focus on the transactions involving the most popular $J$ items during the first $T$ intervals, with $N$ denoting the count of customers who purchased any of the $J$ items within these periods. In each time period $t$, let $y_{ijt}$ be a binary indicator of purchase such that $y_{ijt}= 1 $ if individual $i$ purchased item $j$ and $y_{ijt} = 0$ otherwise. The missing indicator, $r_{it}$, is set to $0$ when the $i$th customer did not purchase any item, including those outside the $J$ item list. 

We introduce a covariate vector $\mathbf{x}_i = (x_{i1}, x_{i2}, x_{i3}, x_{i4})^{\top}$ capturing household sizes and income levels through dummy variables. Here, \(x_{i1} = 1\) indicates two-member households, \(x_{i2} = 1\) for three or more members, \(x_{i3} = 1\) for incomes between \$35,000 and \$74,999, and \(x_{i4} = 1\) for incomes above \$75,000. The baseline level with \(x_{i1} = x_{i2} = 0\) for size and \(x_{i3} = x_{i4} = 0\) for income represents single-member households earning below \$35,000.

\subsection{Statistical Inference}\label{subsect: Statistical Inference}
We first focus on inferring the effects of covariates on customers' shopping behavior. In this analysis, we focus on the most popular 100 items, i.e.,  
$J=100$. The number of observations is $N = 800$. We set the candidate set $\mathcal K = \{1, 2, ..., 15\}$ when selecting the number of factors. Using the proposed information criterion, we obtain $\hat{K}= 8$. We perform statistical inference under the eight-factor model. 

We start with an overall significance test for all the covariates to see if any of the covariates are associated with customers' shopping behavior. That is, we test the null hypothesis of $B=0_{J \times 4}$.
We use $\|\hat{\BB}\|_{\mathrm{F}}$ as the test statistic and perform a permutation test to obtain its reference distribution under the null hypothesis. Specifically, we perform 500 random permutations indexed by $l =1, ..., 500$. In each permutation $l$, we randomly shuffle customers' covariates and then estimate the model parameters. Let the estimate of $\BB$ be denoted by  $\hat{\BB}^{(l)}$. The reference distribution for the test statistic is then obtained by the empirical distribution for $\Vert \hat{\BB}^{(l)}\Vert_F$, $l=1, ..., 500$. 
This results in a $p$-value $<0.001$, suggesting that these covariates are significantly associated with customers' shopping behavior. 

We move on to assess the influences of covariates on individual items. For each item $j$, we calculate the $p$-values associated with the null hypotheses of ${\beta}_{j1} = {\beta}_{j2}=0$ and ${\beta}_{j3} = {\beta}_{j4}=0$, respectively, for all $j$. These hypotheses test the effects of household income and size on the likelihood of purchasing item \(j\), respectively. $P$-values are derived through Wald tests, utilizing the estimated coefficients and the asymptotic variance \(\hat{\SigSig}_{\EE,j}\), as elaborated in Remark \ref{rmk: Asymptotic variance estimation}. To account for multiple testing, we adjust the $p$-values using the BY procedure for FDR control, as discussed in Section \ref{sect: Simulation Study}. This adjustment is carried out separately for the covariates of household income and size, enabling the identification of items significantly associated with each at the predetermined FDR threshold of 5\%.

Our analysis examines the influence of household income and size on the purchase patterns of grocery items, where the items are categorized into six groups -- Vegetables, Dairy and Eggs, Beverages, Fruits, Bakery and Miscellaneous items. The results are reported in Section \ref{app: subsect Additional Results in Real Data Analysis} of the online supplementary material. Specifically, \hyperref[tab: data analysis - result with size ]{Table \ref{tab: data analysis - result with size }} gives the items selected by the BY procedure for the covariate household size, the corresponding regression coefficients, p-value, BY-adjusted p-value (adj p-value), category, subcategory, average price, and package size. \hyperref[tab: data analysis - result with income ]{Table \ref{tab: data analysis - result with income }} is similar to 
\hyperref[tab: data analysis - result with size ]{Table \ref{tab: data analysis - result with size }} but gives the results for household income. Item details absent in the original dataset are marked as NA. The subcategory column represents the lowest level classification available within the dataset, and the price column represents the average unit price derived from all recorded transactions. 

We examine the results about household size. As presented in \hyperref[tab: data analysis - result with size ]{Table \ref{tab: data analysis - result with size }}, the coefficients in columns $\hat{\bbb}_{1}$ and $\hat{\bbb}_{2}$—particularly those corresponding to $\hat{\boldsymbol{\beta}}_{2}$, which denote the coefficients for households comprising three or more individuals—suggest an overall increase in the likelihood of purchasing items compared to the baseline scenario of single-person households. This trend aligns with the expectation that households with a greater number of occupants tend to have higher consumption needs. 

We then explore the effect of household income on consumer behavior, as revealed in \hyperref[tab: data analysis - result with income ]{Table \ref{tab: data analysis - result with income }}. Recall that $\hat{\bbb}_3$ and $\hat{\bbb}_4$ are the estimated coefficients for the dummy variables of
middle and high household incomes, respectively. 
Notably, most coefficients in the fruits category are positive, suggesting a heightened health consciousness among these households in comparison to their lower-income counterparts. This hypothesis is consistent with the data in the Beverages category, where most soft drinks are associated with negative coefficients. Although there are exceptions, the vegetable category mostly displays positive coefficients, further reinforcing the trend toward healthier dietary preferences. For the Bakery category, a consistent negative trend across coefficients suggests that higher-income households 
are generally accepted to be less inclined to consume breakfast at home. These observations match our knowledge about how income levels are associated with dietary choices and lifestyle habits (see, e.g. \citealp{French_etal-2019-BMC}).

On the other hand, divergent preferences across income levels are observed in the Dairy and Egg category. Specifically, we observe two opposite trends for milk: one subset exhibits positive and increasing coefficients across $\hat{\bbb}_3$ and $\hat{\bbb}_4$, signifying a preference among higher-income households, while the other shows negative and diminishing coefficients, indicating the contrary. Notably, price and size do not account for these trends, as evidenced by the table. Further investigation is needed to explore the cause, such as brand differentiation, that drives these preferences. These observations may offer insight for further investigations to explain the differences in preferences uncovered by the current exploratory model. 

\subsection{Model Comparison and Prediction Performance}\label{subsect: Prediction}
Beyond inference for coefficients of covariates, a natural application of such models is for predictions and recommendations. In particular, we can estimate the probabilities for the outcome variables at time $T+1$ given $\hat \XiXi$, assuming that the model \eqref{eq:model} still holds at $t = T+1$. Due to the absence of an estimate for the time-dependent intercept \(\gamma_{j,T+1}\), we substitute \(\hat{\gamma}_{j,T}\) in practice.  More specifically, we predict the occurrence of outcome variable $j$ at time $T+1$ based on the predicted probability
$1/(1+\exp(-(\hat{\gamma}_{jT} + \hat{\aa}_j^{\top}\hat{\thth}_i + \hat{\bbb}^{\top}_j\xx_i))).$ To assess the performance of this approach under varied settings, besides the setting where $J=100$, we also consider $J=200,300$ and $400$, with all scenarios having \(N=800\). 

\begin{figure}[h]
\caption{Residual deviances $D^{\text{res}}_{t}$ over time}
    \label{fig:res_dev}
    \centering
    \includegraphics[width=0.8\linewidth]{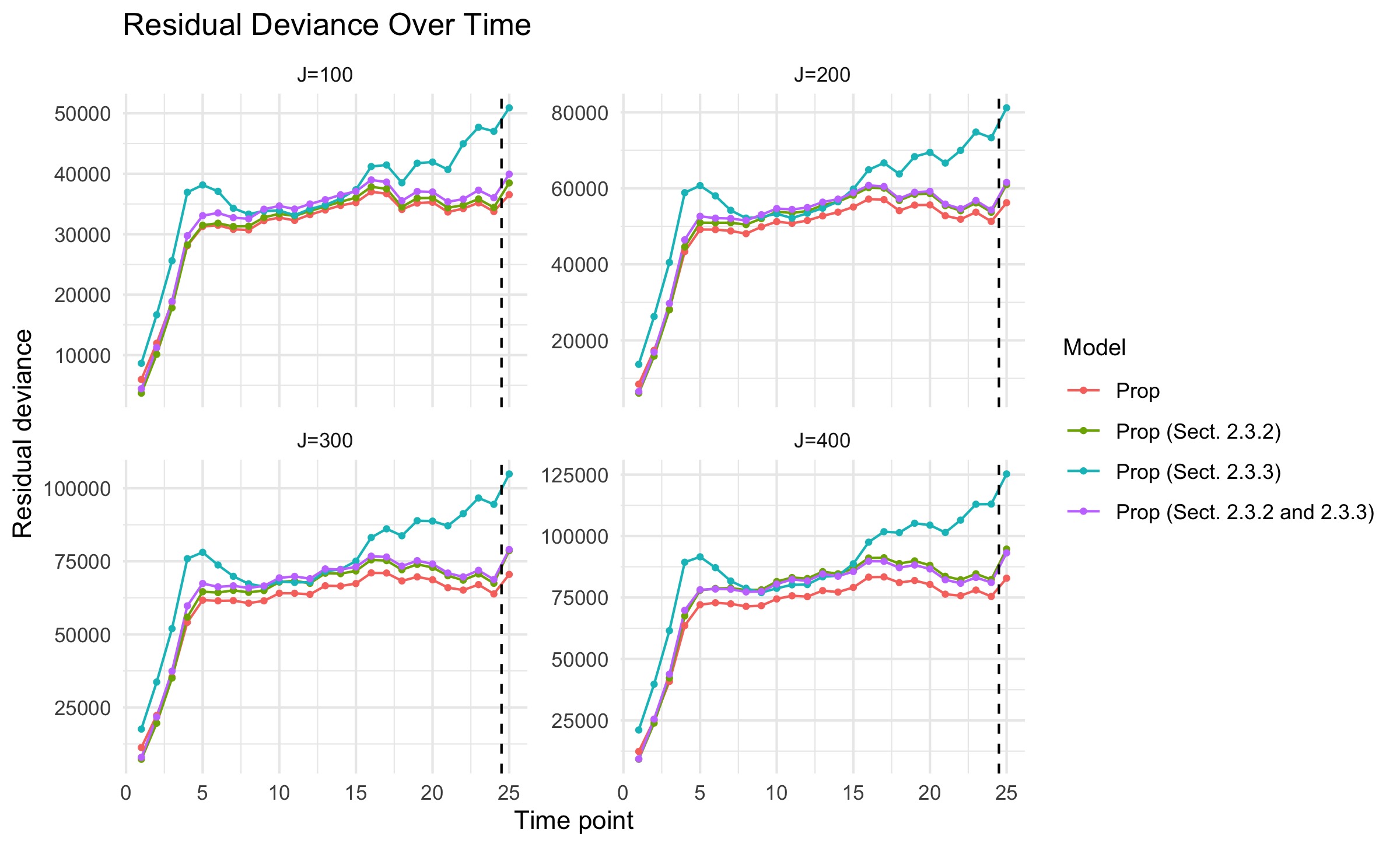}
\end{figure}

We compare the proposed method with its variants introduced in Sections \ref{subsect: ext 2} (Prop Sect 2.3.2), \ref{subsect: gamma_structure} (Prop Sect 2.3.3) and the version combining both extensions (Prop Sect 2.3.2 \& 2.3.3), in terms of both in-sample 
$(t = 1, \dots, T)$ and out-of-sample $(t =T+1)$ fittings. 
Specifically, for $t = 1, \dots, T+1$, we compute the residual deviance defined as $D^{\text{res}}_t =  \sum_{j=1}^{J} D^{\text{res}}_{jt}$, where $D^{\text{res}}_{jt} = \sum_{i=1}^{N} -2 r_{it}\big\{ y_{ijt}\log(\hat{p}_{ijt}) + (1 - y_{ijt})\log\{1 - \hat{p}_{ijt}\} \big\}$, with $\hat{p}_{ijt}$ being the estimated/predicted probability of $y_{ijt} = 1$ under each model. As shown in Figure~\ref{fig:res_dev}, the proposed method (Prop), for which $\hat K = 8$ when  $J = 100$ and $200$ and $\hat K = 11$ when $J =  300$ and 400,  
consistently achieves the best fit across almost all time points, both in-sample  and out-of-sample. Given this superior performance of Prop,  we focus on the results from this method in the rest of this section. Additional results and discussions about the model variants and their comparison with Prop are provided in Section~\ref{app: subsect Additional Results in Real Data Analysis} of the online supplementary material.

Given the nature of the dataset, our evaluations focus on recommendation performance. In particular, we compute the sensitivity, namely, the number of actual purchases in the recommendations divided by the total number of actual purchases. We compare four strategies for making recommendations. Prop ranks recommendations based on the sorted predicted probabilities of the $J$ items from corresponding model estimates. Hist ranks recommendations by the sorted cumulative purchasing frequency for each individual, resorting to random selection when ties occur. Hist-Prop follows the ranking of Hist but employs sorted predicted probabilities from Prop to resolve ties. Lastly, Hist-Hist, like Hist, ranks recommendations but uses the overall cumulative frequency of items across individuals to break ties. \hyperref[tab: Sensitivity based on Number of Recommendations]{Table \ref{tab: Sensitivity based on Number of Recommendations}}  displays the results for \(10\), \(20\), \(30\), and \(40\) recommendations across different values of \(J\) for all methods.  

We observe that Hist generally outperforms Prop. This is not surprising as there is strong tendency for consumers to purchase the same products repeatedly in grocery shopping data (see e.g. \citealp{wan2018representing}), which is captured effectively by the Hist method. Nevertheless, Hist-Prop emerges as the most proficient approach, indicating that our model is beneficial for improving recommendations, especially when customer information is sparse. This shows the capability of our method to borrow information from similar customers and reflect their preference for previously not purchased products. By offering personalized recommendations, this method outperforms Hist-Hist, which merely suggests the most popular items to individuals when there are insufficient individual history data. Finally, we highlight that it is possible to devise more advanced approaches based on our method to further enhance recommendations performances, especially for suggesting relevant new products to customers. For example, instead of using the proposed method to resolve ties only, we could develop more sophisticated criteria to allocate the proportions of recommendations using individual cumulative frequency and sorted predicted probability, respectively.

\begin{table}[!t]
\centering
\caption{Sensitivity Based on Number of Recommendations}
\label{tab: Sensitivity based on Number of Recommendations}
\small
\begin{tabular}{@{}llcccccccc@{}}
\toprule
& & \multicolumn{4}{c}{10 recommendations} & \multicolumn{4}{c}{20 recommendations} \\
\cmidrule(lr){3-6} \cmidrule(lr){7-10}
& & J=100 & J=200 & J=300 & J=400 & J=100 & J=200 & J=300 & J=400 \\
\midrule
&Hist                    & 0.448    & 0.348   & 0.297   & 0.264   & 0.652    & 0.518   & 0.447   & 0.404   \\
&Prop                & 0.352    & 0.256   & 0.223   & 0.199   & 0.533    & 0.394   & 0.340   & 0.304   \\
&Hist-Hist               & 0.451    & 0.350   & 0.299   & 0.266   & 0.654    & 0.519   & 0.450   & 0.406   \\
&Hist-Prop               & 0.456    & 0.352   & 0.301   & 0.269   & 0.659    & 0.525   & 0.453   & 0.407   \\
\midrule
& & \multicolumn{4}{c}{30 recommendations} & \multicolumn{4}{c}{40 recommendations} \\
\cmidrule(lr){3-6} \cmidrule(lr){7-10}
& & J=100 & J=200 & J=300 & J=400 & J=100 & J=200 & J=300 & J=400 \\
\midrule
&Hist                    & 0.774    & 0.627   & 0.546   & 0.496   & 0.848    & 0.705   & 0.618   & 0.565   \\
&Prop               & 0.658    & 0.490   & 0.425   & 0.382   & 0.752    & 0.570   & 0.496   & 0.447   \\
&Hist-Hist               & 0.775    & 0.629   & 0.549   & 0.499   & 0.852    & 0.709   & 0.621   & 0.570   \\
&Hist-Prop               & 0.781    & 0.635   & 0.555   & 0.505   & 0.860    & 0.714   & 0.626   & 0.575  \\
\bottomrule
\end{tabular}
\end{table}

\section{Discussions}\label{sect: Discussions}
This paper concerns the analysis of high-dimensional multivariate longitudinal data. A flexible modeling framework is proposed to account for between-variable and across-time dependence by latent variables. Statistical inference procedures are developed for parameter estimation and model selection, with statistical consistency and asymptotic normality results established. The method's application to customer grocery shopping records demonstrates its ability to identify demographic influences on purchasing patterns and improve recommendation precision, revealing its value for analytical and predictive uses in practical contexts. In particular, we find a positive association between household size and the likelihood of most purchases, whereas income level is positively associated with the consumption probabilities of healthy food and inversely with soft drinks. Moreover, our model's ability to capture information from other customers' purchase behavior allows improved recommendation performance, when combined with the information from one's purchase history. 

The current research may be extended in several directions besides the extensions discussed in Section~\ref{subsect: model extensions}. First, the current analysis focuses on the regression coefficients. In many applications, especially in applications of social sciences,   the substantive interpretation of the factors may be of interest. Section~\ref{app: Normalization algorithm} of the supplementary material presents normalization criteria that allow identification of the latent factors. These results are further supported by additional simulation studies presented in Section~\ref{app: Sensitivity Analysis} of the online supplementary material. These identification criteria are not unique; rotation techniques (e.g., \citealp{Liu_etal-2023-Psychometrika} and \citealp{Rohe_Zeng-2023-JRSSB}) and regularized estimation approaches (e.g., \citealp{Zhu_etal-2016-JASA}) may be employed to derive more interpretable factors. A theoretical examination of these method-specific criteria within the framework of our model lies beyond the present study’s scope and constitutes a valuable avenue for future research.

Second, as presented in the extension in Section~\ref{subsect: ext 2}, which permits the factor loadings $\aa_j$ to vary temporally, the static factors $\thth_i$ can be generalized to time-varying factors $\thth_{it}$. This alteration would not significantly change the estimation method but would require adjustments to the normalization criteria and assumptions to ensure the identification of the parameters $\bbb_j$, as well as $\vv_j$ in the model with time-dependent covariates. In this direction, it is of particular interest to consider a change-point setting 
that assumes the time-dependent factors $\ttt_{it}$ to have a piece-wise constant structure, allowing for individual-specific change points. This model allows us to detect structural changes within each individual, based on which adaptive interventions may be made (e.g., individualized marketing strategies). In addition, by controlling for the maximum number of change points, this change point model enables us to find a balance between model flexibility and parsimony, which leads to high prediction accuracy. 
Finally, the computational cost for the proposed estimator becomes high or even infeasible when some or all of $N$, $J$, $T$ and $p$ are large. In such cases, stochastic optimization algorithms may be developed to efficiently obtain approximation solutions, and further, central limit theorems may be established for the approximate solutions to facilitate statistical inference. 

\section*{Data availability}
The data that support the findings of this study are openly available at \\
https://www.dunnhumby.com/source-files.

\section*{Supplementary material}
The supplementary material includes estimation procedure, normalization algorithm, additional conditions and theorems for extension, technical proofs for main theorems and additional simulation and real data analysis results.

\appendix
\renewcommand{\theequation}{S.\arabic{equation}}
\setcounter{equation}{0}

\newtheorem{extassump}{Assumption}
\renewcommand{\theextassump}{\arabic{extassump}\ensuremath{^{\prime}}}

\newpage  
\vspace*{2cm}  
\begin{center}
    \Large \textbf{Supplementary Materials for: A Latent Variable Approach to Learning
 High-dimensional Multivariate longitudinal Data} \\
\end{center}
\vspace{1cm}  

\setcounter{theorem}{4} 
\setcounter{remark}{2} 
\setcounter{assump}{7}
\setcounter{extassump}{2}
\setcounter{table}{0}
\renewcommand{\thetable}{S\arabic{table}} 
\renewcommand{\thefigure}{S\arabic{figure}}

\section{Estimation Method}\label{app: Estimation Method}
Recall that the log-likelihood function in \eqref{eq: estimator} was defined as 
\begin{align*}
    l(\XiXi) &= \sum_{i=1}^N\sum_{j=1}^J\sum_{t=1}^T r_{it} \left\{y_{ijt}(\gamma_{jt} + \aa_j^{\top}\thth_i + \bbb_j^{\top}\xx_i )  - b_j(\gamma_{jt} + \aa_j^{\top}\thth_i + \bbb_j^{\top}\xx_i)\right\}\\
    &= \sum_{i=1}^N\sum_{j=1}^J\sum_{t=1}^T r_{it} \left\{y_{ijt}(\uu_j^{\top}\ee_{it})  - b_j(\uu_j^{\top}\ee_{it})\right\}.   
\end{align*}
 We further define 
 \begin{align*}
     \rho_{ijt}(t) &=  y_{ijt}(\uu_j^{\top}\ee_{it}) - b_j(\uu_j^{\top}\ee_{it})  \text{ and } \varrho_{ijt}(t)   = r_{it}\rho_{ijt}(t), \text{ such that }\\
     \rhrh_{ij}(\uu_j,\thth_i) &= \sum_{t=1}^{T} r_{it}\rho_{ijt}(\uu_j^{\top}\ee_{it}) = \sum_{t=1}^{T}\varrho_{ijt}(\uu_j^{\top}\ee_{it}). 
 \end{align*}

We now define the following objective functions:
\begin{align*}
&l_{NJ}(\XiXi) = \frac{1}{NJ} l(\XiXi) = \frac{1}{NJ}\sum_{i=1}^{N} \sum_{j=1}^{J}\rhrh_{ij}(\uu_j,\thth_i), \\
        &l_{i,J}(\thth_i,\UU) = 
        \frac{1}{J}\sum_{j=1}^{J} \rhrh_{ij}(\uu_j,\thth_i), \quad  l_{j,N}(\uu_j, \ThTh)= \frac{1}{N}\sum_{i=1}^{N} \rhrh_{ij}(\uu_j,\thth_i).
\end{align*}
Define $ \nabla l_{i,J}(\thth,\UU), \nabla^2 l_{i,J}(\thth,\UU)$ as the gradient and Hessian matrix of $l_{i,J}(\thth,\UU)$ with respect to $\thth$, and $ \nabla l_{j,N}(\uu, \ThTh), \nabla^2 l_{j,N}(\uu, \ThTh)$ as the gradient and Hessian matrix of $l_{j,N}(\uu, \ThTh)$ with respect to $\uu$.
To handle the constraint in \eqref{eq: estimator}, we introduce the projection operator 
\[Prox_c(\yy) = \argmin_{\xx : \|\xx\|\leq c }\| \yy-\xx \|^2=
\begin{cases} 
\yy                     &\text{if } \|\yy\|\leq c;\\
c\|\yy\|^{-1}\yy    &\text{otherwise}.
\end{cases}
\]

Recall that we defined $P$ as the length of $\uu_j$. We propose the following iterative algorithm to estimate the parameters of interest $\XiXi$. 

\begin{enumerate}
    \item \textbf{Initialization:} Choose initial starting parameters $\ThTh^{(0)}, \UU^{(0)}$.
      \item \textbf{Parameter Update:} For $l= 1,\dots ,L$, perform
          \begin{itemize}
            \item Given $\ThTh^{(l-1)},\UU^{(l-1)}$, update 
            $$\uu_j^{(l)} = Prox_c\left(\uu_j^{(l-1)}  -\alpha \left(\nabla^2 l_{j,N}\left(\uu^{(l-1)}, \ThTh^{(l-1)}\right)\right)^{-1}  \nabla l_{j,N}\left(\uu_j^{(l-1)}, \ThTh^{(l-1)}\right)  \right)$$ 
            for $j=1, \dots, J$, where $c = 5\sqrt{P}$,  $\alpha>0$ is a step size chosen by line search.  
            \item Given $\ThTh^{(l-1)},\UU^{(l)}$, update 
            $$\thth_i^{(l)} =  Prox_c\left(\thth_i^{(l-1)}  - \alpha \left(\nabla^2 l_{i,J}\left(\thth^{(l-1)},\UU^{(l)}\right)\right)^{-1} \nabla l_{i,J}\left(\thth^{(l-1)},\UU^{(l)}\right)  \right)  , $$ for $i = 1, \dots, N$, where $c=5\sqrt{K}$,  $\alpha>0$ is a step size chosen by line search.
        \end{itemize}
    \item \textbf{Convergence Check:} Stop iteration when $l_{NJ}(\XiXi^{(L)})$ approximates $ l_{NJ}(\XiXi^{(L-1)})$ closely. 
\end{enumerate}
This algorithm guarantees that the log likelihood increases in each iteration, when the step size $\alpha$ is properly chosen in line search. Readers may refer to \cite{chen_Li_Zhang-2019-Psychometrika} and \cite{Parikh_Boyd-2014-Optimization} for further details regarding the properties of projection operator. Moreover, when applying this estimation approach to simulation studies and data analysis, we develop a singular value decomposition (SVD) based algorithm to choose a good starting point in step 1, as presented in Section \ref{subsect: SVD algoritm}. This algorithm can be generalized easily for mixed types of data. 
\begin{remark}
The estimation procedure described above is flexible and can accommodate the extensions proposed in Sections \ref{subsect: ext} and \ref{subsect: ext 2}, as well as the constraint on $\ggg_{j}$ discussed in Section \ref{subsect: gamma_structure}. For each extension, the definitions of \(\uu_j\) and \(\ee_{it}\) are modified accordingly and are summarized in Table \ref{tab:uu_ee_extensions}, where the definition of the parameter vector \(\XiXi = \left({\uu_1}^{\top}, \dots, {\uu_J}^{\top}, {\thth_1}^{\top}, \dots, {\thth_{N}}^{\top}\right)^{\top}\) remains unchanged. 
\begin{table}[ht]
\footnotesize
\centering
\renewcommand{\arraystretch}{1.5} 
\caption{Definitions of \(\uu_j\) and \(\ee_{it}\) under different model specifications}
\begin{tabular}{|l|l|l|}
\hline
\textbf{Model} & \(\uu_j\) & \(\ee_{it}\) \\ \hline
Main model& 
\((\ggg_j^{\top}, \bbb_j^{\top}, \aa_j^{\top})^{\top}\) & 
\((\DD_{it}^{\top}, \xx_i^{\top}, \thth_i^{\top})^{\top}\) \\ \hline

Section \ref{subsect: ext} & 
\((\ggg_j^{\top}, \bbb_j^{\top}, \vv_j^{\top}, \aa_j^{\top})^{\top}\) & 
\( (\DD_{it}^{\top}, \xx_i^{\top}, \zz_{it}^{\top}, \thth_i^{\top})^{\top}\) \\ \hline

Section \ref{subsect: ext 2} & 
\((\ggg_j^{\top}, \bbb_{j1}^{\top},\dots, \bbb_{jT}, \vv_j^{\top}, \aa_{j1}^{\top}, \dots, \aa_{jT}^{\top})^{\top}\) & 
\( (\DD_{it}^{\top},  D_{it1} \xx_i^{\top},\dots, D_{itT} \xx_i^{\top}, \zz_{it}^{\top}, D_{it1}\thth_i^{\top}, \dots, D_{itT}\thth_i^{\top})^{\top}\) \\ \hline
\end{tabular}
\caption*{For each model specified above, when the restriction \(\gamma_{jt} = t\gamma_j\) for all \(j \in \{1, \dots, J\}\) discussed in Section \ref{subsect: gamma_structure} is imposed, we replace \(\ggg_j\) with \(\gamma\) in \(\uu_j\), and substitute the component \(\DD_{it}\) in \(\ee_{it}\) with \(t\).}
\label{tab:uu_ee_extensions}
\end{table}

\end{remark}

\subsection{SVD-based algorithm for obtaining initial values}\label{subsect: SVD algoritm}
The following algorithm, based on the work from \cite{chen_Li_Zhang-2019-Psychometrika} and \cite{Zhang_etal-2020-Psychometrika}, works for binary variables when $N \geq J$. Modifications are needed when $J > N$. We present the algorithm in the context of the extension presented in Section \ref{subsect: ext}, noting that the main model is a special case when time-dependent covariates are absent.
\begin{enumerate}
    \item Input responses $y_{ijt}$, missing indicators $r_{it}$, dimension $K$ of latent space, and tolerance $\epsilon$. 
      \item Compute $ \hat{p}_t = (\sum_{i=1}^N r_{it})/N $ as the proportion of observed responses for $t = 1, \dots, T$. 
      
      \item Let $L_t = (l_{ijt})_{N \times J}, $ where 
      \begin{align*}
           l_{ijt}= \begin{cases}
               2y_{ijt}-1, &\text{if } r_{it}=1, \\
               0, & \text{otherwise}.
           \end{cases}
      \end{align*}
      
      \item Apply singular value decomposition to matrices $\LL_{t}, t = 1, \dots, T$ and obtain $\LL_t = \sum_{j=1}^{J} \sigma_{jt}\qq_{jt} \hh_{jt}^\top $, where for each $t$, $\sigma_{1t} \geq \dots, \geq \sigma_{Jt}$ are the singular values and $\qq_{jt}$s and $\hh_{jt}$s are the left and right singular vectors. 

      \item Let $\tilde{\LL}_{t} = (\tilde{l}_{ijt})_{N \times J} = \sum_{k=1}^{\tilde{K}} \sigma_{kt} \qq_{kt} \hh_{kt}^\top$, where $\tilde{K} = \max\{K+1, \argmax_{k} \sigma_{kt} \geq 2\sqrt{N}\hat{p}_t \}. $

      \item Let $\MM_t = (m_{ijt})_{N \times J}, $ \text{ where }
            \begin{align*}
                m_{ijt} = \begin{cases}
                    \xi^{-1}(\epsilon) & \text{if } \tilde{l}_{ijt} < -1 + \epsilon,\\
                    \xi^{-1}(0.5(\tilde{l}_{ijt} +1  )   )  & \text{if } -1 + \epsilon \leq \tilde{l}_{ijt} \leq 1 - \epsilon,\\
                    \xi^{-1}(1 - \epsilon)  & \text{if } \tilde{l}_{ijt} > 1-\epsilon.
                \end{cases}
            \end{align*}

        \item Set $\GG^{(0)}_{t} = (\gamma_{1t}^{(0)}, \dots, \gamma_{Jt}^{(0)}  )^{\top}$, where $\gamma_{jt}^{(0)}  = (\sum_{i=1}^{N} m_{ijt} )/N$. 

        \item Apply singular value decomposition to matrix $\tilde{\MM} =   \left(\sum_{t=1}^{T} (m_{ijt} - \gamma_{jt}^{(0)})/T \right)_{N \times J}  $ and obtain $\tilde{\MM}  = \sum_{j=1}^{J} \tilde{\sigma}_{j} \tilde{\qq}_{j} \tilde{\hh}_{j}^\top $, where $\tilde{\sigma}_{1} \geq \dots \geq  \tilde{\sigma}_{J} $ are the singular values and $\tilde{\qq}_{j}$s and $\tilde{\hh}_{j}$s are the left and right singular vectors. 

        \item Set $\mathbf{\Theta}_{k}^{(0)} = \sqrt{N} \tilde{\qq}_{k} $ and $\mathbf{A}_{k}^{(0)} = \tilde{\sigma}_{k} \tilde{\hh}_{k}/\sqrt{N} ,  k = 1, \dots, K$.

        \item Plug in $\GG^{(0)}$, $\ThTh^{(0)} = (\mathbf{\Theta}_{1}^{(0)}, \dots, \mathbf{\Theta}_{K}^{(0)} )$  and $\AAA^{(0)} = (\mathbf{A}_{1}^{(0)}, \dots, \mathbf{A}_{K}^{(0)})$ in \eqref{eq:loglik} and set $\BB^{(0)}$ and $\VV^{(0)}$ such that the log-likelihood is maximized. This can be done using the \texttt{glm} function in R.

        \item Output $\UU^{(0)} = (\GG^{(0)}, \BB^{(0)}, \VV^{(0)}, \AAA^{(0)})$ and $\Theta^{(0)}$ as starting point. 
\end{enumerate}
The tolerance $\epsilon$ is a positive constant that is close to 0. A default value $\epsilon=0.01$ is used in the analysis of this paper. 

\begin{remark}
The above procedures can be easily adapted to obtain initial values for the models introduced in Sections~\ref{subsect: ext 2} and \ref{subsect: gamma_structure}. In particular, in the extension considered in Section~\ref{subsect: ext 2}, we can stack the $T$ matrices into 
$L = (L_1, \dots, L_T)$ in step 3. In step 4, we perform singular value decomposition on $L$, and in step 5, obtain $\tilde{L}$ by replacing $\hat{p}_t$ with the average $\sum_{t=1}^{T} \hat{p}_t / T$. Step 6 proceeds as before to compute $M = (M_1, \dots, M_T)$.  In step 7, each of $\gamma_{jt}$ is computed from the corresponding block $M_t$ from $M$, and we compute singular value decomposition in step 8 on 
\begin{align*}
    \tilde{M} = \left( (m_{ij1} - \gamma_{j1}^{(0)})_{N \times J}, \dots, (m_{ijT} - \gamma_{jT}^{(0)})_{N \times J}  \right).
\end{align*}
We can then obtain $\mathbf{\Theta}^{(0)}$ and $\AAA_{1}^{(0)} \dots, \AAA_T^{(0)}$ in step 9 and the rest of the procedure follows.

When the restriction \(\gamma_{jt} = t\gamma_j\) is imposed, we compute $\Gamma^{(0)} = (\gamma_{1}^{(0)}, \dots,\gamma_{J}^{(0)} )^{\top}, $ where $\gamma_{j}^{(0)}$ is the ordinary least square estimate given by $\gamma_{j}^{(0)} = (\sum_{t=1}^{T} \sum_{i=1}^{N} tm_{ijt} )/(N\sum_{t=1}^{T}t^2)$ in step 7 and define $\tilde{\MM}= ( (\sum_{t=1}^{T} m_{ijt} - t\gamma_{j}^{(0)}) / T )_{N \times J}$ in Step 8. The rest of the steps follow.

Finally, when the extension Section~\ref{subsect: ext 2} is considered and the restriction \(\gamma_{jt} = t\gamma_j\) is imposed, we use a similar procefure, setting $$\tilde{\MM}  = \left( \left( m_{ij1} - \gamma_{j}^{(0)} \right)_{N \times J} , \dots, \left( m_{ijT} - T\gamma_{j}^{(0)} \right)_{N \times J}  \right)$$
using the ordinary least square estimate for $\Gamma$. 
\end{remark}

\section{Normalization criteria and algorithm}\label{app: Normalization algorithm}
In this paper, we imposed the normalization condition as outlined in \eqref{eq: normalization for beta}, which is fundamental for the identification of $\BB$. To ensure that the matrices $\GG$, $\ThTh$, and $\AAA$ are also identifiable, we can introduce additional normalization constraints as follows:
\begin{align} \label{eq: Normalization}
	&\frac{1}{J} \AAA^{\top}\AAA  =I_{K}, \nonumber\\
	&\frac{1}{N} \ThTh^{\top}\ThTh  \text{ is diagonal with non-increasing diagonal elements.}\nonumber\\
	&\ThTh^{\top}\mathbf{1_{N}} = \mathbf{0_{K}},
\end{align} 
where $\mathbf{0_{K}}$ denotes a zero vector of length $K$. The normalization criteria outlined above, along with \eqref{eq: normalization for beta}, are sufficient to ensure the identifiability of the parameters in both the main model and the extension discussed in Section \ref{subsect: ext}. Specifically, the parameter $\VV$ associated with the time-dependent covariate is identifiable, provided there is sufficient variability in $\ZZ_{t}$ across any two time points, as detailed in Assumption \ref{assp:8} in Section \ref{app: Assumptions and Additional Theoretical Results}.  The following algorithm ensures the estimated parameters satisfy the normalization criteria imposed in \eqref{eq: normalization for beta} and \eqref{eq: Normalization}:

Suppose we have initial estimates $\GG$, $\AAA$, $\BB$ and $\ThTh$. For $t = 1, \dots, T$, define $\GG_{t} = (\gamma_{1t}, \dots, \gamma_{Jt})^{\top}$. Define $I\XX = (\mathbf{1_{N}},\XX )$, $\tilde{\ThTh} = \ThTh - I\XX (I\XX^{\top} I\XX)^{-1}{I\XX}^{\top} \ThTh$ and
$L = (I\XX^{\top} I\XX)^{-1}{I\XX}^{\top} \ThTh \AAA^{\top}$. We can set $\hat{\GG} = (\hat{\GG}_1, \dots, \hat{\GG}_T)$ and $\hat{\BB}$ such that  
\begin{align*}
    \hat{\GG}_{t} &= \GG_{t}  + L_{[1, 1:J] }^{\top}, t=1,\dots T\\
\hat{\BB}  &= \BB + L_{[2:(p+1),1:J]}^{\top}.
\end{align*}
Now define $\SigSig_{JA} = \AAA^{\top}\AAA/J$ and $\SigSig_{NT}= \tilde{\ThTh}^{\top} \tilde{\ThTh}/N$.
We apply eigendecomposition
such that 
$$\SigSig_{JA}^{1/2} \SigSig_{NT} \SigSig_{JA}^{1/2} = LDL^{-1},$$
where $D$ is a diagonal matrix and $L$ is the matrix whose columns are the eigenvectors of $D$. 
Then by setting
\begin{align} \label{eq: Theta and H norm step}
    \hat{\ThTh} = \tilde{\ThTh} H^{-1}\text{ and } \hat{\AAA} = \AAA H^{\top} ,
\end{align}
where $H = (\SigSig_{JA}^{-1/2} L)^{\top}$ would satisfy the required constraints. 

It is simple to adjust the algorithm for the special case where only the identifiability of $\BB$ is concerned. In this case we can set $L = (\XX^{\top} \XX)^{-1}\XX^{\top} \ThTh \AAA^{\top}$ and update $\hat{\BB} = \BB + L$, $\hat{\ThTh} = \ThTh - \XX (\XX^{\top} \XX)^{-1}{\XX}^{\top} \ThTh$. Similar adjustments apply to the normalization algorithms described in Remarks~\ref{rmk: identification for time-dependent loadings} and~\ref{rmk: identification for gamma structure}, when only the identifiability of $\BB$ is of interest, corresponding to the models discussed in Sections~\ref{subsect: ext 2} and~\ref{subsect: gamma_structure}. Since Section~\ref{app: Sensitivity Analysis} also evaluates the recovery of latent variables, we adopt the full identifiability criterion and corresponding algorithms throughout the simulation studies and real data analysis.

\begin{remark}\label{rmk: eigengap}
    We note that from this updated normalization criteria, Assumption \ref{assp:2} is equivalent to requiring that $N^{-1}\ThTh^{\top}\ThTh= \text{diag}(\sigma_{N1}, \dots, \sigma_{NK})$ with $\sigma_{N1} \geq \sigma_{N2} \dots \geq \sigma_{NK}$, and $\sigma_{Nk} \to \sigma_{k}$ as $N \to \infty$ for all $k$. While this does not affect the identification of \(\hat{\BB}\), it is well known that the latent factors remain unidentifiable without an eigengap that separates different components. Therefore, throughout this supplementary material, we additionally assume that $\infty > \sigma_1 > \sigma_2 > \cdots > \sigma_K > 0.$
\end{remark}

\begin{remark}\label{rmk: identification for time-dependent loadings}
    In the context of the extension discussed in Section \ref{subsect: ext 2}, we need to adjust the normalization constraints \ref{eq: Normalization}. Specifically, the parameters in this extension are identifiable given the following normalization criteria: 
    \begin{align} \label{eq: ext Normalization}
	&\frac{1}{J} \AAA_1^{\top}\AAA_1  =I_{K}, \nonumber\\
	&\frac{1}{N} \ThTh^{\top}\ThTh  \text{ is diagonal with non-increasing diagonal elements.}\nonumber\\
	&\ThTh^{\top}(X,\mathbf{1_{N}}) = 0_{K \times (p+1)},
\end{align} 
where we define $\AAA_t = (a_{jkt})_{J \times K}$ for $ t  \in \{1, \dots, T\}.$ The normalization algorithm described above can be adapted accordingly. Specifically, we set $\tilde{\ThTh} = \ThTh - I\XX (I\XX^{\top} I\XX)^{-1}{I\XX}^{\top} \ThTh$. For $t=1,\dots, T$, we set $L^{(t)} = (I\XX^{\top} I\XX)^{-1}{I\XX}^{\top} \ThTh \AAA_t^{\top}$, such that $\hat{\GG}_t$ and $\hat{\BB}_t$ are given by
\begin{align*}
    \hat{\GG}_{t} &= \GG_{t}  + L^{(t)\top}_{[1, 1:J] }\\
\hat{\BB}_t  &= \BB_t + L^{(t)\top}_{[2:(p+1),1:J]}.
\end{align*}
We can then compute $\hat{\ThTh}$ and $\hat{\AAA}_1$ as in \eqref{eq: Theta and H norm step} and set $\hat{\AAA}_t  = \AAA_t H^{\top} $ for $t= 2, \dots, T$. 
\end{remark}

\begin{remark}\label{rmk: identification for gamma structure}
When the constraint $\gamma_{jt} = t\gamma_j$ is imposed, the condition \(\ThTh^{\top} \mathbf{1}_N = \mathbf{0}_K\) is no longer required in either the main model or the extensions in Sections~\ref{subsect: ext} and~\ref{subsect: ext 2}, as the remaining conditions are sufficient to ensure identifiability. The normalization algorithm can be modified accordingly by setting \(\tilde{\ThTh} = \ThTh - X(X^{\top}X)^{-1}X \ThTh\), and defining \(L = (\XX^{\top} \XX)^{-1} \XX^{\top} \ThTh \AAA^{\top}\) for the main model. Similarly, for the extension described in Section~\ref{subsect: ext 2}, we use \(L^{(t)} = (\XX^{\top} \XX)^{-1} \XX^{\top} \ThTh \AAA_t^{\top}\). In both cases, there is no need to normalize \(\GG = (\gamma_1, \dots, \gamma_J)^{\top}\). The parameters \(\hat{\BB}\) or \(\hat{\BB}_t\) can then be updated accordingly, and the remainder of the algorithm remains unaffected by this restriction.
\end{remark}

\section{Additional Assumptions and Theoretical Results}\label{app: Assumptions and Additional Theoretical Results} %
In this section, we present the additional assumptions and theoretical results corresponding to the extensions introduced in Sections~\ref{subsect: ext} and~\ref{subsect: ext 2},  along with the necessary adjustments to accommodate the constraint $\gamma_{jt} = t \gamma_j$ discussed in Section~\ref{subsect: gamma_structure}. The definitions of $\uu_j$ and $\ee_{it}$ follow those provided in Table~\ref{tab:uu_ee_extensions}, according to the relevant extension.
\subsection{Extension in Section \ref{subsect: ext}}
Define $\ZZ_t = (\zz_{1t},\zz_{2t}, \dots, \zz_{Nt})$ and $\UU_{t} = (\GG_t,\AAA,\VV,\BB)$ for $t=1 , \dots T$. To establish asymptotic properties under this extension, the following additional assumptions are necessary:

\begin{assump}\label{assp:7} 
    $\zz_{it} \in \mathcal{Z}$ for all $i$ and $t$, where $\mathcal{Z} \subset \mathbb{R}^{p_z}$ is compact.
\end{assump}

\begin{assump}\label{assp:8}
      For some $\kappa_4>0$, there exists $t_1, t_2 \in \{1, \dots , T\}$ such that  
    \begin{align*}
        &\liminf_{N\to \infty}\pi_{\min}\left(\left(\ZZ_{t_1}-\ZZ_{t_2}, \mathbf{1_{N}} \right)^{\top}\left(\ZZ_{t_1}-\ZZ_{t_2}, \mathbf{1_{N}} \right)\right)/N \geq \kappa_4. 
    \end{align*}
\end{assump}

\begin{assump}\label{assp:9}
    There exists $\kappa_5 >0$ such that the minimum eigenvalue of the matrix 
\begin{align*}
     \begin{pmatrix}
        \text{diag} \left(  \left\{(NJ)^{-1/2}  \sum_{i=1}^{N}\sum_{t=1}^{T} \ee^*_{it}{\ee^*_{it}}^{\top}   \right\}_{j \leq J}  \right)& \left\{ (NJ)^{-1/2}\sum_{t=1}^{T}\left(\begin{array}{c}
        \boldsymbol{0}_{T+p}\\
        \zz_{it}   \\
        \boldsymbol{0}_{K^*}  
    \end{array} \right){\aa_{j}^{*}}^{\top} \right\}_{j \leq J, i \leq N}\\
          (NJ)^{-1/2}\left\{ \sum_{t=1}^{T}\aa_{j}^{*}\left(\boldsymbol{0}_{T+p}^{\top} ,\zz_{it}^{\top} ,\boldsymbol{0}_{K^*}^{\top} \right) \right\}_{i \leq N, j \leq J}&\text{diag} \left(  \left\{(NJ)^{-1/2}  T\sum_{j=1}^{J} \aa^*_j{\aa^*_{j}}^{\top}   \right\}_{i \leq N}  \right)
    \end{pmatrix}
\end{align*}
is greater that $\kappa_5$. Here $diag( \{Q_{j}\}_{j \leq J} )$ represents a block-diagonal matrix whose $j$th block on the diagonal is $Q_j$, for any series of matrices $\{Q_j\}_{j \leq J}$. Similarly $\{Q_{ij} \}_{j \leq J, i \leq N}$ represents the block matrix where the $\{i,j\}$th block is $Q_{ij}$.
\end{assump}
Assumption \ref{assp:7} extends Assumption \ref{assp:1} to cater for the time-dependent covariate $\zz_{it}$. Assumption \ref{assp:8} asserts sufficient variability in $\ZZ_{t}$ across at least two time periods, ensuring the identifiability of $\vv_{j}$. Assumption \ref{assp:9} is a working assumption for the proof of asymptotic normality. As discussed in Section~\ref{subsect: gamma_structure}, when the constraint $\gamma_{jt} = t\gamma_j$ is imposed, the intercept term $\gamma_{jt} \mathbf{1_N}$ is effectively replaced by $\gamma_j (t \mathbf{1_N})$, making $t\mathbf{1_N}$ a component of the time-dependent covariate vector $\ZZ_t$. In this setting, Assumptions~\ref{assp:7}–\ref{assp:9} can be naturally adapted to account for this change, and the subsequent theorems remain valid under the constraint. For instance, Assumption~\ref{assp:8} can be rewritten as $\liminf_{N\to \infty}\pi_{\min}\left((\ZZ_{t_1} - \ZZ_{t_2})^{\top}(\ZZ_{t_1} - \ZZ_{t_2})\right)/N \geq \kappa_4$.

We now present the the following theorems under the normalization criteria discussed in \eqref{eq: normalization for beta} and \eqref{eq: Normalization}:
\begin{theorem}\label{Thm: convergence of factors } 
Under Assumptions \ref{assp:1} to \ref{assp:4}, \ref{assp:7} and \ref{assp:8}, we have 
\begin{align*} 
&\frac{1}{\sqrt{N}} \| \hat{\ThTh}- {\ThTh^*} \hat{S}_{\AAA}   \|_{F} = O_P(min\{\sqrt{N}, \sqrt{J}\}^{-1}),\\
 &\frac{1}{\sqrt{J}} \| \hat{\UU}_t- {\UU^*_t} \hat{S}_\UU   \|_{F} = O_P(min\{\sqrt{N}, \sqrt{J}\}^{-1}).
\end{align*}
Here, $\hat{S}_{\AAA}$ is defined as $sgn(\hat{\AAA}^{\top}\AAA^*/J)$, where the function $sgn(X)$ yields a diagonal matrix whose diagonal elements are the signs of the diagonal elements of any square matrix $X$. Moreover, $\hat{S}_{\UU}$ is a $(1+p+p_z+K^*)$ by $(1+p+p_z+K^*)$ diagonal matrix whose diagonal elements are set to 1, except for the last $K^*$ diagonal elements which are equal to $\hat{S}_{\AAA}$. 
\end{theorem}
\begin{theorem}\label{Thm: asymptotic normality of factor}
  Under Assumptions \ref{assp:1} to \ref{assp:9}, for $i = 1, \dots, N$ and $j = 1, \dots, J$, we have 
  \begin{align*}
      \sqrt{N}\left(\hat{\uu}_j - \hat{S}_\UU\uu_j^* \right) \overset{d}{\to} \mathcal{N}\left(0, -\PhPh_{j}^{-1}  \right) \text{ and } \sqrt{J}\left(\hat{\thth}_i - \hat{S}_\AAA\thth_i^* \right) \overset{d}{\to} \mathcal{N}\left(0, -\PsPs_{i}^{-1}  \right).
  \end{align*}
\end{theorem}
\begin{theorem}\label{Thm: Selection of factor under extension}
    Suppose that Assumptions \ref{assp:1} to \ref{assp:4}, \ref{assp:7} and \ref{assp:8} hold and $K^* \in \mathcal K$. If the penalty term $\Lambda_{NJ}$ satisfies  $\max{\left\{N, J\right\}} \lesssim \Lambda_{NJ} \lesssim NJ $, then  
  $$\lim_{N, J \rightarrow \infty} P(\hat{K} = K^*) = 1.$$  
\end{theorem}
Theorem \ref{Thm: convergence of factors } provides the average rate of convergence of $\hat{\ThTh}$ and $\hat{\UU}_{t}, t = 1, \dots, T$. The sign matrices $\hat{S}_{\AAA}$ and $\hat{S}_\UU$ are introduced to address the inherent sign indeterminacy in factors and loadings, that is, the factor structure remains unchanged if both the factors and loadings are multiplied by $-1$.
Theorem \ref{Thm: asymptotic normality of factor} establishes asymptotic normality for event-specific parameters. Similar results can also be obtained for the individual parameters $\hat{\theta}_i$, with a modified form of asymptotic variance due to the construction of the estimator. Theorem \ref{Thm: Selection of factor under extension} extends Theorem \ref{Thm: selection of factors}, proving that the number of latent factors $K$ can be consistently estimated with time-dependent covariates. The proofs of these Theorems will be presented in Section \ref{app: Proof}.

We point out that Theorems \ref{Thm: convergence of factors } and \ref{Thm: asymptotic normality of factor} hold without Assumptions \ref{assp:7} to \ref{assp:9} when only static covariates are considered. They can also be viewed as more general versions of Theorems \ref{thm: consistency} to \ref{thm: Normality}. We will make that connection explicit in Section \ref{app: Proof}, where we present the proofs of these theoretical results.

\subsection{Extension in Section \ref{subsect: ext 2}}\label{app: subsect ext 2}
Define $\AAA_{t} = (a_{jkt})_{J \times K}$ and $\UU_{t} = (\GG_t,\AAA_t,\VV,\BB)$ for $t = 1 ,\dots, T$ in this context, with the understanding that $\GG_t =\GG = (\gamma_1, \dots, \gamma_J)^{\top}$ for all $t$ when $\gamma_{jt} = t \gamma_j$ is imposed. To establish asymptotic properties under this extension, the following assumptions are necessary:
\begin{extassump}\label{extassump: 3}
For $t= 1, \dots, T$, $J^{-1}{\AAA_t^*}^{\top}\AAA_t^*$ converges to a positive definite matrix as $J$ tends to infinity. Also, $N^{-1}{\ThTh^*}^{\top}\ThTh^*$ converge to a positive definite matrix as $N$ tends to infinity.
\end{extassump}
\setcounter{extassump}{9}
\begin{extassump}\label{extassump: 10}
    There exists $\kappa_6 >0$ such that the minimum eigenvalue of the matrix 
\begin{align*}
   (NJ)^{-1/2}\sum_{t=1}^{T}\begin{pmatrix}
        \text{diag} \left(  \left\{  \sum_{i=1}^{N} \ee^*_{it}{\ee^*_{it}}^{\top}   \right\}_{j \leq J}  \right)& \left\{ \left(\begin{array}{c}
        \boldsymbol{0}_{T+Tp}\\
        \zz_{it}   \\
        \boldsymbol{0}_{K^*} \\
        D_{it2}\thth^*_{i}\\
        \vdots \\
        D_{itT}\thth^*_{i}
    \end{array} \right){\aa_{jt}^{*}}^{\top} \right\}_{j \leq J, i \leq N}\\
          \left\{ \aa_{jt}^{*}\left(\boldsymbol{0}_{T+Tp}^{\top} , \zz_{it}^{\top} ,\boldsymbol{0}_{K^*}^{\top},D_{it2}{\thth^*_{i}}^{\top}, \dots, D_{itT}{\thth^*_{i}}^{\top} \right) \right\}_{i \leq N, j \leq J}&\text{diag} \left(  \left\{  \sum_{j=1}^{J}  \aa^*_{jt}{\aa^*_{jt}}^{\top}   \right\}_{i \leq N}  \right)
    \end{pmatrix}
\end{align*}
is greater that $\kappa_6$. When $\gamma_{jt} = t \gamma_j$, we require the minimum eigenvalue of the matrix 
\begin{align*}
     (NJ)^{-1/2}\sum_{t=1}^{T}\begin{pmatrix}
        \text{diag} \left(  \left\{  \sum_{i=1}^{N} \ee^*_{it}{\ee^*_{it}}^{\top}   \right\}_{j \leq J}  \right)& \left\{ \left(\begin{array}{c}
        t\\
        \boldsymbol{0}_{Tp}\\
        \zz_{it}   \\
        \boldsymbol{0}_{K^*} \\
        D_{it2}\thth^*_{i}\\
        \vdots \\
        D_{itT}\thth^*_{i}
    \end{array} \right){\aa_{jt}^{*}}^{\top} \right\}_{j \leq J, i \leq N}\\
          \left\{ \aa_{jt}^{*}\left(t,\boldsymbol{0}_{Tp}^{\top} , \zz_{it}^{\top} ,\boldsymbol{0}_{K^*}^{\top},D_{it2}{\thth^*_{i}}^{\top}, \dots, D_{itT}{\thth^*_{i}}^{\top} \right) \right\}_{i \leq N, j \leq J}&\text{diag} \left(  \left\{  \sum_{j=1}^{J}  \aa^*_{jt}{\aa^*_{jt}}^{\top}   \right\}_{i \leq N}  \right)
    \end{pmatrix}
\end{align*}
to be greater than $\kappa_6$. 
\end{extassump}

\begin{assump}\label{assump: 11}
For some $\kappa_7 >0$, there exists $t_1, t_2 \in \{1, \dots, T\}$ such that 
    \begin{align*}
&\liminf_{N \to \infty}\pi_{\min}((\ThTh^*,\XX, \mathbf{1_N},\ZZ_{t_1},\ZZ_{t_2})^{\top}(\ThTh^*,\XX, \mathbf{1_N},\ZZ_{t_1},\ZZ_{t_2}))/N \geq \kappa_7.
\end{align*}
In the case $\gamma_{jt}^* = t \gamma_{j}$, we impose additional condition that 
\begin{align*}
    \liminf_{J \to \infty} \pi_{\min}((\GG^*,\AAA_{t_1}^*, \AAA_{t_1}^*)^{\top}(\GG^*,\AAA_1^*, \AAA_2^*)) \geq \kappa_7.
\end{align*}
\end{assump}

Assumptions \ref{extassump: 3} and \ref{extassump: 10} replaces Assumption \ref{assp:2} and \ref{assp:9} under this extension, respectively. Assumption \ref{assump: 11} replaces Assumptions \ref{assp:4} and \ref{assp:8} and deserves more explanation. The condition involving $(\ThTh^*, \XX, \mathbf{1}_N, \ZZ_{t_1}, \ZZ_{t_2})$ implies that the static factor $\ThTh^*$ must not lie in the span of $(\ZZ_{t_1}, \ZZ_{t_2})$. We illustrate the necessity of such assumption through the following simple example: Suppose $K=1$ and $T=2$, and we are given $\thth, \aa_1, \aa_2, \vv, \zz_1 $ and $\zz_2$ such that $\thth = \zz_1 + \zz_2$ and $\aa_2 = -2\aa_1$. We can verify that 
\begin{align*}
    \thth \aa_t^{\top} + \zz_t \vv^{\top} = \tilde{\thth} \tilde{\aa}_t^{\top} + \zz_t \tilde{\vv}^{\top}, \text{ for $t = 1,2$, where }
\end{align*}
$\tilde{\thth} = 2(\zz_1 +0.5\zz_2), \tilde{\aa}_1 = \aa_1, \tilde{\aa}_2 = 0.5 \aa_2$ and $\tilde{\vv} = -\aa_1 + \vv$.
Similarly, the additional condition when $\gamma_{jt}^* = t \gamma^*$ ensures that $\GG = (\gamma_1, \dots, \gamma_J)$ does not lie in the span of $\AAA^*_{t_1}$ and $\AAA_{t_2}^*$. We note that although $\GG$ could be absorbed into $\vv$ and $t\mathbf{1}_N$ treated as part of $\ZZ_t$ for estimation, the original assumption no longer ensures identification, as $t_1\mathbf{1}_N$ and $t_2\mathbf{1}_N$ become linearly dependent for all $t_1, t_2 \in \{1, \dots, T\}$. The additional condition is therefore necessary and specifically tailored to address this scenario.
Define $\UU_{t} = (\GG_t,\AAA_t,\VV,\BB)$ for $t = 1, \dots, T$. The following theorems, analogous to Theorems \ref{Thm: convergence of factors }, \ref{Thm: asymptotic normality of factor} and \ref{Thm: Selection of factor under extension} hold for this extension. 
\begin{theorem}\label{Thm: ext convergence of factors } 
Under Assumptions \ref{assp:1}, \ref{assp: 1.5}, \ref{extassump: 3}, \ref{assp:3}, \ref{assp:7} and \ref{assump: 11}, we have 
\begin{align*} 
&\frac{1}{\sqrt{N}} \| \hat{\ThTh}- {\ThTh^*} \hat{S}_{\AAA}   \|_{F} = O_P(min\{\sqrt{N}, \sqrt{J}\}^{-1}),\\
 &\frac{1}{\sqrt{J}} \| \hat{\UU}_t- {\UU^*_t} \hat{S}_\UU   \|_{F} = O_P(min\{\sqrt{N}, \sqrt{J}\}^{-1}).
\end{align*}
Here, $\hat{S}_{\AAA}$ is defined as $sgn(\hat{\AAA}_1^{\top}\AAA_1^*/J)$, where the function $sgn(X)$ yields a diagonal matrix whose diagonal elements are the signs of the diagonal elements of any square matrix $X$. Moreover, $\hat{S}_{\UU}$ is a $(1+p+p_z+K^*)$ by $(1+p+p_z+K^*)$ diagonal matrix whose diagonal elements are set to 1, except for the last $K^*$ diagonal elements which are equal to $\hat{S}_{\AAA}$. 
\end{theorem}
\begin{theorem}\label{Thm: ext asymptotic normality of factor}
  Under Assumptions \ref{assp:1}, \ref{assp: 1.5}, \ref{extassump: 3}, \ref{assp:3}, \ref{assp:5} to  \ref{assp:7},\ref{extassump: 10} and \ref{assump: 11}, for $i = 1,\dots, N$ and $j = 1, \dots, J$, we have 
  \begin{align*}
      \sqrt{N}\left(\hat{\uu}_j - \hat{S}_\UU\uu_j^* \right) \overset{d}{\to} \mathcal{N}\left(0, -\PhPh_{j}^{-1}  \right) \text{ and } \sqrt{J}\left(\hat{\thth}_i - \hat{S}_\AAA\thth_i^* \right) \overset{d}{\to} \mathcal{N}\left(0, -\PsPs_{i}^{-1}  \right).
  \end{align*}
\end{theorem}
\begin{theorem}\label{Thm: ext Selection of factor under extension}
    Suppose that Assumptions \ref{assp:1}, \ref{assp: 1.5}, \ref{extassump: 3}, \ref{assp:3}, \ref{assp:7} and \ref{assump: 11} hold and $K^* \in \mathcal K$. If the penalty term $\Lambda_{NJ}$ satisfies  $\max{\left\{N, J\right\}} \lesssim \Lambda_{NJ} \lesssim NJ $, then  
  $$\lim_{N, J \rightarrow \infty} P(\hat{K} = K^*) = 1.$$  
\end{theorem}
In practice, we set $\Lambda_{NJ} = {\max\{N,TJ\}} \times \log\left(\max\{N,TJ\}^{-1}J\sum_{i=1}^{N}\sum_{t=1}^{T} r_{it} \right) $ to reflect that under this extension, each additional latent dimension introduces $TJ$ new parameters for each item at each time point.

The proofs of these theorems will be presented in Section \ref{app: Proof}. We conclude this section by noting that results analogous to those stated in Theorems~\ref{thm: consistency} to \ref{thm: Normality} follow from the proof of the above theorems, when the focus is on the regression coefficients $\BB^*$.
  
\section{Proofs}\label{app: Proof}
In this section, we present the proofs of the theoretical results discussed in our paper. Specifically, Sections \ref{subsect: Proof of convergence of factors}, \ref{subsect: Proof of asymptotic normalities}, \ref{subsect: proof of selection of factors}, \ref{subsect: Proof of ext convergence of factors} and \ref{subsect: Proof of ext asymptotic normality of factor} are dedicated to the proofs of Theorems \ref{Thm: convergence of factors }, \ref{Thm: asymptotic normality of factor}, \ref{Thm: Selection of factor under extension}, \ref{Thm: ext convergence of factors } and \ref{Thm: ext asymptotic normality of factor}, respectively. Moreover, Section \ref{subsect: Proof of consistency results for asymptotic variance } establishes the consistency of the estimator introduced in Remark \ref{rmk: Asymptotic variance estimation}, for estimating the asymptotic variance of $\hat{\bbb}_j$, where $j = 1, \dots ,J$. The argument is general and also applies to the estimators of asymptotic variances under the extensions discussed in Sections \ref{subsect: ext} and \ref{subsect: ext 2}, as well as to the asymptotic variance estimators for $\hat{\uu}_j$ and $\hat{\thth}_i$ using the same approach.

Furthermore, it is essential to recognize that Theorem~\ref{thm: consistency} emerges directly as a corollary of Lemma~\ref{lm: 4}, which is introduced and proved in Section \ref{subsect: Proof of asymptotic normalities}. Subsequently, Theorem \ref{Thm: convergence of Betas } follows from the proof of Theorem \ref{Thm: convergence of factors }.  Additionally, Theorem \ref{thm: Normality} follows from Theorem \ref{Thm: asymptotic normality of factor}. Lastly, Theorem \ref{Thm: selection of factors} is a special case of Theorem \ref{Thm: Selection of factor under extension}, and Theorem \ref{Thm: ext Selection of factor under extension} follows by an analogous argument, so the detailed derivation is omitted.

To simplify notation, the proofs of the theorems and results in Sections~\ref{subsect: Proof of convergence of factors} to \ref{subsect: Proof of consistency results for asymptotic variance } are conducted under the assumption that the scale parameters $\phi_j$ are known and set to 1. Notably, the derivatives of $l(\XiXi)$ with respect to the parameters $\uu_j$ and $\thth_i$ are equivalent to the derivatives of the full joint log-likelihood function.  It is easy to see that Theorems \ref{Thm: convergence of factors }, \ref{Thm: asymptotic normality of factor}, \ref{Thm: ext convergence of factors } and \ref{Thm: ext asymptotic normality of factor} hold when $\phi_j$ are unknown as the estimated parameters in these theorems are unaffected by the scale parameters. Similarly, Theorem \ref{Thm: Selection of factor under extension} remains valid since the information criterion considered in the work does not depend on the scale parameter. Finally, it is easy to see that estimator introduced in Remark \ref{rmk: Asymptotic variance estimation} remains consistent when $\phi_j$ is unknown as long as the estimate of $\hat \phi_j$ is consistent.

\subsection{Proof of Theorem \ref{Thm: convergence of factors } } \label{subsect: Proof of convergence of factors}

Throughout this section, $\delta_0, \delta_1, \delta_2, \dots$ denote positive constants that do not depend on $N, J$. For any random variable $Y$, define the Orlicz norm $\| Y\|_{\Psi}$ as 
$$\|Y\|_{\Psi} = \inf\left\{C>0: E\Psi(|Y|/C) \leq  1 \right\},$$
where $\Psi$ is a non-decreasing, convex function with $\Psi(0) =0$. We write the norm as $\|Y\|_{\Psi_2}$ when $\Psi(x) = \exp(x^2)-1$. We use $\|\cdot \|_{S}$ to denote the spectral norm. Additionally, we define $C(\cdot, g, \mathcal{G}) $ to denote the covering number of space $\mathcal{G}$ endowed with semimetric $g$.  
We further define $M_t=(m_{ijt})$, where $m_{ijt} = \uu_j^{\top}\ee_{it}, t=1 \dots T$. For any $\XiXi^{(a)}, \XiXi^{(b)}\in \mathcal{H}^{K^*}$, define $d(\XiXi^{(a)}, \XiXi^{(b)}) = \max_{t:t=1,\dots T}\|M^{(a)}_{t} - M^{(b)}_t  \|_{F}/\sqrt{NJ}$. Let $\rhrh_{ij}=\rhrh_{ij}(\uu_j^*,\thth^*_i)$, $ \rho_{ijt} = \rho_{ijt}({\uu_j^*}^{\top}\ee^*_{it}) $, $ w_{ij}(\uu_j, \thth_i) =  \rhrh_{ij} -\rhrh_{ij}(\uu_j,\thth_i)  $ and define
    \begin{align*}
        l^*_{NJ}(\XiXi) &= \frac{1}{NJ} \sum_{i=1}^{N}\sum_{j=1}^{J}w_{ij}(\uu_j, \thth_i),&    \bar{l}^*_{NJ}(\XiXi) &= \frac{1}{NJ} \sum_{i=1}^{N}\sum_{j=1}^{J}E\left(w_{ij}(\uu_j, \thth_i)\right).
    \end{align*}
We further define
\begin{align*}
    \mathbb{W}_{NJ}(\XiXi) =  l^*_{NJ}(\XiXi) - \bar{l}^*_{NJ}(\XiXi)  = \frac{1}{NJ} \sum_{i=1}^{N}\sum_{j=1}^{J} \left(  w_{ij}(\uu_j, \thth_i) - E\left(w_{ij}(\uu_j, \thth_i)\right) \right).
\end{align*}

We prove the following three Lemmas. Theorem \ref{Thm: convergence of factors } then follows from the proof of Theorem 1 in \cite{Chen_etal-2021-Econometrica}.  
\begin{lemma}\label{lemma: 1}
Under Assumptions \ref{assp:1},\ref{assp: 1.5}, \ref{assp:3} and \ref{assp:7}, $d(\hat{\XiXi},\XiXi^*) = o_p(1)$ as $J,N \to \infty.$
\end{lemma}
Proof: Pick any $\XiXi$ from the set $\mathcal{H}^{K^*}$. By taking $m_{ijt} = \uu_j^{\top}\ee_{it}$ and expanding around $m_{ijt}^*={\uu_j^*}^{\top}\ee^*_{it}$ for  $t = 1, \dots ,T$, we have 
\begin{align*}
    &E\left( \rhrh_{ij} -\rhrh_{ij}(\uu_j,\thth_i)   \right) \\
    =&E\Bigg( \rhrh_{ij} - \rhrh_{ij}- \sum_{t=1}^{T} r_{it}(m_{ijt} - m_{ijt}^*)\rho_{ijt}^{'}(m_{ijt}^*) - 0.5 \sum_{t=1}^{T} r_{it}(m_{ijt} - m_{ijt}^*)^2\rho_{ijt}^{''}(\tilde{m}_{ijt}) \Bigg)\\
    =&E\left(  -0.5 \sum_{t=1}^{T}r_{it}(m_{ijt} - m_{ijt}^*)^2\rho_{ijt}^{''}(\tilde{m}_{ijt}) \right)\\
    =&0.5 \sum_{t=1}^{T} \left\{(m_{ijt} - m_{ijt}^*)^2  b_j^{''}(\tilde{m}_{ijt})P(r_{it} = 1) \right\},
\end{align*}
where $\tilde{m}_{ijt}$ lies between $m_{ijt}$ and $m_{ijt}^*, t=1, \dots T$. Therefore, we have
\begin{align}\label{eq: lm 1- barl}
E\left( \rhrh_{ij} -\rhrh_{ij}(\uu_j,\thth_i)   \right)  &\gtrsim  (m_{ijt} - m_{ijt}^*)^2, t=1, \dots T \text{ and }\nonumber\\
    \bar{l}^*_{NJ}(\XiXi)  & \gtrsim d^2(\XiXi,\XiXi^*). 
\end{align}
by Assumptions \ref{assp: 1.5} and \ref{assp:3}. Also, by the definition of $\hat{\XiXi}$, we have $l^*_{NJ}(\hat{\XiXi}) = l_{NJ}(\XiXi^*) - l_{NJ}(\hat{\XiXi})\leq 0, $ or equivalently $\mathbb{W}_{NJ}(\hat{\XiXi}) + \bar{l}^*_{NJ}(\hat{\XiXi})\leq 0.$ Combining it with \eqref{eq: lm 1- barl}, we have 
$$ 0 \leq d^2(\hat{\XiXi},\XiXi^*) \lesssim \bar{l}^*_{NJ}(\hat{\XiXi}) \leq \sup_{\XiXi \in \mathcal{H}^{K^*}}\left| \mathbb{W}_{NJ}(\XiXi) \right|. $$
So it remains to show that 
\begin{align}\label{eq: lm1 - sup W}
    \sup_{\XiXi \in \mathcal{H}^{K^*}}\left| \mathbb{W}_{NJ}(\XiXi) \right| = o_p(1).
\end{align}
By Assumption \ref{assp:1}, we can choose $\delta_1>1$ large enough such that $ \|\uu_j\|, \|\thth_{i}\| \leq \delta_1$ for all $i,j$, $\uu_j \in \mathcal{U}$ and $\thth_i \in \boldsymbol{\Theta}$. Let $B_{d}(\delta_1)$ denote a Euclidean ball in $\mathbb{R}^{d}$ with radius $\delta_1$ for any positive integer $d$. For any $\epsilon >0$, let $\uu_{(1)}, \dots, \uu_{(q_{P})}$ be a maximal set of points in $B_{P}(\delta_1)$ such that $\|\uu_{(h)} - \uu_{(k)}\|> \epsilon/\delta_1$ for any $h \neq k$. Here ``maximal'' signifies that no point can be added without violating the validity of the inequality. Similarly, let $\thth_{(1)}, \dots, \thth_{(q_K)}$ be a maximal set of points in $B_K(\delta_1)$ such that $\| \thth_{(h)} - \thth_{(k)}\| > \epsilon/\delta_1$ for any $h\neq k$. It is well known that $Q_{P}$ and $Q_{K^*} $, the packing numbers of $B_{P}(\delta_1)$ and  $B_{K^*}(\delta_1)$, respectively,  are bounded by $\delta_2(\delta_1/\epsilon)^{P}$.  
For any $\XiXi \in \mathcal{H}^{K^*},$ define $\bar{\XiXi} = (\bar{\uu}_1^{\top}, \dots, \bar{\uu}_J^{\top}, \bar{\thth}_1^{\top}, \dots, \bar{\thth}_N^{\top} )^{\top}$, where $\bar{\uu}_j = \{ \uu_{(q_j)}: q_j = \min\{  q : q\leq Q_{P}, \|\uu_{(q)} - \uu_{j} \|\leq \epsilon/\delta_1\} \}$ and $ \bar{\thth}_i = \{\thth_{(q_i)}: q_i = \min\{q :  q\leq Q_{K^*}, \|\thth_{(q)} - \thth_{i} \|\leq \epsilon/\delta_1\} \}$.
This definition ensures that each $\XiXi \in \HH^{K^*}$ is uniquely sent to a $\bar{\XiXi} \in \HH^{K^*}$ comprised of the maximal sets of points defined previously. Thus, we can write 
$$\mathbb{W}_{NJ}(\XiXi) = \mathbb{W}_{NJ}(\bar{\XiXi}) + \mathbb{W}_{NJ}(\XiXi) - \mathbb{W}_{NJ}(\bar{\XiXi}).$$
Define $\bar{\ee}_{it} = (\DD_{it}^{\top}, \xx_i^{\top}, \zz_{it}^{\top}, \bar{\thth}_i^{\top})^{\top}$. By Assumption \ref{assp: 1.5}, there exists $\delta_0 >0 $ such that  
\begin{align}\label{eq: lm 1 upper bound}
    &\left| \rhrh_{ij}(\uu_j,\thth_i) -\rhrh_{ij}(\bar{\uu}_j,\bar{\thth}_i)  \right| \nonumber\\
    \leq&\sum_{t=1}^{T} \left|\rho_{ijt}(\uu_j^{\top}\ee_{it} ) - \rho_{ijt}(\bar{\uu}_j^{\top}\bar{\ee}_{it} )\right|\nonumber\\
    \leq& \delta_0 \sum_{t=1}^{T} \left| \uu_j^{\top}\ee_{it} - \bar{\uu}_j^{\top} \bar{\ee_{it}} \right|\nonumber\\
    \leq& \delta_0 \sum_{t=1}^{T}\left(  | \gamma_{jt} - \bar{\gamma}_{jt}| +|\bbb_{j}^{\top}\xx_i  - \bar{\bbb}_{j}^{\top}\xx_i  | +  |\vv_j^{\top}\zz_{it} -{\bar{\vv}_j}^{\top}\zz_{it} |  + |{\aa_j}^{\top}\thth_i- {\bar{\aa}_j}^{\top}\bar{\thth}_i| \right)\nonumber\\
    \leq& \delta_0T\left( \|\ggg_{j} - \bar{\ggg}_j\| + \|\aa_j\|\|\thth_i - \bar{\thth}_i\|  + \|\bar{\thth}_i\|\|\aa_j - \bar{\aa}_j\| + \|\xx_{i}\|\|\bbb_j - \bar{\bbb}_j\|+ \|\zz_{it}\|\|\vv_j - \bar{\vv}_j\|\right)\nonumber\\
    \leq& 5\delta_0T\epsilon.
\end{align}
Thus, we have
\begin{align}\label{eq: lm 1- sup W diff}
    \sup_{\XiXi \in \mathcal{H}^{K^*}}\|  \mathbb{W}_{NJ}(\XiXi) - \mathbb{W}_{NJ}(\bar{\XiXi})\| \leq 10 \delta_0 T\epsilon.
\end{align}
Also, note that 
\begin{align*}
    \left|w_{ij}(\bar{\uu}_j, \bar{\thth}_i)\right| &=  \left|\rhrh_{ij} -\rhrh_{ij}(\bar{\uu}_j,\bar{\thth}_i) \right| \leq \delta_0 \sum_{t=1}^{T} \left| {\uu_j^*}^{\top}\ee^*_{it} - \bar{\uu}_j^{\top}\bar{\ee}_{it} \right|.
\end{align*}
By Cauchy-Schwarz inequality,
 \begin{align*}
     \left(\sum_{t=1}^{T} \left|   {\uu_j^*}^{\top}\ee^*_{it} - \bar{\uu}_j^{\top}\bar{\ee}_{it}\right|\right)^2 \leq T\sum_{t=1}^{T} \left|  {\uu_j^*}^{\top}\ee^*_{it} - \bar{\uu}_j^{\top}\bar{\ee}_{it}\right|^2.
 \end{align*}
 By Hoeffding's inequality, we have 
$$P\left( \left|  \sqrt{NJ}\mathbb{W}_{NJ}(\bar{\XiXi})  \right|>c  \right) \leq 2 \exp\left( - \frac{2c^2}{\delta_0^2T^3\cdot d^2(\bar{\XiXi}, \XiXi^*) }\right), $$
and by Lemma 2.2.1 of \cite{Van_etal-1996-weak}, it follows that $\|\mathbb{W}_{NJ}(\bar{\XiXi}) \|_{\Psi_2} \lesssim d(\bar{\XiXi}, \XiXi^*)/\sqrt{NJ}.$ Since $\bar{\XiXi}$ can take at most $Q_{P}^{J}\times Q_{K^*}^{N} \lesssim \left(\delta_1/\epsilon\right)^{P(N + J)}$ different values, and $d(\bar{\XiXi}, \XiXi^*) \lesssim \delta_1$, it follows from Lemma 2.2.2 of \cite{Van_etal-1996-weak} that 
\begin{align}\label{eq: lm1 - EsupW}
    E\left(\sup_{\XiXi \in \mathcal{H}^{K^*}}\left| \mathbb{W}_{NJ}(\bar{\XiXi}) \right| \right) &\leq \left\| \sup_{\XiXi \in \mathcal{H}^{K^*}}\left| \mathbb{W}_{NJ}(\bar{\XiXi}) \right|   \right\|_{\Psi_2}\nonumber\\
    &\lesssim \sqrt{\log(\delta_1/\epsilon)}\sqrt{P(N + J)}/\sqrt{NJ}\nonumber\\
    &\lesssim \sqrt{\log(\delta_1/\epsilon)}\min\left\{\sqrt{N},\sqrt{J}\right\}^{-1}.
\end{align}
Finally, by Markov's inequality and \eqref{eq: lm 1- sup W diff}, for any $c>0$,
\begin{align*}
P\left( \sup_{\XiXi \in \mathcal{H}^{K^*}}\left| \mathbb{W}_{NJ}(\XiXi) \right| > c\right)&\leq
P\left( \sup_{\XiXi \in \mathcal{H}^{K^*}}\left| \mathbb{W}_{NJ}(\bar{\XiXi}) \right| >  \frac{c}{2}\right) + P\left( \sup_{\XiXi \in \mathcal{H}^{K^*}}\left| \mathbb{W}_{NJ}(\XiXi)-\mathbb{W}_{NJ}(\bar{\XiXi}) \right| >  \frac{c}{2}\right)\\
&\leq \frac{2}{c}E\left( \sup_{\XiXi \in \mathcal{H}^{K^*}}\left| \mathbb{W}_{NJ}(\bar{\XiXi}) \right| \right) + P\left(10 \delta_0 T\epsilon > \frac{c}{2}\right).
\end{align*}
Thus, by choosing $c = 30\delta_0T\epsilon$, \eqref{eq: lm1 - sup W} follows from \eqref{eq: lm1 - EsupW} and the fact that $\epsilon$ is arbitrary, which concludes the proof.
\begin{lemma}\label{lemma 2}
    Define $\mathcal{H}^{K^*}(c) = \left\{ \XiXi \in \mathcal{H}^{K^*}: d(\XiXi, \XiXi^*) \leq c\right\}.$ Under Assumptions \ref{assp:1}-\ref{assp:4} and \ref{assp:7}-\ref{assp:8}, for sufficiently small $c>0$ and sufficiently large $N$ and $J$,  for any $\XiXi \in \mathcal{H}^{K^*}(c),$ it holds that 
    $$\| \ThTh- {\ThTh^*} S_\AAA   \|_{F}/\sqrt{N} +  \| \UU_t- \UU^*_t S_\UU \|_{F}/\sqrt{J}  \leq \delta_3c, t=1, \dots, T,  $$
    where $S_{\AAA} = sgn\left(\AAA^{\top}\AAA^*/J\right)$ \text{ and } $S_\UU$ is a $(1+p + p_z+K^*)\times (1+p +p_z+K^*)$ diagonal matrix whose diagonal elements are 1, except for the last $K^*$ diagonal elements which are equal to $S_{\AAA}$.
\end{lemma}
Proof: 
We first prove that $ \left\| \VV  -\VV^{*}\right\|_{F}/\sqrt{J} \lesssim d(\XiXi, \XiXi^*).$
By Assumption \ref{assp:8}, there exists $t_1, t_2 \in \{1, \dots , T\}$ such that  $\pi_{\min}\left(\left(\ZZ_{t_1}-\ZZ_{t_2}, \mathbf{1_{N}} \right)^{\top}\left(\ZZ_{t_1}-\ZZ_{t_2}, \mathbf{1_{N}} \right)\right)/N \geq \kappa_4/2$ for sufficiently large $N$. Without loss of generality, assume $t_1=1$ and $ t_2=2$.
We have
\begin{align*}
    &\frac{1}{\sqrt{NJ}}\left\|  \left(\ZZ_{1}-\ZZ_{2}\right)\left( {\VV}^{\top}  -{\VV^{*}}^{\top}\right)+ \mathbf{1_{N}} \left\{\left({\GG_{1}}   - {\GG^*_{1}}\right)-\left({\GG_{2}}   - {\GG^*_{2}}\right)\right\}^{\top} \right\|_{F}\\
    \leq&\frac{1}{\sqrt{NJ}}\sum_{t=1}^{2}\left\| \ThTh \AAA^{\top} - \ThTh^* \AAA^* + \XX\left(\BB - \BB^* \right)^{\top}+ \ZZ_t\left( {\VV}^{\top}  -{\VV^{*}}^{\top}\right)+ \mathbf{1_{N}} \left({\GG_{t}}   - {\GG^*_{t}}\right)^{\top} \right\|_{F} \leq 2d(\XiXi, \XiXi^*).
\end{align*}
Let $Q =\left( {\VV}  -{\VV^{*}}, \left({\GG_{1}}   - {\GG^*_{1}}\right)-\left({\GG_{2}}   - {\GG^*_{2}}\right)\right) $, we have 
\begin{align*}
    &\frac{1}{\sqrt{NJ}}\left\|  \left(\ZZ_{1}-\ZZ_{2}\right)\left( {\VV}^{\top}  -{\VV^{*}}^{\top}\right)+ \mathbf{1_{N}} \left\{\left({\GG_{1}}   - {\GG^*_{1}}\right)-\left({\GG_{2}}   - {\GG^*_{2}}\right)\right\}^{\top} \right\|_{F}\\
      =&\frac{1}{\sqrt{NJ}}\left\|  \left(\ZZ_{1}-\ZZ_{2}, \mathbf{1_{N}} \right)Q^{\top} \right\|_{F}\\
     =&\frac{1}{\sqrt{NJ}}\left\|  \left(\ZZ_{1}-\ZZ_{2}, \mathbf{1_{N}} \right)Q^{\top}Q(Q^{\top}Q)^{-1/2}\right\|_{F}\\
     =&\frac{1}{\sqrt{NJ}}\left\|  \left(\ZZ_{1}-\ZZ_{2}, \mathbf{1_{N}} \right)\left(Q^{\top}Q\right)^{1/2}\right\|_{F}\\
     =&\frac{1}{\sqrt{NJ}} \sqrt{tr\left(\left(\ZZ_{1}-\ZZ_{2}, \mathbf{1_{N}} \right)\left(Q^{\top}Q\right)^{1/2}\left(Q^{\top}Q\right)^{1/2} \left(\ZZ_{1}-\ZZ_{2}, \mathbf{1_{N}} \right)^{\top}\right)}\\
     \geq&\frac{1}{\sqrt{J}}\sqrt{\frac{\pi_{\min}\left(\left(\ZZ_{1}-\ZZ_{2}, \mathbf{1_{N}} \right)^{\top}\left(\ZZ_{1}-\ZZ_{2}, \mathbf{1_{N}} \right)\right)}{N}} \sqrt{tr(Q^{\top}Q)}\\
     =&\frac{1}{\sqrt{J}}\sqrt{\frac{\pi_{\min}\left(\left(\ZZ_{1}-\ZZ_{2}, \mathbf{1_{N}} \right)^{\top}\left(\ZZ_{1}-\ZZ_{2}, \mathbf{1_{N}} \right)\right)}{N}} \left\| Q\right\|_{F}
\end{align*}
Therefore, we have 
\begin{align}\label{eq: Bound for B}
   \frac{1}{\sqrt{J}}\left\| Q\right\|_{F} =  \frac{1}{\sqrt{J}}\left\| \VV  -\VV^{*}\right\|_{F}+ \frac{1}{\sqrt{J}}\left\| \left({\GG_{1}}   - {\GG^*_{1}}\right)-\left({\GG_{2}}   - {\GG^*_{2}}\right) \right\|_{F} \lesssim d(\XiXi, \XiXi^*). 
\end{align}
By triangle inequality, we have
\begin{align}\label{eq: Remove B}
    &\frac{1}{\sqrt{NJ}}\left\| \ThTh \AAA^{\top}  + X \BB^{\top} +\ZZ_t{\VV}^{\top} + \mathbf{1_{N}} \GG_{t}^{\top}  -\ThTh^* {\AAA^*}^{\top} - X {\BB^*}^{\top} -\ZZ_t {\VV^{*}}^{\top} -\mathbf{1_{N}} {\GG^*_{t}}^{\top} \right\|_{F} \leq d(\XiXi, \XiXi^*),\nonumber\\
    &\frac{1}{\sqrt{NJ}}\left\| \ThTh \AAA^{\top}  + X \BB^{\top} + \mathbf{1_{N}} \GG_{t}^{\top}  -\ThTh^* {\AAA^*}^{\top} - X {\BB^*}^{\top} -\mathbf{1_{N}} {\GG^*_{t}}^{\top} \right\|_{F}  \leq d(\XiXi, \XiXi^*)  +  \frac{1}{\sqrt{NJ}}\left\|\ZZ_t\left(\VV - \VV^*\right)^{\top} \right\|_{F}.
\end{align}
Let $\pi_{\max}(\cdot)$ refers to the largest eigenvalue of a matrix. Since $\mathcal{Z}$ is bounded, we have
\begin{align}\label{eq: Bound for Z(B-B*)}
    \frac{1}{\sqrt{NJ}}\left\|\ZZ_t\left(\VV - \VV^*\right)^{\top} \right\|_{F}=&\frac{1}{\sqrt{NJ}} \sqrt{ tr(\ZZ_t\left(\VV - \VV^*\right)^{\top}\left(\VV - \VV^*\right)\ZZ_t^{\top} )}\nonumber\\
    \leq&\frac{1}{\sqrt{J}} \sqrt{ tr(\ZZ_t^{\top}\ZZ_t/N \cdot \left(\VV - \VV^*\right)^{\top}\left(\VV - \VV^*\right) )}\nonumber\\
    \leq& \frac{1}{\sqrt{J}} \pi_{\max}\left(\sqrt{ tr(\ZZ_t^{\top}\ZZ_t/N)}\right)\left\| \left(\VV - \VV^*\right)^{\top}\right\|_{F}\nonumber\\
    \lesssim& d(\XiXi, \XiXi^*). 
\end{align}
Thus by \eqref{eq: Remove B} and \eqref{eq: Bound for Z(B-B*)}, we have 
\begin{align*}
    &\frac{1}{\sqrt{NJ}}\left\| \ThTh \AAA^{\top}  + X \BB^{\top} + \mathbf{1_{N}} \GG_{t}^{\top}  -\ThTh^* {\AAA^*}^{\top} - X {\BB^*}^{\top} -\mathbf{1_{N}} {\GG^*_{t}}^{\top} \right\|_{F} \lesssim d(\XiXi, \XiXi^*).
\end{align*}
 Define $H_1 =\ThTh \AAA^{\top} - \ThTh^* {\AAA^*}^{\top} $ and $H_2 = X (\BB- {\BB^*})^{\top}  + \mathbf{1_{N}} ({\GG_{t}}-{\GG^*_{t}})^{\top},$
\begin{align*}
    &\left\| \ThTh \AAA^{\top}  + X \BB^{\top} + \mathbf{1_{N}} {\GG_{t}}^{\top}  -\ThTh^* {\AAA^*}^{\top} - X {\BB^*}^{\top} -\mathbf{1_{N}} {\GG^*_{t}}^{\top} \right\|^2_{F}  \\
    =& \left\|H_1 \right\|^2_{F} +\left\| H_2 \right\|^2_{F} + 2tr( H_1^{\top}H_2)\\
    =&\left\|H_1 \right\|^2_{F} +\left\| H_2 \right\|^2_{F}
\end{align*}
because ${\ThTh^*}^{\top}(\mathbf{1_{N}}, \XX) ={\ThTh}^{\top}(\mathbf{1_{N}}, \XX) =\mathit{0}_{K^* \times (1 + p)}.$
Therefore, we have 
\begin{align}    
&\frac{1}{\sqrt{NJ}}\left\| \ThTh \AAA^{\top} - \ThTh^* {\AAA^*}^{\top}\right\|_{F} \lesssim d(\XiXi, \XiXi^*) \label{eq: Bound for ThetaA}\\
\text{ and }&\frac{1}{\sqrt{NJ}}\left\| \XX (\BB- {\BB^*})^{\top}  + \mathbf{1_{N}} ({\GG_{t}}-{\GG^*_{t}})^{\top}\right\|_{F} \lesssim d(\XiXi, \XiXi^*). \label{eq: Bound for XV and IGamma}
\end{align}
From \eqref{eq: Bound for ThetaA} and the arguments in Lemma 2 of \cite{Chen_etal-2021-Econometrica}, we can prove that 
\begin{align}\label{eq: Bound for Theta and A}
    \left\|\ThTh - \ThTh^* S_{\AAA}\right\|_{F}/\sqrt{N} + \left\| \AAA - \AAA^* S_{\AAA}  \right\|_{F}/\sqrt{J} \lesssim d(\XiXi, \XiXi^*).
\end{align}
Finally, from \eqref{eq: Bound for XV and IGamma}, 
\begin{align*}
    \frac{1}{\sqrt{N}}\left\| \XX (\BB- {\BB^*})^{\top}  + \mathbf{1_{N}} ({\GG_{t}}-{\GG^*_{t}})^{\top}\right\|_{F} &\geq \sqrt{\frac{\pi_{\min}\left(\XX,\mathbf{1_{N}} \right)^{\top}\left(\XX,\mathbf{1_{N}} \right) }{N}} \left\|  (\BB,{\GG_{t}})- ({\BB^*},{\GG^*_{t}})\right\|_{F}.
\end{align*}
Hence 
\begin{align}\label{eq: Bound for V and gammar}
    \left\|  (\BB,{\GG_{t}})- ({\BB^*},{\GG^*_{t}})\right\|_{F} /\sqrt{J}\lesssim d(\XiXi, \XiXi^*)
\end{align}
for sufficiently large $N$ and $J$ by Assumption \ref{assp:4}. The proof of Lemma \ref{lemma 2} is thus complete by \eqref{eq: Bound for B}, \eqref{eq: Bound for Theta and A} and \eqref{eq: Bound for V and gammar}. 
\begin{lemma}\label{lemma 3}
    Under Assumptions \ref{assp:1}-\ref{assp:4} and \ref{assp:7}-\ref{assp:8}, for sufficiently small $c$ and sufficiently large $N$ and $J$, it holds that 
    $$E\left( \sup_{\XiXi \in \mathcal{H}^{K^*}(c)} \left|\mathbb{W}_{NJ}(\XiXi)\right| \right) \lesssim \frac{c}{\min\left\{\sqrt{N}, \sqrt{J}\right\}}.$$
\end{lemma}
Proof: In the proof of Lemma \ref{lemma: 1} we have shown that, for $\XiXi^{(a)}, \XiXi^{(b)} \in \mathcal{H}^{K^*},$
$$  \left\| \sqrt{NJ}\left|\mathbb{W}_{NJ}\left(\XiXi^{(a)}\right) - \mathbb{W}_{NJ}\left(\XiXi^{(b)}\right) \right| \right\|_{\Psi_2} \lesssim d\left(\XiXi^{(a)},\XiXi^{(b)}\right). $$
Since the process $\mathbb{W}_{NJ}(\XiXi)$ is separable, it follows from Theorem 2.2.4 of \cite{Van_etal-1996-weak} that 
\begin{align*}
    \sqrt{NJ}E\left( \sup_{\XiXi \in \mathcal{H}^{K^*}(c)} \left|\mathbb{W}_{NJ}(\XiXi)\right| \right) \lesssim \sqrt{NJ}\left\| \sup_{\XiXi \in \mathcal{H}^{K^*}(c)} \left|\mathbb{W}_{NJ}(\XiXi)\right| \right\|_{\Psi_2} \lesssim \int_{0}^{c} \sqrt{\log D(\epsilon, d, \mathcal{H}^{K^*}(c)})d\epsilon. 
\end{align*}
Thus, it remains to be shown that 
\begin{align}\label{eq: lm3-first bound}
    \int_{0}^{c}\sqrt{\log D(\epsilon, d, \mathcal{H}^{K^*}(c)) }d\epsilon = O\left( \sqrt{N+J} c\right).
\end{align}
To prove \eqref{eq: lm3-first bound}, first note that Lemma \ref{lemma 2} implies that 
$$ \mathcal{H}^{K^*}(c)  \subset \bigcup_{E \in \mathcal{S}} \mathcal{H}^{K^*}(c;E),  $$ 
where 
\begin{align*}
   \mathcal{S} = \bigg\{& E \in \mathbb{R}^{(1+p+p_z+K^*) \times (1+p+p_z+K^*)}: E= \text{diag}(1,\dots,1 ,u_{(2+p+p_z)},\dots, u_{(1+p+p_z+K^*)}), u_k \in \{-1,1\}\\ &\text{ for } k =2+p+p_z, \dots ,1+p+p_z+K^* \bigg\} 
\end{align*}
is the set of diagonal matrices $E$ that has ones on the diagonal in all positions except for the $(2+p+p_z)$th to $(1+p+p_z+K^*)$th positions, which can take either $+1$ or $-1$, and 
\begin{align*}
    \mathcal{H}^{K^*}(c;E) = \bigg\{\XiXi \in \mathcal{H}^{K^*} : &\|\ThTh - \ThTh^* E_{[(2+p+p_z):(1+p+p_z+K^*),(2+p+p_z):(1+p+p_z+K^*)]} \|_{F}/\sqrt{N} \\
    &+ \max_{t=1, \dots, T}\| \UU_{t} - \UU^*_t E \|_{F}/\sqrt{J} \leq \delta_3 c  \bigg\}.
\end{align*}

Since there are $2^{K^*}$ elements in $\mathcal{S}$ and $K^*$ is fixed, it suffices to show that $$\int_{0}^{c}\sqrt{\log D(\epsilon, d, \mathcal{H}^{K^*}(c;E)) }d\epsilon =O\left( \sqrt{N+J} c\right) $$ for each $E \in \mathcal{S}$. Without loss of generality, we focus on the case $E = I_{1+K^*+p+p_z}$. \\
Second, for any $\XiXi^{(a)}, \XiXi^{(b)}\in \mathcal{H}^{K^*}$, 
\begin{align*}
    d(\XiXi^{(a)},\XiXi^{(b)}) =& \frac{1}{\sqrt{NJ}}\max_{t=1, \dots T} \left\| (\mathbf{1_N},\XX,\ZZ_t, \ThTh^{(a)}){\UU^{(a)}_{t}}^{\top}   - (\mathbf{1_N}, \XX,\ZZ_t,\ThTh^{(b)}){\UU^{(b)}_{t}}^{\top}    \right\|_{F}\\
    \leq& \frac{1}{\sqrt{NJ}}\max_{t=1, \dots T} \left\| (\mathbf{1_N},\XX,\ZZ_t, \ThTh^{(a)}){\UU^{(a)}_{t}}^{\top} - (\mathbf{1_N}, \XX,\ZZ_t,\ThTh^{(b)}){\UU^{(a)}_{t}}^{\top}\right\|_{F}\\
    &+\frac{1}{\sqrt{NJ}}\max_{t=1, \dots T}\left\|(\mathbf{1_N}, \XX,\ZZ_t,\ThTh^{(b)}){\UU^{(a)}_{t}}^{\top}   - (\mathbf{1_N}, \XX,\ZZ_t,\ThTh^{(b)}){\UU^{(b)}_{t}}^{\top}    \right\|_{F}\\
    \leq& \frac{1}{\sqrt{NJ}}\left\|\left(\ThTh^{(a)}-\ThTh^{(b)}\right){\AAA^{(a)}}^{\top}\right\|_{F} + \frac{\|(\mathbf{1_N}, \XX,\ZZ_t,\ThTh^{(b)})\|_{F}}{\sqrt{N}}\cdot \max_{t=1, \dots, T}\frac{\left\|\UU^{(a)}_{t} -  \UU^{(b)}_{t}\right\|_{F}}{\sqrt{J}}\\
    \leq & \delta_4\left( \frac{\left\|\ThTh^{(a)}-\ThTh^{(b)}\right\|_{F}}{\sqrt{N}}  + \max_{t=1, \dots, T}\frac{\left\|\UU^{(a)}_{t} -  \UU^{(b)}_{t}\right\|_{F}}{\sqrt{J}}\right)\\
    \leq & \delta_4\left( \frac{\left\|\ThTh^{(a)}-\ThTh^{(b)}\right\|_{F}}{\sqrt{N}}  + \frac{\left\|\left(\GG^{(a)},\BB^{(a)},\VV^{(a)},\AAA^{(a)}\right) -\left(\GG^{(b)},\BB^{(b)},\VV^{(b)},\AAA^{(b)}\right)  \right\|_{F}}{\sqrt{J}}\right).
\end{align*}
Now define
\begin{align*}
    d^*(\XiXi^{(a)}, \XiXi^{(b)}) = 2\delta_4 \sqrt{\frac{\left\|\ThTh^{(a)}-\ThTh^{(b)}\right\|^2_{F}}{N}  + \frac{\left\|\left(\GG^{(a)},\BB^{(a)},\VV^{(a)},\AAA^{(a)}\right) -\left(\GG^{(b)},\BB^{(b)},\VV^{(b)},\AAA^{(b)}\right)  \right\|^2_{F}}{J}}.
\end{align*}
It follows from $\sqrt{x} + \sqrt{y} \leq 2\sqrt{x+y}$ that $d(\XiXi^{(a)}, \XiXi^{(b)})\leq  d^*(\XiXi^{(a)}, \XiXi^{(b)})$. Moreover, for any $\XiXi \in \mathcal{H}^{K^*}(c;I_{1+p+p_z+K^*}),$ we have 
$$\left( \frac{\left\|\ThTh-\ThTh^{*}\right\|_{F}}{\sqrt{N}}  + \frac{\left\|\left(\GG,\AAA,\BB,\VV\right) -\left(\GG^{*},\AAA^*,\BB^*,\VV^*\right)  \right\|_{F}}{\sqrt{J}}\right)\leq T\delta_3 c.$$
Thus it follows from $\sqrt{x+y}\leq \sqrt{x} + \sqrt{y}$ that 
$$\mathcal{H}^{K^*}(c;I_{1+K^*+p+p_z}) \subset \mathcal{H}^{K*}(c) =  \left\{ \XiXi \in \mathcal{H}^{K^*}: d^*(\XiXi, \XiXi^*) \leq \delta_5 c \right\}$$
where $\delta_5 = 2T\delta_3\delta_4$. The remainder of the proof follows from the argument presented in lemma 3 of \cite{Chen_etal-2021-Econometrica}. 
\newpage
\subsection{Proof of Theorem \ref{Thm: asymptotic normality of factor}} \label{subsect: Proof of asymptotic normalities}
It suffices to prove the result for $\uu_j$, as the argument for $\thth_i$ is symmetric. Without loss of generality, we assume that $\hat{S}_\AAA = I_{K^*}$ to simplify the notation. 
Define  
\begin{align*}
    l^*_{j,N}(\uu_j, \ThTh)      &= \frac{1}{N}\sum_{i=1}^{N} \left(  \rhrh_{ij}(\uu^*_j,\thth^*_i)  -\rhrh_{ij}(\uu_j,\thth_i) \right), \quad
    \bar{l}^*_{j,N}(\uu_j, \ThTh)= \frac{1}{N}\sum_{i=1}^{N} E\left(  \rhrh_{ij}(\uu^*_j,\thth^*_i) -\rhrh_{ij}(\uu_j,\thth_i) \right),\\
     \Breve{\varrho}_{ijt}(\cdot )  &= \varrho_{ijt}(\cdot ) - E(\varrho_{ijt}(\cdot )), \quad  \Breve{\varrho}_{ijt} = \Breve{\varrho}_{ijt}({\uu_j^*}^{\top}\ee^*_{it}) . 
\end{align*}
We first prove the following Lemmas:
\begin{lemma}\label{lm: 4}
    Under Assumptions \ref{assp:1}-\ref{assp:4} and \ref{assp:7}-\ref{assp:8}, we have 
    $$\left\|\hat{\uu}_j- \uu^*_j \right\|_{F} = o_P(1) \text{ for each } j.$$ 
\end{lemma}
Proof: Note that 
$ \hat{\uu}_j = \argmin_{\uu_j \in \mathcal{U}} l^*_{j,N}(\uu_j, \hat{\ThTh}).$
First, we show that 
\begin{align}\label{eq: lm4 - first op1}
     \sup_{\uu_j \in \mathcal{U}} \left|  l^*_{j,N}(\uu_j, \hat{\ThTh}) -  \bar{l}^*_{j,N}(\uu_j, \ThTh^*) \right| = o_P(1).
\end{align}
Note that 
\begin{align*}
     \sup_{\uu_j \in \mathcal{U}} \left|  l^*_{j,N}(\uu_j, \hat{\ThTh}) -  \bar{l}^*_{j,N}(\uu_j, \ThTh^*) \right|
     \leq\sup_{\uu_j \in \mathcal{U}} \left|  l^*_{j,N}(\uu_j, \hat{\ThTh}) - l^*_{j,N}(\uu_j, \ThTh^*)  \right| 
     + \sup_{\uu_j \in \mathcal{U}} \left|  l^*_{j,N}(\uu_j, \ThTh^*) - \bar{l}^*_{j,N}(\uu_j, \ThTh^*)  \right|.
\end{align*}
It is easy to show that 
\begin{align*}
    \sup_{\uu_j \in \mathcal{U}} \left|  l^*_{j,N}(\uu_j , \hat{\ThTh}) - l^*_{j,N}(\uu_j , \ThTh^*)  \right|&\lesssim  \sup_{\uu_j \in  \mathcal{U}} \|\aa_j \|\cdot \frac{1}{N}\sum_{i=1}^{N}\left\|\hat{\thth}_i - \thth_i^*\right\| \\
    &\lesssim \left\|\hat{\ThTh} - \ThTh^*\right\|_{F}/\sqrt{N}=O_P\left(\min\{\sqrt{N}, \sqrt{J}\}^{-1}\right),\\
    \sup_{\uu_j \in \mathcal{U}} \left|  l^*_{j,N}(\uu_j, \ThTh^*) - \bar{l}^*_{j,N}(\uu_j, \ThTh^*)  \right|&= o_P(1),
\end{align*}
thus showing that \eqref{eq: lm4 - first op1} holds.\\
Second, we can show that for any $\epsilon >0$, and $B_j(\epsilon)  = \left\{ \uu_j \in \mathcal{U}: \left\| \uu_j - \uu^*_j  \right\| \leq \epsilon \right\},$
\begin{align}\label{eq: lm 4 - lower bound }
    \inf_{\uu_j \in B_j^C(\epsilon)} \bar{l}^*_{j,N}(\uu_j, \ThTh^*) > \bar{l}^*_{j,N}(\uu^*_j, \ThTh^*)=0,
\end{align}
where $B_j^C(\epsilon)$ denotes the complement of $B_j(\epsilon)$. It follows from standard argument by noting that $\bar{l}^*_{j,N}(\uu_j, \ThTh^*)$ is convex and differentiable, holding $\ThTh^*$ fixed, for example, the proof of Proposition 3.1 of \cite{Galvao_Kato-2016-JoE}. \\
Finally, given \eqref{eq: lm4 - first op1} and \eqref{eq: lm 4 - lower bound }, the proof is complete by standard consistency proof of M-estimator(see Theorem 2.1 of \cite{Newey_Mcfadden-1994-HoE}.)
\begin{lemma}\label{lm: 5}
    Under Assumptions \ref{assp:1}-\ref{assp:8}, we have $$\left\|  \hat{\uu}_j - \uu^*_j \right\| = O_p(N^{-1/2}) \text{ for each } j.$$
\end{lemma}
Proof:
For any fixed  $\uu_j \in \mathcal{U}$ and $\thth_i \in \boldsymbol{\Theta}$, expanding  $\rho^{'}_{ijt}(\uu_{j}^{\top}\ee_{it} ) \ee_{it}$ gives 
\begin{align*}
    &\rho^{'}_{ijt}(\uu_{j}^{\top}\ee_{it} ) \ee_{it}\\
    =&\rho^{'}_{ijt}( {\uu^*_{j}}^{\top}\ee_{it})\ee_{it}
      + \rho^{''}_{ijt}( {\uu^*_{j}}^{\top}\ee_{it} )\ee_{it}\left\{\ee_{it}^{\top} (\uu_{j} - \uu^*_{j})  \right\} + 0.5\rho^{'''}_{ijt}( {\tilde{\uu}_{j}}^{\top}\ee_{it} )\ee_{it}\left\{\ee_{it}^{\top} (\uu_{j} - \uu^*_{j})  \right\}^2\\
    =&\left\{\rho^{'}_{ijt} + \rho^{''}_{ijt}( {\uu^*_{j}}^{\top}\tilde{\ee}_{it}){\uu^*_{j}}^{\top}\left(\ee_{it}-  \ee^*_{it}\right)           \right\}\ee_{it}\\
     &+ \left\{\rho^{''}_{ijt} + \rho^{'''}_{ijt}( {\uu^*_{j}}^{\top}\tilde{\ee}_{it}){\uu^*_{j}}^{\top}\left(\ee_{it}-  \ee^*_{it}\right)          \right\}   \ee_{it}\left\{\ee_{it}^{\top} (\uu_{j} - \uu^*_{j})  \right\}\\
     &+0.5\rho^{'''}_{ijt}( {\tilde{\uu}_{j}}^{\top}\ee_{it} )\ee_{it}\left\{\ee_{it}^{\top} (\uu_{j} - \uu^*_{j})  \right\}^2\\
     =&\rho^{'}_{ijt}\ee^*_{it} + \rho^{'}_{ijt}\left(  \ee_{it} -  \ee^*_{it}   \right) 
     +\rho^{''}_{ijt}( {\uu^*_{j}}^{\top}\tilde{\ee}_{it})\ee_{it}{\uu^*_{j}}^{\top}\left(\ee_{it}-  \ee^*_{it}\right)
     + \rho^{''}_{ijt}\ee_{it}\left\{\ee_{it}^{\top} (\uu_{j} - \uu^*_{j})  \right\}\\
     &+ \rho^{'''}_{ijt}( {\uu^*_{j}}^{\top}\tilde{\ee}_{it})\ee_{it}{\uu^*_{j}}^{\top}\left(\ee_{it}-  \ee^*_{it}\right)   \left\{\ee_{it}^{\top} (\uu_{j} - \uu^*_{j})  \right\}
     +0.5\rho^{'''}_{ijt}( {\tilde{\uu}_{j}}^{\top}\ee_{it} )\ee_{it}\left\{\ee_{it}^{\top} (\uu_{j} - \uu^*_{j})  \right\}^2,
\end{align*}
where $\tilde{\uu}_{j}$ lies between $\uu_{j}$ and $\uu_{j}^*$ and $\tilde{\ee}_{it}$ lies between $\ee_{it}$ and $\ee_{it}^*$. Taking expectations on both sides of the above equation, and setting $\uu_{j} = \hat{\uu}_{j}, \ee_i= \hat{\ee}_i$, it follows that
\begin{align*}
    &\frac{1}{N}\sum_{i=1}^{N} E\left( \sum_{t=1}^{T} r_{it}\rho^{'}_{ijt}(\hat{\uu}_{j}^{\top}\hat{\ee}_{it} ) \hat{\ee}_{it}\right) \\
    =&\frac{1}{N}\sum_{i=1}^{N} \sum_{t=1}^{T}E\left( \varrho^{'}_{ijt}(\hat{\uu}_{j}^{\top}\hat{\ee}_{it} ) \right)\hat{\ee}_{it} \\ 
    =&\frac{1}{N}\sum_{i=1}^{N} \sum_{t=1}^{T}E\left(\varrho^{'}_{ijt}\right)\ee^*_{it} +  \frac{1}{N}\sum_{i=1}^{N}\sum_{t=1}^{T}\left(E\left(\varrho^{''}_{ijt}\right)\hat{\ee}_{it}\hat{\ee}_{it}^{\top} \right)(\hat{\uu}_{j} - \uu^*_{j})  \\
    &+ O_P\left(N^{-1/2}\left\| \hat{\ThTh} - \ThTh^* \right\|_{F}\right) + O_P\left( \left\|  \hat{\uu}_{j} - \uu_{j}^*  \right\| \right) \cdot  O_P\left( N^{-1/2} \left\| \hat{\ThTh} - \ThTh^*   \right\|_{F} \right) + O_P(\left\|  \hat{\uu}_{j} - \uu_{j}^*  \right\|^2).
\end{align*}
Note that we have $N^{-1}\sum_{i=1}^{N} \sum_{t=1}^{T} E\left(\varrho^{'}_{ijt}\right)\ee^*_{it}=0.$ Also, by Theorem \ref{Thm: convergence of factors }, 
\begin{align*}
    \frac{1}{N}\sum_{i=1}^{N}\sum_{t=1}^{T} E\left(\varrho^{''}_{ijt}\right)\hat{\ee}_{it}\hat{\ee}_{it}^{\top}
    &=\frac{1}{N}\sum_{i=1}^{N}\sum_{t=1}^{T} E\left(\varrho^{''}_{ijt}\right)\ee^*_{it}{\ee^*_{it}}^{\top} 
    +\frac{1}{N}\sum_{i=1}^{N}\sum_{t=1}^{T}E\left(\varrho^{''}_{ijt}\right)\left(\hat{\ee}_{it}\hat{\ee}_{it}^{\top}- \ee^*_{it}{\ee^*_{it}}^{\top} \right)\\
    &=\PhPh_{j} + o_p(1).
\end{align*}
Then, by Theorem \ref{Thm: convergence of factors } and Lemma \ref{lm: 4}, we have 
\begin{align}\label{eq: lm5 pre bound for uhat - ustar}
    \PhPh_{j} (\hat{\uu}_{j} - \uu^*_{j}) + o_p\left( \left\| \hat{\uu}_{j} - \uu^*_{j}\right\|  \right)&=  O_P\left(\min\left\{\sqrt{N}, \sqrt{J}\right\}^{-1}\right) + \frac{1}{N}\sum_{i=1}^{N}\sum_{t=1}^{T} E\left(  \varrho^{'}_{ijt}(\hat{\uu}_{j}^{\top}\hat{\ee}_{it} )\right) \hat{\ee}_{it}.
\end{align}
 Note that we can write 
 \begin{align*}
      &\frac{1}{N}\sum_{i=1}^{N}\sum_{t=1}^{T} E\left(  \varrho^{'}_{ijt}(\hat{\uu}_{j}^{\top}\hat{\ee}_{it} )\right) \hat{\ee}_{it}\\
      =&-\frac{1}{N}\sum_{i=1}^{N}\sum_{t=1}^{T}  \Breve{\varrho}^{'}_{ijt}\ee^*_{it}
      - \frac{1}{N}\sum_{i=1}^{N}\sum_{t=1}^{T} \left(  \Breve{\varrho}^{'}_{ijt}(\hat{\uu}_{j}^{\top}\hat{\ee}_{it} ) \hat{\ee}_{it} -  \Breve{\varrho}^{'}_{ijt} \ee^*_{it} \right)\\
    =&-\frac{1}{N}\sum_{i=1}^{N}\sum_{t=1}^{T}  \Breve{\varrho}^{'}_{ijt}\ee^*_{it} 
   - \frac{1}{N}\sum_{i=1}^{N}\sum_{t=1}^{T} \left(  \Breve{\varrho}^{'}_{ijt}(\hat{\uu}_{j}^{\top}\hat{\ee}_{it} ) \hat{\ee}_{it} -   \Breve{\varrho}^{'}_{ijt}(\hat{\uu}_{j}^{\top}\ee^*_{it} ) \ee^*_{it} \right)\\
    &- \frac{1}{N}\sum_{i=1}^{N}\sum_{t=1}^{T} \left( \Breve{\varrho}^{'}_{ijt}(\hat{\uu}_{j}^{\top}\ee^*_{it} ) -  \Breve{\varrho}^{'}_{ijt}\right)\ee^*_{it}. 
 \end{align*}
 The first term of the RHS of the above equation is clearly $O_P(N^{-1/2})$ by central limit theorem. For the second term on the RHS of the equation, we have 
 \begin{align}\label{eq: lm5 breve rho mvt}
     &\frac{1}{N}\sum_{i=1}^{N}\sum_{t=1}^{T} \left(  \Breve{\varrho}^{'}_{ijt}(\hat{\uu}_{j}^{\top}\hat{\ee}_{it} ) \hat{\ee}_{it} -   \Breve{\varrho}^{'}_{ijt}(\hat{\uu}_{j}^{\top}\ee^*_{it} ) \ee^*_{it} \right)\nonumber\\
     =&\frac{1}{N}\sum_{i=1}^{N} \sum_{t=1}^{T}  \Breve{\varrho}^{'}_{ijt}(\hat{\uu}_{j}^{\top}\ee^*_{it} )\left( \hat{\ee}_{it} -\ee^*_{it}\right) 
     + \frac{1}{N}\sum_{i=1}^{N}\sum_{t=1}^{T} \Breve{\varrho}^{''}_{ijt}(\hat{\uu}_{j}^{\top}\tilde{\ee}_{it} ) \hat{\ee}_{it} \hat{\uu}_{j}^{\top}\left( \hat{\ee}_{it} -\ee^*_{it}\right), 
 \end{align}
 where $\tilde{\ee}_{it}$ lies between $\hat{\ee}_{it}$ and $\ee^*_{it}.$ The first term of the RHS of \eqref{eq: lm5 breve rho mvt} is $O_P\left(1/\min\left\{  \sqrt{N},\sqrt{J}\right\}\right)$ due to the boundedness of $ \Breve{\varrho}^{'}_{ijt}$, and by Theorem \ref{Thm: convergence of factors }, $N^{-1}\sum_{i=1}^{N}\|\hat{\ee}_{it} -\ee^*_{it} \| = O_P\left(1/\min\left\{  \sqrt{N},\sqrt{J}\right\}\right)$. Similarly, the second term on the RHS of \eqref{eq: lm5 breve rho mvt} is also $O_P\left(1/\min\left\{  \sqrt{N},\sqrt{J}\right\}\right)$. Finally, we can show that
 \begin{align*}
     &\frac{1}{N}\sum_{i=1}^{N} \sum_{t=1}^{T}\left( \Breve{\varrho}^{'}_{ijt}(\hat{\uu}_{j}^{\top}\ee^*_{it} ) -  \Breve{\varrho}^{'}_{ijt} \right)\ee^*_{it}\\
     =&\frac{1}{N}\sum_{i=1}^{N}\sum_{t=1}^{T} \left\{ \Breve{\varrho}^{'}_{ijt} + \Breve{\varrho}^{''}_{ijt}( {\tilde{\uu}_{j}}^{\top}\ee^*_{it} ){\ee^*_{it}}^{\top}\left( \hat{\uu}_{j}-  \uu^*_{j} \right) -  \Breve{\varrho}^{'}_{ijt} \right\}\ee^*_{it}\\
     =&\frac{1}{N}\sum_{i=1}^{N} \sum_{t=1}^{T}\left\{ \Breve{\varrho}^{''}_{ijt}( {\tilde{\uu}_{j}}^{\top}\ee^*_{it} ){\ee^*_{it}}^{\top}\left( \hat{\uu}_{j}-  \uu^*_{j} \right) \right\}\ee^*_{it}\\
     =&O_P( 1/\sqrt{N}) \cdot o_p(\|\hat{\uu}_{j}-  \uu^*_{j}   \| )
 \end{align*}
 by central limit theorem, where $\tilde{\uu}_{j}$ lies between $\hat{\uu}_{j}$ and $\uu^*_{j}.$ Combining the above results yields 
 \begin{align}\label{eq: lm5 convergence of E breve rho}
     \frac{1}{N}\sum_{i=1}^{N}\sum_{t=1}^{T} E\left(  \varrho^{'}_{ijt}(\hat{\uu}_{j}^{\top}\hat{\ee}_{it} )\right) \hat{\ee}_{it} = O_P\left(1/\min\left\{  \sqrt{N},\sqrt{J}\right\} \right) +  o_p(\|\hat{\uu}_{j}-  \uu^*_{j}   \| ).
 \end{align}
 Thus the desired result follows from \eqref{eq: lm5 pre bound for uhat - ustar}, \eqref{eq: lm5 convergence of E breve rho} and Assumption \ref{assp:5}.
To derive the asymptotic distribution of $\hat{\uu}_j$, it is essential to obtain the stochastic expansion of $\hat{\thth}_i$. Define 
\begin{align*}
   &\PhPh_{N,j} = \frac{1}{N}\sum_{i=1}^{N}\sum_{t=1}^{T} E\left(\varrho^{''}_{ijt}\right)\ee^*_{it}{\ee^*_{it}}^{\top},\quad
    \PsPs_{J,i} = \frac{1}{J}\sum_{j=1}^{J}\sum_{t=1}^{T} E\left(\varrho^{''}_{ijt}\right)\aa^*_{j}{\aa_{j}^*}^{\top}, \\
    &\mathbb{P}_{NJ}(\XiXi) = T \delta \Biggl\{\frac{1}{2J}\sum_{l=1}^{K^*} \sum_{q>l}^{K^*}\biggl( \sum_{j=1}^{J} a_{jl}a_{jq} \biggr)^2 + \frac{1}{2N} \sum_{l=1}^{K^*} \sum_{q>l}^{K^*}\biggl( \sum_{i=1}^{N} \theta_{il}\theta_{iq} \biggr)^2    + \frac{1}{8J} \sum_{k=1}^{K^*} \biggl(\sum_{j=1}^{J} a_{jk}^2 - J \biggr)^2 \\
    &\quad \quad \quad \quad \quad \quad + \frac{1}{2N}\sum_{l=1}^{K^*} \biggl(\sum_{i=1}^{N} \theta_{il} \biggr)^2   + \frac{1}{2N}\sum_{k=1}^{K^*}\sum_{l=1}^{p} \biggl( \sum_{i=1}^{N}\theta_{ik}x_{il}  \biggr) \Biggr\}
\end{align*} 
for some $\delta>0$. We further define
\begin{align*}
    &S^*(\XiXi) \\
=&\left(\underbrace{\dots, \frac{1}{\sqrt{NJ}}  \sum_{i=1}^{N}\sum_{t=1}^{T} E\left(\varrho^{'}_{ijt}(\uu_j^{\top}\ee_{it} )\right)\ee_{it}^{\top}  ,\dots}_{1\times JP}, \underbrace{\dots,\frac{1}{\sqrt{NJ}}  \sum_{j=1}^{J}\sum_{t=1}^{T} E\left(\varrho^{'}_{ijt}(\uu_j^{\top}\ee_{it})\right)\aa_j^{\top} , \dots}_{1 \times NK^*}   \right)^{\top} , 
\end{align*} 
\begin{align*}
    &S(\XiXi) = S^*(\XiXi) + \partial \mathbb{P}_{NJ}(\XiXi)/\partial \XiXi,& \HHH(\XiXi) =  \partial S^*(\XiXi)/ \partial \XiXi^{\top}  + \partial \mathbb{P}_{NJ}(\XiXi)/\partial \XiXi \partial \XiXi^{\top}
\end{align*}
and let $\HHH = \HHH(\XiXi^*).$
Expanding $S(\hat{\XiXi})$ around $S(\XiXi^*)$ gives 
\begin{align}\label{eq: S hat eta expansion}
    S(\hat{\XiXi}) = S(\XiXi^*) + \HHH\cdot (\hat{\XiXi} - \XiXi^*) + 0.5 \mathcal{R}(\hat{\XiXi}), 
\end{align}
where 
$$ \mathcal{R}(\hat{\XiXi}) = \left\{ \sum_{m=1}^{JP + NK^*} \partial \HHH(\tilde{\XiXi})/ \partial \XiXi_{m} \cdot (\hat{\XiXi}_m - \XiXi^*_m)  \right\} (\hat{\XiXi} - \XiXi^*) ,$$
$\tilde{\XiXi}$ lies between $\hat{\XiXi}$ and $\XiXi^*$. 
Further, define 
\begin{align*}
    \HHH_d = 
    \begin{pmatrix}
      \HHH_d^\mathcal{U}  & 0\\
      0         & \HHH_d^\ThTh
    \end{pmatrix} 
    ,  \HHH_d^\mathcal{U}  = \frac{\sqrt{N}}{\sqrt{J}} \text{diag}\left(\PhPh_{N,1}, \dots,   \PhPh_{N,J}  \right) ,  \HHH_d^\ThTh =\frac{\sqrt{J}}{\sqrt{N}} \text{diag}\left(\PsPs_{J,1}, \dots,   \PsPs_{J,N}  \right),
\end{align*}
we have the following lemma. 
\begin{lemma}\label{lemma 6}
    Under Assumptions \ref{assp:1}-\ref{assp:9} and the condition that $\sum_{i=1}^{N}\xx_i = \boldsymbol{0}_{p}$ and $\sum_{i=1}^{N} x_{ik}x_{il} =0$ for $l,k \in \{1, \dots, p\}, l\neq k$, the matrix $\HHH$ is invertible and $\left\| \HHH^{-1} - \HHH_d^{-1} \right\|_{\max} = O(1/N)$.
\end{lemma}
 Proof: We assume $p=T=K^*=2$ for simplicity, which can be generalized easily. We consider 
  \begin{align*}
    \mathbb{P}_{NJ}(\XiXi) = 2 \delta \Biggl\{&   \frac{1}{2N}  \biggl(\sum_{i=1}^{N} \theta_{i1}\theta_{i2}  \biggr)^2 +  \frac{1}{2J}  \biggl(\sum_{j=1}^{J} a_{j1}a_{j2}  \biggr)^2 + \frac{1}{8J} \biggl( \sum_{j=1}^{J} a_{j1}^2-J \biggr)^2+ \frac{1}{8J} \biggl( \sum_{j=1}^{J} a_{j2}^2-J \biggr)^2  \\
    & +  \frac{1}{2N} \biggl(\sum_{i=1}^{N} \theta_{i1} \biggr)^2 +    \frac{1}{2N} \biggl(\sum_{i=1}^{N} \theta_{i2} \biggr)^2+  \frac{1}{2N}\sum_{k=1}^{2}\sum_{p=1}^{2} \biggl(\sum_{i=1}^{N} \theta_{ik}x_{ip} \biggr)^2\Biggr\}.
\end{align*}

Then we can define 
\begin{align*}
   \mumu_1 &=(( \boldsymbol{0}_{4+p_z}^{\top},a^*_{11},0), \dots,(\boldsymbol{0}_{4+p_z}^{\top},a^*_{J1},0),\boldsymbol{0}_{2N}^{\top}   )^{\top} /\sqrt{J}, \\
    \mumu_2 &=((\boldsymbol{0}_{4+p_z}^{\top},0,a^*_{12}), \dots,(\boldsymbol{0}_{4+p_z}^{\top},0,a^*_{J2}),\boldsymbol{0}_{2N}^{\top}    )^{\top}/\sqrt{J}, \\
    \mumu_3 &=((\boldsymbol{0}_{4+p_z}^{\top},a^*_{12},a^*_{11}), \dots,( \boldsymbol{0}_{4+p_z}^{\top},a^*_{J2},a^*_{J1}),\boldsymbol{0}_{2N}^{\top}    )^{\top}/\sqrt{J}, \\
    \mumu_4 &=(\boldsymbol{0}_{PJ}^{\top},(\theta^*_{12},\theta^*_{11}), \dots,(\theta^*_{N2},\theta^*_{N1})   )^{\top}/\sqrt{N}\\
     \mumu_5 &=(\boldsymbol{0}_{PJ}^{\top},(1,0), \dots,(1,0)   )^{\top}/\sqrt{N},\\
    \mumu_6 &=(\boldsymbol{0}_{PJ}^{\top},(0,1), \dots,(0,1)   )^{\top}/\sqrt{N},\\
    \mumu_7 &=(\boldsymbol{0}_{PJ}^{\top},(x_{11},0), \dots,(x_{N1},0)   )^{\top}/\sqrt{N},\\
    \mumu_8 &=(\boldsymbol{0}_{PJ}^{\top},(0,x_{11}), \dots,(0,x_{N1})   )^{\top}/\sqrt{N},\\
     \mumu_9 &=(\boldsymbol{0}_{PJ}^{\top},(x_{12},0), \dots,(x_{N2},0)   )^{\top}/\sqrt{N},\\
    \mumu_{10} &=(\boldsymbol{0}_{PJ}^{\top},(0,x_{12}), \dots,(0,x_{N2})   )^{\top}/\sqrt{N},
\end{align*}
such that  $\partial \mathbb{P}_{NJ}(\XiXi^*) / \partial \XiXi \partial \XiXi^{\top} = 2 \delta \left(\sum_{m=1}^{10}\mumu_m\mumu_{m}^{\top} \right)$.
We further define

\begin{align*}
    \omom_{1,1}& =  \left(  \underbrace{(\boldsymbol{0}_{4+p_z}^{\top} ,a^*_{11}/\sqrt{J},0), \dots,(\boldsymbol{0}_{4+p_z}^{\top} ,a^*_{J1}/\sqrt{J},0) }_{\omom_{1\mathcal{U},1}^{\top}}, \underbrace{(-\theta^*_{11}/\sqrt{N}, 0), \dots,(-\theta^*_{N1}/\sqrt{N}, 0 )}_{\omom_{1\ThTh,1}^{\top}}  \right)^{\top}, \\
     \omom_{2,1}& =  \left( \underbrace{(\boldsymbol{0}_{4+p_z}^{\top},0,a^*_{12}/\sqrt{J}), \dots, (\boldsymbol{0}_{4+p_z}^{\top} ,0,a^*_{J2}/\sqrt{J})}_{\omom_{2\mathcal{U},1}^{\top}}, \underbrace{(0,-\theta^*_{12}/\sqrt{N}), \dots,(0,-\theta^*_{N2}/\sqrt{N})}_{\omom_{2\ThTh,1}^{\top}}  \right)^{\top}, \\
      \omom_{3,1}& =  \left(  \underbrace{(\boldsymbol{0}_{4+p_z}^{\top},a^*_{12}/\sqrt{J},0), \dots, (\boldsymbol{0}_{4+p_z}^{\top},a^*_{J2}/\sqrt{J},0) }_{\omom_{3\mathcal{U},1}^{\top}}, \underbrace{(0,-\theta^*_{11}/\sqrt{N}), \dots,(0,-\theta^*_{N1}/\sqrt{N} )}_{\omom_{3\ThTh,1}^{\top}}  \right)^{\top}, \\
       \omom_{4,1}& =  \left( \underbrace{(\boldsymbol{0}_{4+p_z}^{\top} ,0,a^*_{11}/\sqrt{J}), \dots, (\boldsymbol{0}_{4+p_z}^{\top} ,0,a^*_{J1}/\sqrt{J})}_{\omom_{4\mathcal{U},1}^{\top}}, \underbrace{(-\theta^*_{12}/\sqrt{N}, 0), \dots,(-\theta^*_{N2}/\sqrt{N}, 0 )}_{\omom_{4\ThTh,1}^{\top}}  \right)^{\top}, \\
       \omom_{5,1}&=\left( \underbrace{(a^*_{11}/\sqrt{J},\boldsymbol{0}_{5+p_z}^{\top}), \dots, (a^*_{J1}/\sqrt{J},\boldsymbol{0}_{5+p_z}^{\top})}_{\omom_{5\mathcal{U},1}^{\top}}, \underbrace{(-1/\sqrt{N}, 0), \dots,(-1/\sqrt{N}, 0 )}_{\omom_{5\ThTh,1}^{\top}}  \right)^{\top}, \\
           \omom_{6,1}&=\left( \underbrace{(a^*_{12}/\sqrt{J},\boldsymbol{0}_{5+p_z}^{\top}), \dots, (a^*_{J2}/\sqrt{J},\boldsymbol{0}_{5+p_z}^{\top})}_{\omom_{6\mathcal{U},1}^{\top}}, \underbrace{(0,-1/\sqrt{N}), \dots,(0,-1/\sqrt{N} )}_{\omom_{6\ThTh,1}^{\top}}  \right)^{\top},\\
       \omom_{7,1}&=\left( \underbrace{(0,0,a^*_{11}/\sqrt{J},\boldsymbol{0}_{3+p_z}^{\top}), \dots, (0,0,a^*_{J1}/\sqrt{J},\boldsymbol{0}_{3+p_z}^{\top})}_{\omom_{7\mathcal{U},1}^{\top}}, \underbrace{(-x_{11}/\sqrt{N}, 0), \dots,(-x_{N1}/\sqrt{N}, 0 )}_{\omom_{7\ThTh,1}^{\top}}  \right)^{\top}, \\
      \omom_{8,1}&=\left( \underbrace{(0,0,a^*_{12}/\sqrt{J},\boldsymbol{0}_{3+p_z}^{\top}), \dots, (0,0,a^*_{J2}/\sqrt{J},\boldsymbol{0}_{3+p_z}^{\top})}_{\omom_{8\mathcal{U},1}^{\top}}, \underbrace{(0,-x_{11}/\sqrt{N}), \dots,(0,-x_{N1}/\sqrt{N} )}_{\omom_{8\ThTh,1}^{\top}}  \right)^{\top},\\
       \omom_{9,1}&=\left( \underbrace{(0,0,0,a^*_{11}/\sqrt{J},\boldsymbol{0}_{2+p_z}^{\top}), \dots, (0,0,0,a^*_{J1}/\sqrt{J},\boldsymbol{0}_{2+p_z}^{\top})}_{\omom_{9\mathcal{U},1}^{\top}}, \underbrace{(-x_{12}/\sqrt{N}, 0), \dots,(-x_{N2}/\sqrt{N}, 0 )}_{\omom_{9\ThTh,1}^{\top}}  \right)^{\top}, \\
      \omom_{10,1}&=\left( \underbrace{(0,0,0,a^*_{12}/\sqrt{J},\boldsymbol{0}_{2+p_z}^{\top}), \dots, (0,0,0,a^*_{J2}/\sqrt{J},\boldsymbol{0}_{2+p_z}^{\top})}_{\omom_{10\mathcal{U},1}^{\top}}, \underbrace{(0,-x_{12}/\sqrt{N}), \dots,(0,-x_{N2}/\sqrt{N} )}_{\omom_{10\ThTh,1}^{\top}}  \right)^{\top},
\end{align*}
and $W_1= (\omom_{1,1}, \omom_{2,1}, \dots, \omom_{10,1})$.
It is easy to check that $\omom^{\top}_{p,1}\omom_{q,1}=0$ for $p\neq q$. Moreover, we have 
\begin{align}\label{eq: W1W1 top}
    &W_1 W_1^{\top} = \sum_{k=1}^{10} \omom_{k,1}\omom_{k,1}^{\top}\nonumber \\
     &=  \left(
\begin{array}{cc}
    \sum_{k=1}^{10}\omom_{k\mathcal{U},1}\omom_{k\mathcal{U},1}^{\top}  & -(NJ)^{-1/2} \left\{ \left(\begin{array}{c}
        \DD_{i1}\\
        \xx_i   \\
        \boldsymbol{0}_{p_z}\\
        \theta^*_{i}
    \end{array}  \right){\aa_{j}^{*}}^{\top}  \right\}_{j \leq J, i \leq N} \\
    -(NJ)^{-1/2}\left\{\aa_{j}^{*}\left(\DD_{i1}^{\top},\xx_i^{\top},\boldsymbol{0}_{p_z}^{\top},{\theta^*_{i}}^{\top}\right) \right\}_{i \leq N, j \leq J}  & \sum_{k=1}^{10}\omom_{k\ThTh,1} \omom_{k\ThTh,1}^{\top}
\end{array}
\right).
\end{align}
Further, it is easy to see that under our normalization, 
\begin{align*}
    W_1^{\top} W_1 =
    \text{diag}\bigg(&
        \sigma_{N1}+1,
        \sigma_{N2}+1,
        \sigma_{N1}+1 ,
        \sigma_{N2}+1,
        2,
        2, \\
        &1 + N^{-1}\sum_{i=1}^{N}x_{i1}^2,
         1 + N^{-1}\sum_{i=1}^{N}x_{i1}^2,
         1 +N^{-1}\sum_{i=1}^{N}x_{i2}^2,
          1 +N^{-1}\sum_{i=1}^{N}x_{i2}^2
\bigg).
\end{align*}
Next, we project $\mumu_{k}$ onto $W_1$, and write $\mumu_k = W_1 \sss_{k,1} + \zeze_{k,1}$ for $k = 1, \dots 10$, where $\sss_{k,1} = (W_1^{\top}W_1)^{-1}W_1^{\top}\mumu_k$. In particular, 

\begin{align*}
    &\sss_{1,1} = \begin{pmatrix}
        \frac{1}{\sigma_{N1}+1}\\
        0\\
        0\\
        0\\
        0\\
        0\\
        0\\
        0\\
        0\\
        0
    \end{pmatrix}, 
    \sss_{2,1}  = \begin{pmatrix}
        0\\
        \frac{1}{\sigma_{N2}+1}\\
        0\\
        0\\
        0\\
        0\\
        0\\
        0\\
        0\\
        0
    \end{pmatrix}
    \sss_{3,1}= \begin{pmatrix}
        0\\
        0\\
        \frac{1}{\sigma_{N1}+1}\\
        \frac{1}{\sigma_{N2}+1}\\
        0\\
        0\\
        0\\
        0\\
        0\\
        0
    \end{pmatrix},   
    \sss_{4,1} = \begin{pmatrix}
        0\\
        0\\
        -\frac{\sigma_{N1}}{\sigma_{N1}+1}\\
        -\frac{\sigma_{N2}}{\sigma_{N2}+1}\\
        0\\
        0\\
        0\\
        0\\
        0\\
        0
    \end{pmatrix}, 
     \sss_{5,1} = \begin{pmatrix}
        0\\
        0\\
        0\\
        0\\
       -0.5\\
        0\\
        0\\
        0\\
        0\\
        0
    \end{pmatrix}, 
     \sss_{6,1} = \begin{pmatrix}
        0\\
        0\\
        0\\
        0\\
        0\\
        -0.5\\
        0\\
        0\\
        0\\
        0
    \end{pmatrix}, \\
    &\sss_{7,1} = \begin{pmatrix}
        0\\
        0\\
        0\\
        0\\
        0\\
        0\\
        \frac{-N^{-1}\sum_{i=1}^{N}x_{i1}^2}{1 + N^{-1}\sum_{i=1}^{N}x_{i1}^2}\\
        0\\
        0\\
        0
    \end{pmatrix}, 
     \sss_{8,1} = \begin{pmatrix}
        0\\
        0\\
        0\\
        0\\
        0\\
        0\\
        0\\
         \frac{-N^{-1}\sum_{i=1}^{N}x_{i1}^2}{1 + N^{-1}\sum_{i=1}^{N}x_{i1}^2}\\
        0\\
        0
    \end{pmatrix}, 
     \sss_{9,1} = \begin{pmatrix}
        0\\
        0\\
        0\\
        0\\
        0\\
        0\\
        0\\
        0\\
         \frac{-N^{-1}\sum_{i=1}^{N}x_{i2}^2}{1 + N^{-1}\sum_{i=1}^{N}x_{i2}^2}\\
        0
    \end{pmatrix}, 
     \sss_{10,1} = \begin{pmatrix}
        0\\
        0\\
        0\\
        0\\
        0\\
        0\\
        0\\
        0\\
        0\\
         \frac{-N^{-1}\sum_{i=1}^{N}x_{i2}^2}{1 + N^{-1}\sum_{i=1}^{N}x_{i2}^2}
    \end{pmatrix}. 
\end{align*}

Define $\mathcal{S}_{N,1} = \sum_{k=1}^{10}\sss_{k,1} \sss_{k,1}^{\top} $. 
Now we have 
\begin{align*}
    &\mathcal{S}_{N,1}  =
\begin{pmatrix}
\begin{array}{cccc}
\frac{1}{(1+\sigma_{N1})^2} & 0 & 0 & 0\\
0 & \frac{1}{(1+\sigma_{N2})^2} & 0 & 0\\
0 & 0 & \frac{1 + \sigma_{N1}^2}{(1+\sigma_{N1})^2} & \frac{1 + \sigma_{N1}\sigma_{N2}}{(\sigma_{N1}+1)(\sigma_{N2}+1)} \\
0 & 0 & \frac{1 + \sigma_{N1}\sigma_{N2}}{(\sigma_{N1}+1)(\sigma_{N2}+1) } & \frac{1 + \sigma_{N2}^2}{(1+\sigma_{N2})^2}
\end{array} & \mathit{0}_{4 \times 6}\\
\mathit{0}_{6 \times 4} & \text{diag}\left(\begin{array}{c}
      0.25\\
      0.25\\
      \frac{\left(N^{-1}\sum_{i=1}^{N}x_{i1}^2\right)^2}{\left(1 + N^{-1}\sum_{i=1}^{N}x_{i1}^2\right)^2}\\ \frac{\left(N^{-1}\sum_{i=1}^{N}x_{i1}^2\right)^2}{\left(1 + N^{-1}\sum_{i=1}^{N}x_{i1}^2\right)^2}\\ \frac{\left(N^{-1}\sum_{i=1}^{N}x_{i2}^2\right)^2}{\left(1 + N^{-1}\sum_{i=1}^{N}x_{i2}^2\right)^2}\\ \frac{\left(N^{-1}\sum_{i=1}^{N}x_{i2}^2\right)^2}{\left(1 + N^{-1}\sum_{i=1}^{N}x_{i2}^2\right)^2}) 
\end{array}\right)
\end{pmatrix}.
\end{align*}
It is easy to show  that there exists $\underline{\pi}>0$ such that $\pi_{\min}(\mathcal{S}_{N,1}) >\underline{\pi} $ for all large $N$ as long as $\sigma_{N1} - \sigma_{N2}$ is bounded below by a positive constant for all large $N$, which is true under our assumption that $\sigma_{N1} \to \sigma_1$, $\sigma_{N2} \to \sigma_2$, and $\sigma_1 > \sigma_2$ as well as Assumption \ref{assp:4}. Likewise, we can define $\omom_{1,2}, \dots, \omom_{10,2}$ and $W_2= (\omom_{1,2}, \omom_{2,2}, \dots, \omom_{10,2})$ such that 
\begin{align*}
\omom_{5,2}&=\left( \underbrace{(0,a^*_{11}/\sqrt{J},\boldsymbol{0}_{4+p_z}^{\top}), \dots, (0,a^*_{J1}/\sqrt{J},\boldsymbol{0}_{4+p_z}^{\top})}_{\omom_{5\mathcal{U},2}^{\top}}, \underbrace{(-1/\sqrt{N}, 0), \dots,(-1/\sqrt{N}, 0 )}_{\omom_{5\ThTh,2}^{\top}}  \right)^{\top}, \\
    \omom_{6,2}&=\left( \underbrace{(0,a^*_{12}/\sqrt{J},\boldsymbol{0}_{4+p_z}^{\top}), \dots, (0,a^*_{J2}/\sqrt{J}),\boldsymbol{0}_{4+p_z}^{\top})}_{\omom_{6\mathcal{U},2}^{\top}}, \underbrace{(0,-1/\sqrt{N}), \dots,(0,-1/\sqrt{N} )}_{\omom_{6\ThTh,2}^{\top}}  \right)^{\top},\\
\end{align*}
and $\omom_{k,2} = \omom_{k,1}$ for $k=1,\dots ,4$ and $k= 7, \dots ,10$. We can easily verify that $\omom^{\top}_{p,2}\omom_{q,2}=0$ for $p\neq q$, 
\begin{align}\label{eq: W2W2 top}
    &W_2 W_2^{\top} = \sum_{k=1}^{10} \omom_{k,2}\omom_{k,2}^{\top} \nonumber\\
    &=  \left(
\begin{array}{cc}
    \sum_{k=1}^{10}\omom_{k\mathcal{U},2}\omom_{k\mathcal{U},2}^{\top}  & -(NJ)^{-1/2} \left\{ \left(\begin{array}{c}
     \DD_{i2}\\
        \xx_i   \\
        \boldsymbol{0}_{p_z}\\
        \theta^*_{i}
    \end{array}  \right){\aa_{j}^{*}}^{\top}  \right\}_{j \leq J, i \leq N} \\
    -(NJ)^{-1/2}\left\{\aa_{j}^{*}\left(\DD_{i2}^{\top},\xx_i^{\top},\boldsymbol{0}_{p_z}^{\top},{\theta^*_{i}}^{\top}\right) \right\}_{i \leq N, j \leq J}  & \sum_{k=1}^{10}\omom_{k\ThTh,2} \omom_{k\ThTh,2}^{\top}
\end{array}
\right),
\end{align}
and $W_2^{\top} W_2 = W_1^{\top} W_1$. Similarly, we can write $\mumu_k = W_2 \sss_{k,2} + \zeze_{k,2}$ for $k = 1, \dots 10$, where $\sss_{k,2} = (W_2^{\top}W_2)^{-1}W_2^{\top}\mumu_k$. We can easily verify that $\sss_{k,2} = \sss_{k,1} $ for all $k$ and  thus $\mathcal{S}_{N,2} = \sum_{k=1}^{10}\sss_{k,2} \sss_{k,2}^{\top} = \mathcal{S}_{N,1} $. Therefore, we write $\sss_{k} = \sss_{k,1}=\sss_{k,2}$ and $\mathcal{S}_N = \mathcal{S}_{N,1}=\mathcal{S}_{N,2}$.
It then follows that 
\begin{align}\label{eq: P_NT decomposition}
    \partial \mathbb{P}_{NJ}(\XiXi^*) / \partial \XiXi \partial \XiXi^{\top} 
    &= 2 \delta \left(\sum_{k=1}^{10}\mumu_k\mumu_{k}^{\top} \right)\nonumber\\
    &= \delta\sum_{l=1}^{2}W_{l} \left(\sum_{k=1}^{10}\sss_{k} \sss_{k}^{\top}\right) W_{l}^{\top} + \delta\sum_{l=1}^{2} \left(\sum_{k=1}^{10}  \zeze_{k,l} \zeze_{k,l}^{\top} \right)\nonumber\\
    &= \delta \underline{\pi}\sum_{l=1}^{2} W_{l} W_{l}^{\top} + \delta\sum_{l=1}^{2} W_{l} \left( \mathcal{S}_N -  \underline{\pi} I_{10}  \right) W_{l}^{\top} + \delta\sum_{l=1}^{2} \left(\sum_{k=1}^{10}  \zeze_{k,l} \zeze_{k,l}^{\top} \right).
\end{align}
Note that there exists $\kappa_8>0 $ such that $E(\varrho^{''}_{ijt}(\uu_{j}^{\top}\ee_{it})) > \kappa_8$ by Assumptions \ref{assp:1}, \ref{assp: 1.5} and \ref{assp:3}. Now let $\underline{\delta} = \min\{\kappa_8, \delta \underline{\pi} \} $. Then it follows from  \eqref{eq: P_NT decomposition} that 
\begin{align*}
    \mathbb{H} =& \partial S^*(\XiXi^*)/ \partial \XiXi^{\top}  + \partial \mathbb{P}_{NJ}(\XiXi^*)/\partial \XiXi \partial \XiXi^{\top}\\
               =& \partial S^*(\XiXi^*)/ \partial \XiXi^{\top}  +  \underline{\delta}\sum_{l=1}^{2}W_{l} W_{l}^{\top} + (\delta\underline{\pi} -\underline{\delta})\sum_{l=1}^2  W_{l} W_{l}^{\top} + \delta\sum_{l=1}^2 W_{l} \left( \mathcal{S}_N -  \underline{\pi} I_{10}  \right) W_{l}^{\top} \\
               &+ \delta\sum_{l=1}^2 \left(\sum_{k=1}^{10}  \zeze_{k,l} \zeze_{k,l}^{\top} \right)\\
               \geq& \partial S^*(\XiXi^*)/ \partial \XiXi^{\top}  +  \underline{\delta}\sum_{l=1}^{2} W_{l} W_{l}^{\top}. 
\end{align*}
Moreover, we can write 
\begin{align*}
    &\partial S^*(\XiXi^*)/ \partial \XiXi^{\top}\\
    =& (NJ)^{-1/2}\begin{pmatrix}
      \text{diag} \left(  \left\{  \sum_{i=1}^{N} \sum_{t=1}^{T} E\left(\varrho^{''}_{ijt}\right)\ee^*_{it}{\ee^*_{it}}^{\top}   \right\}_{j \leq J}  \right)&  \left\{  \sum_{t=1}^{T}E\left(\varrho^{''}_{ijt}\right)\ee^*_{it}{\aa_{j}^*}^{\top} \right\}_{j\leq J, i \leq N}\\
       \left\{ \sum_{t=1}^{T} E\left(\varrho^{''}_{ijt}\right)\aa^*_j{\ee_{it}^*}^{\top} \right\}_{i \leq N,j\leq J } &\text{diag} \left(  \left\{  \sum_{j=1}^{J} \sum_{t=1}^{T}E\left(\varrho^{''}_{ijt}\right)\aa^*_j{\aa^*_{j}}^{\top}   \right\}_{i \leq N}\right)
    \end{pmatrix}\\
    =& \underbrace{\underline{\delta} \begin{pmatrix}
        \text{diag} \left(  \left\{(NJ)^{-1/2}  \sum_{i=1}^{N}\sum_{t=1}^{T} \ee^*_{it}{\ee^*_{it}}^{\top}   \right\}_{j \leq J}  \right)& \left\{ (NJ)^{-1/2}\sum_{t=1}^{T}\left(\begin{array}{c}
        \boldsymbol{0}_{4}\\
        \zz_{it}   \\
        \boldsymbol{0}_{2}  
    \end{array} \right){\aa_{j}^{*}}^{\top} \right\}_{j \leq J, i \leq N}\\
          (NJ)^{-1/2}\left\{ \sum_{t=1}^{T}\aa_{j}^{*}\left(\boldsymbol{0}_{4}^{\top} ,\zz_{it}^{\top} ,\boldsymbol{0}_{2}^{\top} \right) \right\}_{i \leq N, j \leq J}&\text{diag} \left(  \left\{(NJ)^{-1/2}  T\sum_{j=1}^{J} \aa^*_j{\aa^*_{j}}^{\top}   \right\}_{i \leq N}  \right)
    \end{pmatrix} }_{I} \\
    &+\underbrace{\underline{\delta} \begin{pmatrix}
         \mathit{0}_{(6+p_z )J \times (6+p_z )J}& (NJ)^{-1/2} \left\{ \sum_{t=1}^{T} \left(\begin{array}{c}
        \DD_{it} \\
        \xx_i\\
        \boldsymbol{0}_{p_z}\\
        \thth^*_{i}
    \end{array} \right) {\aa_{j}^{*}}^{\top}  \right\}_{j \leq J, i \leq N} \\
           (NJ)^{-1/2}\left\{\aa_{j}^{*}\left( \DD_{it}^{\top} ,\xx_i^{\top},\boldsymbol{0}_{p_z}^{\top}, {\thth^*_{i}}^{\top}\right) \right\}_{i \leq N, j \leq J}& \mathit{0}_{2N \times 2N}
    \end{pmatrix} }_{II}\\
    &+(NJ)^{-1/2}\underbrace{\sum_{t=1}^{T}\begin{pmatrix}
      \text{diag} \left(  \left\{  \sum_{i=1}^{N} \left(E\left(\varrho^{''}_{ijt}\right) - \underline{\delta}\right)\ee^*_{it}{\ee^*_{it}}^{\top}   \right\}_{j \leq J}  \right)&  \left\{ \left(E\left(\varrho^{''}_{ijt}\right)- \underline{\delta}\right)\ee^*_{it}{\aa_{j}^*}^{\top} \right\}_{j\leq J, i \leq N}\\
       \left\{ \left( E\left(\varrho^{''}_{ijt}\right)- \underline{\delta} \right)\aa^*_j{\ee_{it}^*}^{\top} \right\}_{i \leq N,j\leq J } &\text{diag} \left(  \left\{  \sum_{j=1}^{J} \left(E\left(\varrho^{''}_{ijt}\right)- \underline{\delta}\right)\aa^*_j{\aa^*_{j}}^{\top}   \right\}_{i \leq N}\right)
    \end{pmatrix} }_{III}.
\end{align*}
For $I$, we have 
\begin{align}\label{eq: I}
    I \geq \kappa_5\cdot I_{(6+ p_z)J+2N}
\end{align} 
by Assumption \ref{assp:9}.
From \eqref{eq: W1W1 top} and \eqref{eq: W2W2 top}, we have 
\begin{align}\label{eq: II}
II + \underline{\delta} \sum_{l=1}^{2} W_{l} W_{l}^{\top} = \underline{\delta}\cdot \begin{pmatrix}
    \sum_{l=1}^{2}\sum_{k=1}^{10}\omom_{k\mathcal{U},l}\omom_{k\mathcal{U},l}^{\top}   &\mathit{0}_{(6+p_z)J \times 2N }  \\
    \mathit{0}_{2N \times (6+p_z)J}  & \sum_{l=1}^{2} \sum_{k=1}^{10}\omom_{k\ThTh,l} \omom_{k\ThTh,l}^{\top}
\end{pmatrix}
    \geq 0_{PJ + 2N, PJ + 2N}. 
\end{align}
For the last term, we have, for $J,N$ large enough, 
\begin{align}\label{eq: III}
    III = \frac{1}{NJ} \sum_{j=1}^{J} \sum_{i=1}^{N} \sum_{t=1}^{T}\left(E\left(\varrho^{''}_{ijt}\right) - \underline{\delta}\right)\varsigma_{ijt}\varsigma_{ijt}^{\top} \geq 0_{PJ + 2N, PJ + 2N}, 
\end{align}
where $\varsigma_{ijt} = \left( \underbrace{0, \dots, {\ee^*_{it}}^{\top},\dots, 0  }_{(6+p_z)J},\underbrace{0, \dots, {\aa^*_{j}}^{\top},\dots, 0  }_{2N}   \right)^{\top}$, by the definition of $\underline{\delta}$. It then follows from \eqref{eq: I}, \eqref{eq: II} and \eqref{eq: III} that 
\begin{align*}
    \mathbb{H} \geq \partial S^*(\XiXi^*)/ \partial \XiXi^{\top}  +  \underline{\delta}\cdot \omom \omom^{\top} = I + II + III +  \underline{\delta}\cdot \omom \omom^{\top} \geq \kappa_5 \cdot I_{(6+p_z)J+2N}, 
\end{align*}
and thus
\begin{align}\label{eq: H inv bound}
    \mathbb{H}^{-1} \leq \kappa_5^{-1} \cdot I_{(6+p_z)J+2N}.
\end{align}
Finally, write $\mathbb{H} = \mathbb{H}_d + \mathcal{C}, $ where 
\begin{align*}
    \mathcal{C} = &\begin{pmatrix}
      \mathit{0}_{(6+p_z )J \times (6+p_z )J}&  (NJ)^{-1/2}\left\{  \sum_{t=1}^{T}E\left(\varrho^{''}_{ijt}\right)\ee^*_{it}{\aa_{j}^*}^{\top} \right\}_{j\leq J, i \leq N}\\
       (NJ)^{-1/2}\left\{  \sum_{t=1}^{T}E\left(\varrho^{''}_{ijt}\right)\aa^*_j{\ee_{it}^*}^{\top} \right\}_{i \leq N,j\leq J } &\mathit{0}_{2N \times 2N} 
    \end{pmatrix} \\
    &+ 2\delta\left(\sum_{k=1}^{10}\mumu_k\mumu_{k}^{\top} \right).
\end{align*}
Note that 
\begin{align*}
    \mathbb{H}^{-1} -  \mathbb{H}_d^{-1} &= -\mathbb{H}_d^{-1}\mathcal{C}\mathbb{H}_d^{-1} + \mathbb{H}_d^{-1}\mathcal{C}\mathbb{H}^{-1} \mathcal{C}\mathbb{H}_d^{-1},
\end{align*}
and thus $\left\|\mathbb{H}^{-1} - \mathbb{H}_d^{-1} \right\|_{\max} \leq \left\|\mathbb{H}_d^{-1}\mathcal{C}\mathbb{H}_d^{-1}\right\|_{\max} + \left\|\mathbb{H}_d^{-1}\mathcal{C}\mathbb{H}^{-1} \mathcal{C}\mathbb{H}_d^{-1}\right\|_{\max}. $ Inequality \eqref{eq: H inv bound} implies that $\mathbb{H}_d^{-1}\mathcal{C}\mathbb{H}^{-1} \mathcal{C}\mathbb{H}_d^{-1}\leq \kappa_5^{-1}\mathbb{H}_d^{-1}\mathcal{C}^2\mathbb{H}_d^{-1}$, and thus the $l$-th diagonal element of $\mathbb{H}_d^{-1}\mathcal{C}\mathbb{H}^{-1} \mathcal{C}\mathbb{H}_d^{-1}$ is smaller than the $l$-th diagonal element of $\kappa_5^{-1}\mathbb{H}_d^{-1}\mathcal{C}^2\mathbb{H}_d^{-1}$. It then follows that $\left\|\mathbb{H}_d^{-1}\mathcal{C}\mathbb{H}^{-1} \mathcal{C}\mathbb{H}_d^{-1} \right\|_{\max} \leq \kappa_5^{-1}\left\|\mathbb{H}_d^{-1}\mathcal{C}^2\mathbb{H}_d^{-1}\right\|_{\max} $ and therefore 
\begin{align*}
    \left\|\mathbb{H}^{-1} - \mathbb{H}_d^{-1} \right\|_{\max} \leq \left\|\mathbb{H}_d^{-1}\mathcal{C}\mathbb{H}_d^{-1}\right\|_{\max} + \kappa_5^{-1}\left\|\mathbb{H}_d^{-1}\mathcal{C}^2\mathbb{H}_d^{-1}\right\|_{\max}
\end{align*}
because the entry with the largest absolute value of a positive semidefinite matrix is always on the diagonal. Since $\mathbb{H}_d^{-1}$ is a block diagonal matrix whose elements are all $O(1)$ from Assumptions \ref{assp:5} and $\|\mathcal{C}  \|_{\max}$ and $\|\mathcal{C}^2  \|_{\max}$ are $O(N^{-1})$, the proof is complete. \\
By this lemma, \eqref{eq: S hat eta expansion}, and the fact that $ \partial \mathbb{P}_{NJ}(\hat{\XiXi})/ \partial \XiXi = \partial \mathbb{P}_{NJ}(\XiXi^*)/ \partial \XiXi=0$, we can write 
\begin{align}\label{eq: eta expansion}
    \hat{\XiXi} - \XiXi^* = \mathbb{H}^{-1}S^*(\hat{\XiXi}) - \mathbb{H}^{-1}S^*(\XiXi^*) - 0.5  \mathbb{H}^{-1} \mathcal{R}(\hat{\XiXi}).
\end{align}
Define 
\begin{align*}
S^*_{NJ}(\XiXi) &= \left(\underbrace{\dots, \frac{1}{\sqrt{NJ}}  \sum_{i=1}^{N}\sum_{t=1}^{T} \varrho^{'}_{ijt}(\uu_j^{\top}\ee_{it} )\ee_{it}^{\top}  ,\dots}_{JK}, \underbrace{\dots,\frac{1}{\sqrt{NJ}}  \sum_{j=1}^{J}\sum_{t=1}^{T} \varrho^{'}_{ijt}(\uu_j^{\top}\ee_{it})\aa_j^{\top} , \dots}_{NK^*}   \right)^{\top},
\end{align*}
$\Breve{S}^*(\XiXi) = S^*_{NJ}(\XiXi) - S^*(\XiXi)$ and $\mathcal{D} = \mathbb{H}^{-1} - \mathbb{H}_d^{-1}$. Note that by the first-order conditions, 
$S^*_{NJ}(\hat{\XiXi})=0$. As a result, we can write 
\begin{align}
    \mathbb{H}^{-1}S^*(\hat{\XiXi}) =& \mathbb{H}_d^{-1}S^*(\hat{\XiXi}) + \mathcal{D}S^*(\hat{\XiXi})\nonumber\\
                                    =&-\mathbb{H}_d^{-1}\Breve{S}^*(\hat{\XiXi}) +  \mathcal{D}S^*(\hat{\XiXi}) \nonumber \\
                                    =&-\mathbb{H}_d^{-1}\Breve{S}^*(\XiXi^*) -\mathbb{H}_d^{-1}\left(  \Breve{S}^*(\hat{\XiXi}) -\Breve{S}^*(\XiXi^*)\right)+  \mathcal{D}S^*(\hat{\XiXi})\label{eq: H inv S hat zeta expansion 1}\\
                                    =&-\mathbb{H}_d^{-1}\Breve{S}^*(\XiXi^*) -\mathbb{H}_d^{-1}\left(  \Breve{S}^*(\hat{\XiXi}) -\Breve{S}^*(\XiXi^*)\right)\nonumber\\
                                      &- \mathcal{D}\Breve{S}^*(\XiXi^*) - \mathcal{D}\left(  \Breve{S}^*(\hat{\XiXi}) -\Breve{S}^*(\XiXi^*)\right) .\label{eq: H inv S hat zeta expansion}
\end{align}
Next, let $\mathcal{R}(\hat{\zeze})_{m}$ denote the vector containing the $\{(m-1)(P)+1\}$th to the $mP$th elements of $\mathcal{R}(\hat{\zeze})$ for $m = 1, \dots J$, and $\{JP + (m-J-1)(K^*)+1\}$th to the $\{JP + (m-J)(K^*)\}$th elements of $\mathcal{R}(\hat{\zeze})$ for $m = J+1 \dots J+N$. We further let $\bar{O}_P(\cdot)$ denote a stochastic order that is uniform in $i$ and $j$. For example, $Q_{ij} = \bar{O}_P(1)$ means that $\max_{i \leq N, j \leq J} \|Q_{ij}\| = O_P(1)$. Then, by the result of Theorem \ref{Thm: convergence of factors }, it can be shown that 
\begin{align}\label{eq: R hat eta m}
    \mathcal{R}(\hat{\XiXi})_m = \bar{O}_P(1)\|\hat{\uu}_j - \uu_j^* \|^2 +  \bar{O}_P(1/\sqrt{N})\|\hat{\uu}_j - \uu_j^* \| + \bar{O}_P(1/N)
\end{align}
for $m = 1, \dots, J$ and 
\begin{align}\label{eq: R hat eta J+m}
    \mathcal{R}(\hat{\XiXi})_{J+m} = \bar{O}_P(1)\|\hat{\thth}_i - \thth_i^* \|^2 +  \bar{O}_P(1/\sqrt{N})\|\hat{\thth}_i - \thth_i^* \| + \bar{O}_P(1/N) 
\end{align}
for $i = 1, \dots, N$. 
We define $\mathcal{D}_{m,s}$ such that 
\begin{align*}
    \mathcal{D}_{m,s} = \begin{cases}
        &\mathcal{D}_{[ (m-1)P:mP, (s-1)P:sP   ] } , \text{if } m,s \in \{1, \dots, J\} \\
        &\mathcal{D}_{[ (m-1)P:mP,JP +  (s-J-1)K^*:JP + (s-J)K^*   ] } , \text{if } m \in \{1, \dots, J\}, s \in \{J+1, \dots N\}\\
        &\mathcal{D}_{[ JP +  (m-J-1)K^*:JP + (m-J)K^*  ,(s-1)P:sP  ] } , \text{if } m \in \{J+1, \dots N\}, s \in \{1, \dots, J\} \\
        &\mathcal{D}_{[ JP +  (m-J-1)K^*:JP + (m-J)K^*  ,JP +  (s-J-1)K^*:JP + (s-J)K^*  ] } , \text{if } m ,s \in \{J+1, \dots N\}
    \end{cases}.
\end{align*} 
Note that $S^*(\XiXi^*)=0$. Then, from equations \eqref{eq: eta expansion} to  \eqref{eq: R hat eta J+m}, recall that we defined  $ \Breve{\varrho}^{'}_{ijt}(\uu_{j}^{\top}\ee_{it} ) = \varrho^{'}_{ijt}({\uu_{j}}^{\top}\ee_{it} ) - E(\varrho^{'}_{ijt}({\uu_{j}}^{\top}\ee_{it} )) $ and $ \Breve{\varrho}^{'}_{ijt} =  \Breve{\varrho}^{'}_{ijt}({{\uu^*_{j}}^{\top}}\ee^*_{it} )$, we can write 
\begin{align}\label{eq: hat thi - star th expansion}
\hat{\thth}_i - \thth^*_i  =& -(\PsPs_{J,i})^{-1}\frac{1}{J} \sum_{j=1}^{J} \sum_{t=1}^{T} \Breve{\varrho}^{'}_{ijt} \aa^*_j\nonumber - \frac{1}{\sqrt{NJ}}\sum_{s=1}^{J}\sum_{m=1}^{N} \mathcal{D}_{J+i, s} \cdot \sum_{t=1}^{T} \Breve{\varrho}^{'}_{mst} \ee^*_{mt}\nonumber \\
                             &- \frac{1}{\sqrt{NJ}}\sum_{s=1}^{J}\sum_{m=1}^{N} \mathcal{D}_{J+i,J+ m} \cdot\sum_{t=1}^{T}  \Breve{\varrho}^{'}_{mst} \aa^*_s \nonumber\\
                                  &- (\PsPs_{J,i})^{-1}\frac{1}{J}\sum_{j=1}^{J}\sum_{t=1}^{T} \left( \Breve{\varrho}^{'}_{ijt}(\hat{\uu}_{j}^{\top}\hat{\ee}_{it} )\hat{\aa}_j-  \Breve{\varrho}^{'}_{ijt}\aa^*_j\right)  \nonumber\\
                              &- \frac{1}{\sqrt{NJ}}\sum_{s=1}^{J}\sum_{m=1}^{N} \mathcal{D}_{J+i, s}\sum_{t=1}^{T}  \left( \Breve{\varrho}^{'}_{mst}(\hat{\uu}_s^{\top}\hat{\ee}_{mt} )\hat{\ee}_{mt}   - \Breve{\varrho}^{'}_{mst}\ee^*_{mt} \right)\nonumber\\
                                &- \frac{1}{\sqrt{NJ}}\sum_{s=1}^{J}\sum_{m=1}^{N} \mathcal{D}_{J+i,J+ m} \sum_{t=1}^{T}\left( \Breve{\varrho}^{'}_{mst}(\hat{\uu}_s^{\top}\hat{\ee}_{mt} )\hat{\aa}_s-  \Breve{\varrho}^{'}_{mst}\aa^*_s\right) \nonumber\\
                                &-0.5(\PsPs_{J,i})^{-1}\mathcal{R}(\hat{\XiXi})_{J+i} - 0.5 \sum_{s=1}^{J}\mathcal{D}_{J+i,s}\mathcal{R}(\hat{\XiXi})_{s} \nonumber\\
                                &-0.5 \sum_{m=1}^{N}\mathcal{D}_{J+i,J+m}\mathcal{R}(\hat{\XiXi})_{J+m}.
\end{align}
\begin{lemma}\label{lemma 7}
    Let $c_1, \dots, c_N$ be a sequence of uniformly bounded constants. Then, under the conditions in Lemma \ref{lemma 6}, we have 
    \begin{align*}
        \frac{1}{N}\sum_{i=1}^{N} c_i(\hat{\thth}_i - \thth^*_i) = O_P\left(\frac{1}{N}\right).
    \end{align*}
\end{lemma}
Proof: Define $d_s = \sqrt{NJ} \cdot N^{-1} \sum_{i=1}^{N} c_i \mathcal{D}_{J+i,s}$, for $s = 1,\dots, N+J$. Lemma \ref{lemma 6} implies that $\max_{1 \leq s \leq N+J}\|d_s\|$ is bounded. From \eqref{eq: hat thi - star th expansion}, we have 
\begin{align}\label{eq: sum ci theta expansion}
     &\frac{1}{N}\sum_{i=1}^{N} c_i(\hat{\thth}_i - \thth^*_i)\nonumber\\
     =& -\frac{1}{NJ}\sum_{j=1}^{J}  \sum_{i=1}^N\sum_{t=1}^{T} c_i(\PsPs_{J,i})^{-1} \Breve{\varrho}^{'}_{ijt}\aa^*_j\nonumber \\
                            &- \frac{1}{NJ}\sum_{m=1}^{N}\sum_{s=1}^{J}\sum_{t=1}^{T} d_s  \Breve{\varrho}^{'}_{mst}\cdot \ee^*_{mt} - \frac{1}{NJ}\sum_{s=1}^{J} \sum_{m=1}^{N}\sum_{t=1}^{T} d_{J+m}  \Breve{\varrho}^{'}_{mst}\cdot \aa^*_s \nonumber\\
                             &- \frac{1}{NJ}\sum_{j=1}^{J}\sum_{i=1}^{N}\sum_{t=1}^{T}  c_i(\PsPs_{J,i})^{-1} \left( \Breve{\varrho}^{'}_{ijt}(\hat{\uu}_j^{\top}\hat{\ee}_{it} )\hat{\aa}_j  -  \Breve{\varrho}^{'}_{ijt}\aa^*_j\right)\nonumber\\
                              &- \frac{1}{NJ}\sum_{m=1}^{N}\sum_{s=1}^{J}\sum_{t=1}^{T} d_s  \left( \Breve{\varrho}^{'}_{mst}(\hat{\uu}_s^{\top}\hat{\ee}_{mt} )\hat{\ee}_{mt}   - \Breve{\varrho}^{'}_{mst}\ee^*_{mt} \right)\nonumber\\
                                &- \frac{1}{NJ}\sum_{s=1}^{J}\sum_{m=1}^{N}\sum_{t=1}^{T} d_{J+m}\left( \Breve{\varrho}^{'}_{mst}(\hat{\uu}_s^{\top}\hat{\ee}_{mt} )\hat{\aa}_s-  \Breve{\varrho}^{'}_{mst}\aa^*_s\right) -0.5\frac{1}{N}\sum_{i=1}^{N}c_i(\PsPs_{J,i})^{-1}\mathcal{R}(\hat{\zeze})_{J+i} \nonumber\\
                                &- 0.5 \frac{1}{\sqrt{NJ}}\sum_{s=1}^{J}d_s\mathcal{R}(\hat{\XiXi})_{s} -0.5\frac{1}{\sqrt{NJ}} \sum_{m=1}^{N}d_{J+m}\mathcal{R}(\hat{\XiXi})_{J+m}.
\end{align}
First, by Lyapunov's CLT, it is easy to see that the first three terms on the RHS of \eqref{eq: sum ci theta expansion} are all $O_P(1/ \sqrt{NJ}).$
Next, it follows from Theorem \ref{Thm: convergence of factors }, \eqref{eq: R hat eta m} and \eqref{eq: R hat eta J+m} that the last three terms on the RHS of \eqref{eq: sum ci theta expansion} are all $O_P(1/ \min\{N,J\}^2)$. Finally, we will show that the remaining three terms on the RHS of \eqref{eq: sum ci theta expansion} are all $O_P(1/N)$, from which the desired result follows. 
Define 
\begin{align*}
    \mathbb{V}_{NJ}(\XiXi) =  \frac{1}{NJ}\sum_{m=1}^{N}\sum_{s=1}^{J}\sum_{t=1}^T d_s  \left( \Breve{\varrho}^{'}_{mst}(\uu_s^{\top}\ee_{mt})\ee_{mt}   - \Breve{\varrho}^{'}_{mst}\ee^*_{mt} \right),
\end{align*}
and $\Delta_{NJ}(\XiXi^{(a)}, \XiXi^{(b)}) = \sqrt{NJ}\left(\mathbb{V}_{NJ}(\XiXi^{(a)}) - \mathbb{V}_{NJ}(\XiXi^{(b)}) \right).$ Note that 
\begin{align*}
    \Delta_{NJ}(\XiXi^{(a)}, \XiXi^{(b)})=& \underbrace{\frac{1}{\sqrt{NJ}}\sum_{m=1}^{N}\sum_{s=1}^{J}\sum_{t=1}^{T} d_s   \Breve{\varrho}^{'}_{mst}({\uu_s^{(a)}}^{\top} \ee_{mt}^{(a)})(\ee^{(a)}_{mt} - \ee^{(b)}_{mt})}_{\Delta_{1,NJ}(\XiXi^{(a)}, \XiXi^{(b)})}  \\
                                        &+\underbrace{\frac{1}{\sqrt{NJ}}\sum_{m=1}^{N}\sum_{s=1}^{J}\sum_{t=1}^{T} d_s  \left( \Breve{\varrho}^{'}_{mst}({\uu_s^{(a)}}^{\top} \ee_{mt}^{(a)})  -   \Breve{\varrho}^{'}_{mst}({\uu_s^{(b)}}^{\top} \ee_{mt}^{(b)}) \right)\ee^{(b)}_{mt}}_{\Delta_{2,NJ}(\zeze^{(a)}, \zeze^{(b)})}.
\end{align*}
 Note that we have 
 \begin{align*}
     \left\|\sum_{t=1}^{T} d_s   \Breve{\varrho}^{'}_{mst}({\uu_s^{(a)}}^{\top} \ee_{mt}^{(a)})(\ee^{(a)}_{mt} - \ee^{(b)}_{mt})\right\| &\lesssim \left\| \thth^{(a)}_m - \thth^{(b)}_m \right\| \text{ and }\\
     \left|\sum_{t=1}^{T} d_s  \left( \Breve{\varrho}^{'}_{mst}({\uu_s^{(a)}}^{\top} \ee_{mt}^{(a)})  -   \Breve{\varrho}^{'}_{mst}({\uu_s^{(b)}}^{\top} \ee_{mt}^{(b)}) \right) \right|  &\lesssim \left| {\uu_{s}^{(a)}}^{\top}\ee_{m}^{(a)} - {\uu_{s}^{(b)}}^{\top}\ee_{m}^{(b)}   \right|.
 \end{align*}
 By Hoeffding's inequality, Lemma 2.2.1 of \cite{vandervaart1996_weak}, and arguments similar to the proof of Lemma \ref{lemma 2},  we can show that for $d(\XiXi^{(a)}, \XiXi^{(b)}) $ sufficiently small, 
 \begin{align*}
      \left\|  \Delta_{1,NJ}(\XiXi^{(a)}, \XiXi^{(b)})    \right\|_{\Psi_2} &\lesssim \left\|\ThTh^{(a)} - \ThTh^{(b)} \right\|_{F} \lesssim d(\XiXi^{(a)}, \XiXi^{(b)}),\\
      \left\|  \Delta_{2,NJ}(\XiXi^{(a)}, \XiXi^{(b)})  \right\|_{\Psi_2} &\lesssim d(\XiXi^{(a)}, \XiXi^{(b)}).
 \end{align*}
Thus, 
\begin{align*}
   \left\|  \Delta_{NJ}(\XiXi^{(a)}, \XiXi^{(b)}) \right\|_{\Psi_2} \leq \left\|  \Delta_{1,NJ}(\XiXi^{(a)}, \XiXi^{(b)})    \right\|_{\Psi_2} + \left\|  \Delta_{2,NJ}(\XiXi^{(a)}, \XiXi^{(b)})  \right\|_{\Psi_2} \lesssim d(\XiXi^{(a)}, \XiXi^{(b)}).
\end{align*}
Therefore, similar to the proof of Lemma \ref{lemma 3}, we can show that for sufficiently small $c>0$,
\begin{align}\label{eq: V bound}
    E\left( \sup_{\XiXi \in \HH^{K^*}(c)} \left|\mathbb{V}_{NJ}(\XiXi)  \right| \right) \lesssim \frac{c}{\min\{\sqrt{N}, \sqrt{J}\}}.
\end{align}
It then follows from \eqref{eq: V bound} and Theorem \ref{Thm: convergence of factors } that $\mathbb{V}_{NJ}(\hat{\XiXi}) = O_P(1/N)$. Thus the fifth term of the right of \eqref{eq: sum ci theta expansion} is $O_P(1/N)$. Similar results can be obtained for the fourth term and sixth term on the right of \eqref{eq: sum ci theta expansion}, and the desired result follows.
\begin{lemma}\label{lemma 8}
    Under the conditions in lemma \ref{lemma 6}, for each $j$ we have 
    \begin{align*}
      \frac{1}{N}\sum_{i=1}^{N}\sum_{t=1}^{T}  \Breve{\varrho}_{ijt}^{'} (\hat{\ee}_{it} - \ee^*_{it}) = O_P\left(\frac{1}{N}\right) \text{ and }   \frac{1}{N}\sum_{i=1}^{N}\sum_{t=1}^{T}  \Breve{\varrho}_{ijt}^{''}\ee^*_{it} (\hat{\ee}_{it} - \ee^*_{it})^{\top} = O_P\left(\frac{1}{N}\right) .
    \end{align*} 
\end{lemma}
Proof: It suffices to prove that for each $l \in \{ 1, \dots, T\}$,
 \begin{align*}
      \frac{1}{N}\sum_{i=1}^{N}  \Breve{\varrho}_{ijl}^{'} (\hat{\thth}_i - \thth^*_i) = O_P\left(\frac{1}{N}\right) \text{ and }   \frac{1}{N}\sum_{i=1}^{N}  \Breve{\varrho}_{ijl}^{''}\ee^*_{il} (\hat{\thth}_i - \thth^*_i)^{\top} = O_P\left(\frac{1}{N}\right) .
    \end{align*} The proof of the two results are similar, thus we only prove the second one to save space. By \eqref{eq: H inv S hat zeta expansion 1} and \eqref{eq: hat thi - star th expansion}, we have 
\begin{align}\label{eq: lm 8 expansion}
    &\frac{1}{N}\sum_{i=1}^{N}  \Breve{\varrho}_{ijl}^{''}\ee^*_{il} (\hat{\thth}_i - \thth^*_i)^{\top}\nonumber\\
=&  -\frac{1}{NJ} \sum_{i=1}^{N} \sum_{s=1}^{J} \sum_{t=1}^{T} \Breve{\varrho}_{ijl}^{''} \Breve{\varrho}^{'}_{ist} \ee^*_{il}{\aa^*_s}^{\top}(\PsPs_{J,i})^{-1}\nonumber\\
&- \frac{1}{NJ} \sum_{i=1}^{N} \sum_{s=1}^{J} \sum_{t=1}^{T} \Breve{\varrho}_{ijl}^{''}\ee^*_{il}\left( \Breve{\varrho}^{'}_{ist}(\hat{\uu}_{s}^{\top}\hat{\ee}_{it} )\hat{\aa}_s-  \Breve{\varrho}^{'}_{ist}\aa^*_s\right)(\PsPs_{J,i})^{-1}\nonumber\\
&- \frac{1}{\sqrt{NJ}}\sum_{s=1}^{J}\sum_{m=1}^{N}\sum_{t=1}^{T}\left( \frac{1}{N}\sum_{i=1}^{N}  \Breve{\varrho}_{ijl}^{''}\ee^*_{il}    \hat{\ee}^{\top}_{mt}\mathcal{D}_{J+i, s}^{\top} \right)E\left(\varrho^{'}_{mst}({\hat{\uu}_{s}}^{\top}\hat{\ee}_{mt} )\right)\nonumber\\
&- \frac{1}{\sqrt{NJ}}\sum_{s=1}^{J}\sum_{m=1}^{N}\sum_{t=1}^{T}\left( \frac{1}{N}\sum_{i=1}^{N}  \Breve{\varrho}_{ijl}^{''}\ee^*_{il}    \hat{\aa}^{\top}_{s}\mathcal{D}_{J+i, J+m}^{\top} \right)E\left(\varrho^{'}_{mst}({\hat{\uu}_{s}}^{\top}\hat{\ee}_{mt} )\right)\nonumber\\
&-\frac{1}{2N}\sum_{i=1}^{N}  \Breve{\varrho}_{ijl}^{''}\ee^*_{il}\mathcal{R}(\hat{\XiXi})^{\top}_{J+i}(\PsPs_{J,i})^{-1} - \frac{1}{2N}\sum_{i=1}^{N} \sum_{s=1}^{J} \Breve{\varrho}_{ijl}^{''}\ee^*_{il} \mathcal{R}(\hat{\XiXi})_{s}^{\top} \mathcal{D}_{J+i,s}^{\top}\nonumber\\
&- \frac{1}{2N}\sum_{i=1}^{N} \sum_{m=1}^{N}  \Breve{\varrho}_{ijl}^{''}\ee^*_{il} \mathcal{R}(\hat{\XiXi})_{J+m}^{\top}\mathcal{D}_{J+i,J+m}^{\top}.
\end{align}
First, we can write 
\begin{align*}
    &-\frac{1}{NJ} \sum_{i=1}^{N} \sum_{s=1}^{J} \sum_{t=1}^{T} \Breve{\varrho}_{ijl}^{''} \Breve{\varrho}^{'}_{ist} \ee^*_{il}{\aa^*_s}^{\top}(\PsPs_{J,i})^{-1}\\
    =& -\frac{1}{NJ} \sum_{i=1}^{N}  \sum_{t=1}^{T} \Breve{\varrho}_{ijl}^{''} \Breve{\varrho}^{'}_{ijt} \ee^*_{il}{\aa^*_j}^{\top}(\PsPs_{J,i})^{-1} -\frac{1}{NJ} \sum_{i=1}^{N} \sum_{s=1,s \neq j}^{J} \sum_{t=1}^{T} \Breve{\varrho}_{ijl}^{''} \Breve{\varrho}^{'}_{ist} \ee^*_{il}{\aa^*_s}^{\top}(\PsPs_{J,i})^{-1}.
\end{align*}
Since $ \Breve{\varrho}_{ijl}^{''}, \Breve{\varrho}^{'}_{ist}$ are bounded and $\max_{i \leq N}\|(\PsPs_{J,i})^{-1} \|=O(1)$ for large $J$ by Assumption \ref{assp:5}, the first term of the RHS of the above equation is $O_P(J^{-1})$. By Lyapunov's CLT, the second term on the RHS of the above equation can be shown to be $O_P((NJ)^{-1/2})$. Thus, the first term on the RHS of \eqref{eq: lm 8 expansion} is $O_P(N^{-1})$. \\
Second, for the second term on the RHS of \eqref{eq: lm 8 expansion}, it can be written as 
\begin{align*}
    O_P(J^{-1})  - \frac{1}{NJ} \sum_{i=1}^{N} \sum_{s=1, s\neq j}^{J} \sum_{t=1}^{T} \Breve{\varrho}_{ijl}^{''}\ee^*_{il}\left( \Breve{\varrho}^{'}_{ist}(\hat{\uu}_{s}^{\top}\hat{\ee}_{it} )\hat{\aa}_s-  \Breve{\varrho}^{'}_{ist}\aa^*_s\right)(\PsPs_{J,i})^{-1}.
\end{align*}
Similar to the proof of Lemma \ref{lemma 7}, the second term of the above expression can be shown to be $O_P(N^{-1})$. So the second term of the RHS of \eqref{eq: lm 8 expansion} is $O_P(N^{-1})$. 
For the third term on the  RHS of \eqref{eq: lm 8 expansion}, its $p,q$th element is given by 
\begin{align*}
    - \frac{1}{NJ}\sum_{s=1}^{J}\sum_{m=1}^{N}\sum_{t=1}^{T} \chichi_{j,s} E(\varrho^{'}_{mst}({\hat{\uu}_{s}}^{\top}\hat{\ee}_{mt} )) \hat{\ee}_{mt},
\end{align*}
 where $\chichi_{j,s} =  N^{-1}\sum_{i=1}^{N} \left( \sqrt{NJ}   \mathcal{D}_{J+i, s,q}\right) \Breve{\varrho}_{ijl}^{''}\ee^*_{il,p} $, and $\mathcal{D}_{J+i, s,q}$ is the $q$th row of $\mathcal{D}_{J+i, s}$. Therefore, 
 \begin{align*}
     &\left|    - \frac{1}{NJ}\sum_{s=1}^{J}\sum_{m=1}^{N}\sum_{t=1}^{T} \chichi_{j,s} E(\varrho^{'}_{mst}({\hat{\uu}_{s}}^{\top}\hat{\ee}_{mt} )) \hat{\ee}_{mt} \right|\\
     \lesssim& \sqrt{\frac{1}{J}\sum_{s=1}^{J} \|\chichi_{j,s}\|^2 } \cdot \sqrt{\frac{1}{NJ}\sum_{s=1}^{J}\sum_{m=1}^{N}\sum_{t=1}^{T} E(\varrho^{'}_{mst}({\hat{\uu}_{s}}^{\top}\hat{\ee}_{mt} ))^2 \|\hat{\ee}_{mt} \|^2  }.
 \end{align*}
 Since $|\sqrt{NJ}   \mathcal{D}_{J+i, s,q}| $ is uniformly bounded by Lemma \ref{lemma 6}, it can be shown that $E\|\chichi_{j,s}\|^2 $ is $O(N^{-1})$. Moreover, for some $\tilde{c}_{mst}$ between ${{\uu}^*_{s}}^{\top}{\ee}^*_{mt}$ and ${\hat{\uu}_{s}}^{\top}\hat{\ee}_{mt}$, 
 \begin{align*}
     E\left(\varrho^{'}_{mst}({\hat{\uu}_{s}}^{\top}\hat{\ee}_{mt} )\right)^2  &= \left\{ E(\varrho^{'}_{mst}) +  E(\varrho^{''}_{mst}(\tilde{c}_{mst}))( {{\uu}^*_{s}}^{\top}{\ee}^*_{mt}- {\hat{\uu}_{s}}^{\top}\hat{\ee}_{mt}   ) \right\}^2 \\
     &\lesssim ( {{\uu}^*_{s}}^{\top}{\ee}^*_{mt}- {\hat{\uu}_{s}}^{\top}\hat{\ee}_{mt}   )^2.
 \end{align*}
 Thus, 
 \begin{align*}
     \sqrt{\frac{1}{NJ}\sum_{s=1}^{J}\sum_{m=1}^{N}\sum_{t=1}^{T} E(\varrho^{'}_{mst}({\hat{\uu}_{s}}^{\top}\hat{\ee}_{mt} ))^2 \|\hat{\ee}_{mt} \|^2  } \lesssim d(\hat{\XiXi}, \XiXi^*) = O_P(1/\min\{\sqrt{N}, \sqrt{J}\})
 \end{align*}
 by Theorem \ref{Thm: convergence of factors }. So, the third term on the RHS of \eqref{eq: lm 8 expansion} is $O_P(N^{-1})$, and the fourth term of the RHS of \eqref{eq: lm 8 expansion} can be shown to be $O_P(N^{-1})$ in the same way. Finally, it follows from Theorem \ref{Thm: convergence of factors } and \eqref{eq: R hat eta J+m} that the fifth term on the RHS of \eqref{eq: lm 8 expansion}  is $O_P( (\min\{\sqrt{N}, \sqrt{J}\}^{-1})^2) = O_P(N^{-1})$. For the sixth term, the absolute value of the $p,q$th element can be written as 
 \begin{align*}
    \frac{1}{2\sqrt{NJ}} \left| \sum_{s=1}^{J}\chichi_{j,s} \mathcal{R}(\hat{\XiXi})_{s} \right|
    \leq \frac{\sqrt{J}}{2\sqrt{N}} \sqrt{\frac{1}{J} \sum_{s=1}^{J} \|\chichi_{j,s}\|^2 } \sqrt{\frac{1}{J} \sum_{s=1}^{J}  \| \mathcal{R}(\hat{\XiXi})_{s}  \|^2 } 
    =O_P(N^{-3/2} ). 
 \end{align*}
 The same bound for the seventh term on the RHS of \eqref{eq: lm 8 expansion} can be obtained using the same argument. Therefore, we get 
 \begin{align*}
     \frac{1}{N}\sum_{i=1}^{N}  \Breve{\varrho}_{ijl}^{''}\ee^*_{il} (\hat{\thth}_i - \thth^*_i)^{\top} =O_P(N^{-1})
 \end{align*}
 as desired by combining the results.
\paragraph{Proof of Theorem \ref{Thm: asymptotic normality of factor}}
We first consider the case where the conditions in Lemma \ref{lemma 6} hold. From the expansion in the proof of Lemma \ref{lm: 5}, we can derive that 
\begin{align*}
    \PhPh_{N,j}(\hat{\uu}_{j} - \uu^*_{j})=& \frac{1}{N}\sum_{i=1}^{N}\sum_{t=1}^{T} E\left( \varrho^{'}_{ijt}(\hat{\uu}_{j}^{\top}\hat{\ee}_{it} )\right) \hat{\ee}_{it} -  \frac{1}{N}\sum_{i=1}^{N}\sum_{t=1}^{T}E\left( \varrho^{'}_{ijt} \right)\left(  \hat{\ee}_{it} -  \ee^*_{it}   \right)\\
    &-\frac{1}{N}\sum_{i=1}^{N}\sum_{t=1}^{T}E\left( \varrho^{''}_{ijt}( {\uu^*_{j}}^{\top}\tilde{\ee}_{it})\right)\ee^*_{it}{\uu^*_{j}}^{\top}\left(\hat{\ee}_{it}-  \ee^*_{it}\right)    \\
    &+  O_P\left( N^{-1} \left\| \hat{\ThTh} - \ThTh^*   \right\|^2_{F} \right) + o_P(\left\|  \hat{\uu}_{j} - \uu_{j}^*  \right\|).
\end{align*}
Then, it follows from Theorem \ref{Thm: convergence of factors } and Lemma \ref{lemma 7} that 
\begin{align*}
    \PhPh_{N,j}(\hat{\uu}_{j} - \uu^*_{j}) & = \frac{1}{N}\sum_{i=1}^{N}\sum_{t=1}^{T} E\left( \varrho^{'}_{ijt}(\hat{\uu}_{j}^{\top}\hat{\ee}_{it} )\right) \hat{\ee}_{it} +  O_P(N^{-1}) + o_P(\left\|  \hat{\uu}_{j} - \uu_{j}^*  \right\|).
\end{align*}
Note that 
\begin{align*}
     &\frac{1}{N}\sum_{i=1}^{N}\sum_{t=1}^{T} E\left( \varrho^{'}_{ijt}(\hat{\uu}_{j}^{\top}\hat{\ee}_{it} )\right) \hat{\ee}_{it}\\
     =&-\frac{1}{N}\sum_{i=1}^{N}\sum_{t=1}^{T}  \Breve{\varrho}^{'}_{ijt}(\hat{\uu}_{j}^{\top}\hat{\ee}_{it}) \hat{\ee}_{it}\\
     =&-\frac{1}{N}\sum_{i=1}^{N}\sum_{t=1}^{T}  \Breve{\varrho}^{'}_{ijt} \hat{\ee}_{it} 
     - \frac{1}{N}\sum_{i=1}^{N}\sum_{t=1}^{T}  \Breve{\varrho}^{''}_{ijt}\cdot (\hat{\uu}_{j}^{\top}\hat{\ee}_{it} - {\uu^*_{j}}^{\top}\ee^*_{it})\hat{\ee}_{it} 
     - 0.5\frac{1}{N}\sum_{i=1}^{N}\sum_{t=1}^{T}  \Breve{\varrho}^{'''}_{ijt}(\tilde{m}_{ijt} ) (\hat{\uu}_{j}^{\top}\hat{\ee}_{it} - {\uu^*_{j}}^{\top}\ee^*_{it})^2\hat{\ee}_{it},
\end{align*}
where $\tilde{m}_{ijt}$ is between $\hat{\uu}_{j}^{\top}\hat{\ee}_{it}$ and  ${\uu^*_{j}}^{\top}\ee^*_{it}$. \\
By Lemma \ref{lemma 8}, we have 
\begin{align*}
    -\frac{1}{N}\sum_{i=1}^{N}\sum_{t=1}^{T}  \Breve{\varrho}^{'}_{ijt} \hat{\ee}_{it} 
    &=   -\frac{1}{N}\sum_{i=1}^{N}\sum_{t=1}^{T}  \Breve{\varrho}^{'}_{ijt} \ee^*_{it} 
        - \frac{1}{N}\sum_{i=1}^{N}\sum_{t=1}^{T}  \Breve{\varrho}^{'}_{ijt}(\hat{\ee}_{it} - \ee^*_{it})\\
    &=-\frac{1}{N}\sum_{i=1}^{N}\sum_{t=1}^{T}  \Breve{\varrho}^{'}_{ijt} \ee^*_{it}  + O_P(N^{-1}).
\end{align*}
Second, 
\begin{align}\label{eq: Th2 expansion }
    &- \frac{1}{N}\sum_{i=1}^{N}\sum_{t=1}^{T}  \Breve{\varrho}^{''}_{ijt}\cdot (\hat{\uu}_{j}^{\top}\hat{\ee}_{it} - {\uu^*_{j}}^{\top}\ee^*_{it})\hat{\ee}_{it} \nonumber\\
    =& - \frac{1}{N}\sum_{i=1}^{N}\sum_{t=1}^{T}  \Breve{\varrho}^{''}_{ijt}\hat{\ee}_{it}  (\hat{\ee}_{it} - \ee^*_{it})^{\top}\hat{\uu}_{j} -  \frac{1}{N}\sum_{i=1}^{N}\sum_{t=1}^{T}  \Breve{\varrho}^{''}_{ijt}\hat{\ee}_{it}  {\ee^*_{it}}^{\top}(\hat{\uu}_{j} - \uu_j^*)\nonumber\\
     =& - \frac{1}{N}\sum_{i=1}^{N}\sum_{t=1}^{T}  \Breve{\varrho}^{''}_{ijt}\ee^*_{it}  (\hat{\ee}_{it} - \ee^*_{it})^{\top}\hat{\uu}_{j} 
        -  \frac{1}{N}\sum_{i=1}^{N}\sum_{t=1}^{T}  \Breve{\varrho}^{''}_{ijt}(\hat{\ee}_{it}-  \ee^*_{it})(\hat{\ee}_{it}-  \ee^*_{it})^{\top}\hat{\uu}_{j} \nonumber\\
        &- \frac{1}{N}\sum_{i=1}^{N}\sum_{t=1}^{T}  \Breve{\varrho}^{''}_{ijt} \ee^*_{it} {\ee^*_{it}}^{\top}(\hat{\uu}_{j} - \uu_j^*) - \frac{1}{N}\sum_{i=1}^{N}\sum_{t=1}^{T}  \Breve{\varrho}^{''}_{ijt} (\hat{\ee}_{it}-  \ee^*_{it}) {\ee^*_{it}}^{\top}(\hat{\uu}_{j} - \uu_j^*).
\end{align}
It then follows from Lemma \ref{lemma 8} and Theorem \ref{Thm: convergence of factors } that the first two terms on the RHS of \eqref{eq: Th2 expansion } are $O_P(N^{-1})$, respectively. It is easy to show that the last two terms on the right of \eqref{eq: Th2 expansion } are both $o_P(\|\hat{\uu}_{j} - \uu_j^*\|)$. Thus, we have 
\begin{align*}
    - \frac{1}{N}\sum_{i=1}^{N}\sum_{t=1}^{T}  \Breve{\varrho}^{''}_{ijt}\cdot (\hat{\uu}_{j}^{\top}\hat{\ee}_{it} - {\uu^*_{j}}^{\top}\ee^*_{it})\hat{\ee}_{it} = O_P(N^{-1}) + o_P(\|\hat{\uu}_{j} - \uu_j^*\|).
\end{align*}
Next, it is easy to show that 
\begin{align*}
     &\left\|\frac{1}{N}\sum_{i=1}^{N}\sum_{t=1}^{T}  \Breve{\varrho}^{'''}_{ijt}(\tilde{m}_{ijt} ) (\hat{\uu}_{j}^{\top}\hat{\ee}_{it} - {\uu^*_{j}}^{\top}\ee^*_{it})^2\hat{\ee}_{it}\right\| \\
     \lesssim& \|\hat{\uu}_{j} - \uu_j^*  \|^2 \frac{1}{N}\sum_{t=1}^{T}  | \Breve{\varrho}^{'''}_{ijt}(\tilde{m}_{ijt} )| 
                + \frac{1}{N}\sum_{t=1}^{T}  | \Breve{\varrho}^{'''}_{ijt}(\tilde{m}_{ijt} )| \cdot \|\hat{\ee}_{it} - \ee_{it}^*  \|^2 .
\end{align*}
Therefore, from Theorem \ref{Thm: convergence of factors } and Lemma \ref{lm: 5}, we have 
\begin{align}\label{eq: 2nd last for th2 }
    \frac{1}{N}\sum_{i=1}^{N}\sum_{t=1}^{T}  \Breve{\varrho}^{'''}_{ijt}(\tilde{m}_{ijt} ) (\hat{\uu}_{j}^{\top}\hat{\ee}_{it} - {\uu^*_{j}}^{\top}\ee^*_{it})^2\hat{\ee}_{it} = O_P(N^{-1}). 
\end{align}
Finally, combining all the results, we get 
\begin{align*}
    \PhPh_{N,j}(\hat{\uu}_{j} - \uu^*_{j}) =-\frac{1}{N}\sum_{i=1}^{N}\sum_{t=1}^{T}  \Breve{\varrho}^{'}_{ijt} \ee^*_{it} + o_P(\| \hat{\uu}_{j} - \uu^*_{j}\|) + O_P(N^{-1}).
\end{align*}
By Assumption \ref{assp:5}, we can show that 
\begin{align}\label{eq: last for th2 }
     \PhPh_{N,j} \to \PhPh_{j} >0 \text{ and }  -\frac{1}{\sqrt{N}}\sum_{i=1}^{N}\sum_{t=1}^{T}  \Breve{\varrho}^{'}_{ijt} \ee^*_{it} \overset{d}{\to} \mathcal{N} \left(0, \lim_{N \to \infty} \frac{1}{N} \sum_{i=1}^{N} \sum_{t=1}^{T}E\left(  (\varrho^{'}_{ijt})^2  \right)\ee^*_{it} {\ee^*_{it}}^{\top}\right).
\end{align}
Thus, the desired result follows from \eqref{eq: 2nd last for th2 } and \eqref{eq: last for th2 } and the fact that 
\begin{align*}
    \lim_{N \to \infty} \frac{1}{N} \sum_{i=1}^{N} \sum_{t=1}^{T}E\left(  (\varrho^{'}_{ijt})^2  \right)\ee^*_{it} {\ee^*_{it}}^{\top} = -\PhPh_{j}.
\end{align*}
For the general case where the additional assumptions for $\XX$ in Lemma \ref{lemma 6} are not satisfied, we note that by similar arguments in the normalization algorithm mentioned in Section \ref{app: Normalization algorithm}, there exists a $(1+ p +p_z +K^*) \times (1+ p +p_z +K^*)$ matrix $Q$ such that $(\Gamma_1, \XX, \ZZ_t, \Theta ) Q^{\top}$ satisfies the conditions in Lemma \ref{lemma 6}. The corresponding estimates and values of $\UU_t$ will be $\hat{\UU}_{t}Q^{-1}$ and $\UU^*_t Q^{-1}$ accordingly. Define 
$$\PhPh^{Q}_{N,j} = \frac{1}{N}\sum_{i=1}^{N}\sum_{t=1}^{T} E\left(\varrho^{''}_{ijt}\right)Q\ee^*_{it}{\ee^*_{it}}^{\top}Q^{\top} = Q\PhPh_{N,j}Q^{\top}.$$ 
By the previous result, we have 
\begin{align*}
     \PhPh^{Q}_{N,j} {Q^{-1}}^{\top}(\hat{\uu}_{j} - \uu^*_{j}) &=-\frac{1}{N}\sum_{i=1}^{N}\sum_{t=1}^{T}  \Breve{\varrho}^{'}_{ijt} Q \ee^*_{it} + o_P(\| \hat{\uu}_{j} - \uu^*_{j}\|) + O_P(N^{-1}), \text{ and thus }\\
     \PhPh_{N,j} (\hat{\uu}_{j} - \uu^*_{j}) &=-\frac{1}{N}\sum_{i=1}^{N}\sum_{t=1}^{T}  \Breve{\varrho}^{'}_{ijt}  \ee^*_{it} + o_P(\| \hat{\uu}_{j} - \uu^*_{j}\|) + O_P(N^{-1}) 
\end{align*}
by straightforward calculations. This completes the proof of Theorem \ref{Thm: asymptotic normality of factor}. 
\subsection{Proof of Theorem \ref{Thm: Selection of factor under extension}}\label{subsect: proof of selection of factors}
Recall that we defined $M_{t} = (m_{ijt}), $ where $m_{ijt}= \uu_j^{\top}\ee_{it}$, $t = 1, \dots , T$. With a little abuse of notation, for $\XiXi_{K_1} \in \HH^{K_1} $ and $\XiXi_{K_2} \in \HH^{K_2}$, let 
$$ d(\XiXi_{K_1}, \XiXi_{K_2}) = \max_{t: t = 1, \dots, T} \left\|M_{K_1,t} - M_{K_2,t}    \right\|_{F}.  $$
We further define the following functions for $\XiXi_K \in \HH^K$ and the corresponding $\uu_{K,j}$s and $\thth_{K,i}$s:
\begin{align*}
     w_{ij}(\uu_{K,j}, \thth_{K,i}) &=  \rhrh_{ij}(\uu_j^*,\thth^*_i) -\rhrh_{ij}(\uu_{K,j},\thth_{K,i}),  \\
     l^*_{NJ}(\XiXi_K) &= \frac{1}{NJ} \sum_{i=1}^{N}\sum_{j=1}^{J}w_{ij}(\uu_{K,j}, \thth_{K,i}),\\
     \bar{l}^*_{NJ}(\XiXi_K) &= \frac{1}{NJ} \sum_{i=1}^{N}\sum_{j=1}^{J}E\left(w_{ij}(\uu_{K,j}, \thth_{K,i})\right),\\
     \mathbb{W}_{NJ}(\XiXi_K) &=  l^*_{NJ}(\XiXi_K) - \bar{l}^*_{NJ}(\XiXi_K)  = \frac{1}{NJ} \sum_{i=1}^{N}\sum_{j=1}^{J} \left(  w_{ij}(\uu_{K,j}, \thth_{K,i}) - E\left(w_{ij}(\uu_{K,j}, \thth_{K,i})\right) \right).
\end{align*}
Following the proof of \cite{Bai_Ng-2002-Econometrica} and \cite{Chen_etal-2021-Econometrica}, it suffices to show that for some $\delta>0$, 
\begin{align}
    -l_{NJ}(\hat{\XiXi}_K) + l_{NJ}(\hat{\XiXi}_{K^*}) > \delta + o_P(1) \text{ for } K < K^*,\label{eq: thm 3 eqn1} \\
    -l_{NJ}(\hat{\XiXi}_K) + l_{NJ}(\hat{\XiXi}_{K^*}) = O_P(\min\{N,J\}^{-1}  ) \text{ for } K > K^*. \label{eq: thm 3 eqn2}
\end{align}
\paragraph{Proof of \eqref{eq: thm 3 eqn1}} Suppose $K < K^*$. We can write 
\begin{align*}
    -l_{NJ}(\hat{\XiXi}_K) + l_{NJ}(\hat{\XiXi}_{K^*}) = l^*_{NJ}(\hat{\XiXi}_K) - l^*_{NJ}(\hat{\XiXi}_{K^*}) = \bar{l}^*_{NJ}(\hat{\XiXi}_K)+ \mathbb{W}_{NJ}(\hat{\XiXi}_K) - l^*_{NJ}(\hat{\XiXi}_{K^*}).
\end{align*}
By the arguments in the proof of Lemma \ref{lemma: 1}, it can be shown that $|\mathbb{W}_{NJ}(\hat{\XiXi}_K)| = o_P(1)$. Further, since $|\rhrh_{ij}(\uu_j^*,\thth^*_i) -\rhrh_{ij}(\uu_j,\thth_i)|\lesssim \sum_{t=1}^{T}|{\uu_j^*}^{\top}\ee^*_{it} - {\uu_j}^{\top}\ee_{it} |$, it follows from Lemma \ref{lemma: 1} that 
\begin{align*}
   | l^*_{NJ}(\hat{\XiXi}_{K^*})| \lesssim \frac{1}{NJ}\sum_{i=1}^{N}\sum_{j=1}^{J}\sum_{t=1}^{T}|({\uu_j^*})^{\top}\ee^*_{it} - ({\hat{\uu}_{K^*,j}})^{\top}\hat{\ee}_{K^*,it} |\leq T \cdot d(\XiXi^*, \hat{\XiXi}_{K^*}) =  o_P(1).
\end{align*}
Therefore, it remains to show that $\bar{l}^*_{NJ}(\hat{\XiXi}_K) \geq \delta$. 
By the arguments in Lemma \ref{lemma: 1}, we can show that $\bar{l}^*_{NJ}(\hat{\XiXi}_K) \gtrsim d^2(\hat{\XiXi}_K,\XiXi^*)$. Next, similar to the derivation of \eqref{eq: Bound for ThetaA}, we have 
\begin{align*}
    (NJ)^{-1/2}\| \hat{\ThTh}_K ({\hat{\AAA}_K})^{\top} - {\ThTh^*} ({\AAA^*})^{\top} \|_{F} \lesssim d(\hat{\XiXi}_K,\XiXi^*) .
\end{align*}
Then, by the arguments in Lemma 2 of \cite{Chen_etal-2021-Econometrica}, we have 
\begin{align*}
    \left\|\left[I_{J} - \hat{A}_{K}\left\{({\hat{A}_{K}})^{\top}\hat{A}_{K}\right\}^{-1}({\hat{A}_{K}})^{\top}\right]A^* \right\|_{F}/\sqrt{J} \lesssim d(\hat{\XiXi}_K,\XiXi^*) .
\end{align*}
It then follows from the arguments of (A.34) to (A.37) in \cite{Chen_etal-2021-Econometrica} that \eqref{eq: thm 3 eqn1} holds. 
\paragraph{Proof of \eqref{eq: thm 3 eqn2}} Suppose $K > K^*$. we can write 
\begin{align*}
     -l_{NJ}(\hat{\XiXi}_K) + l_{NJ}(\hat{\XiXi}_{K^*}) = \mathbb{W}_{NJ}(\hat{\XiXi}_K) - \mathbb{W}_{NJ}(\hat{\XiXi}_{K^*}) +  \bar{l}^*_{NJ}(\hat{\XiXi}_K) - \bar{l}^*_{NJ}(\hat{\XiXi}_{K^*}).
\end{align*}
Similar to the proof of Lemma \ref{lemma 3} and Theorem \ref{Thm: convergence of factors }, we can show that for sufficiently small $c$ and sufficiently large $N$ and $J$, it holds that 
\begin{align*}
    E\left( \sup_{d(\XiXi_{K}, \XiXi^*) \leq c } \left|\mathbb{W}_{NJ}(\XiXi_K)\right| \right)\lesssim \frac{c}{\min\left\{\sqrt{N}, \sqrt{J}\right\}}.
\end{align*}
and $d(\hat{\XiXi}_K, \XiXi^*) = O_P(\min\{\sqrt{N}, \sqrt{J}\}^{-1} )$. It then follows that 
\begin{align*}
    \mathbb{W}_{NJ}(\hat{\XiXi}_K)  = O_P(\min\{\sqrt{N}, \sqrt{J}\}^{-2}).
\end{align*}
Similarly we can show that 
\begin{align*}
    \mathbb{W}_{NJ}(\hat{\XiXi}_{K^*})  = O_P(\min\{\sqrt{N}, \sqrt{J}\}^{-2}).
\end{align*}
Finally, consider $\bar{l}^*_{NJ}(\hat{\XiXi}_K) - \bar{l}^*_{NJ}(\hat{\XiXi}_{K^*}).$ By Taylor expansion and the fact that $\exp(c)/(1+ \exp(c))^2 $ is bounded above for $c \in \mathbb{R}$, it is easy to show that 
\begin{align*}
    |\bar{l}^*_{NJ}(\hat{\XiXi}_K)| \lesssim d^2(\hat{\XiXi}_K, \XiXi^*) \text{ and } |\bar{l}^*_{NJ}(\hat{\XiXi}_{K^*})| \lesssim d^2(\hat{\XiXi}_{K^*}, \XiXi^*).
\end{align*}
The rest of the proof follows from the remaining arguments in the proof of Theorem 3 in \cite{Chen_etal-2021-Econometrica}. 

\subsection{Proof of Theorem \ref{Thm: ext convergence of factors } }\label{subsect: Proof of ext convergence of factors}
Following the proof of Theorem \ref{Thm: convergence of factors }, it suffices to show the following three lemmas: 
\begin{lemma}\label{lemma: ext 1}
Under Assumptions \ref{assp:1},\ref{assp: 1.5}, \ref{assp:3} and \ref{assp:7}, $d(\hat{\XiXi},\XiXi^*) = o_p(1)$ as $J,N \to \infty.$
\end{lemma}
Proof: The proof largely follows the arguments in the proof of Lemma \ref{lemma: 1}. Therefore, we omit the full details, and instead highlight the key step that in this context, we can derive an upper bound similar to that in \eqref{eq: lm 1 upper bound}, given by 
\begin{align*}
    &\left| \rhrh_{ij}(\uu_j,\thth_i) -\rhrh_{ij}(\bar{\uu}_j,\bar{\thth}_i)  \right| \nonumber\\
    \leq& \delta_6 \sum_{t=1}^{T}\left(  | \gamma_{jt} - \bar{\gamma}_{jt}| +|\bbb_{jt}^{\top}\xx_i  - \bar{\bbb}_{jt}^{\top}\xx_i  | +  |\vv_j^{\top}\zz_{it} -{\bar{\vv}_j}^{\top}\zz_{it} |  + |{\aa_{jt}}^{\top}\thth_i- {\bar{\aa}_{jt}}^{\top}\bar{\thth}_i| \right)\nonumber\\
    \leq& \delta_6T\bigg( \|\ggg_{j} - \bar{\ggg}_j\| + (\max_{t}\|\aa_{jt}\|)\|\thth_i - \bar{\thth}_i\|  + \|\bar{\thth}_i\|(\max_{t}\|\aa_{jt} - \bar{\aa}_{jt}\|) + \|\xx_{i}\|\max_{t} \|\bbb_{jt} - \bar{\bbb}_{jt}\|\\
    &~~~~~~~~+ \|\zz_{it}\|\|\vv_j - \bar{\vv}_j\|\bigg)\nonumber\\
    \leq& 5\delta_6T\epsilon.
\end{align*}

\begin{lemma}
 Define $\mathcal{H}^{K^*}(c) = \left\{ \XiXi \in \mathcal{H}^{K^*}: d(\XiXi, \XiXi^*) \leq c\right\}.$ Under Assumptions \ref{assp:1}, \ref{assp: 1.5}, \ref{extassump: 3}, \ref{assp:3},  \ref{assp:7} and \ref{assump: 11}, for sufficiently small $c>0$ and sufficiently large $N$ and $J$,  for any $\XiXi \in \mathcal{H}^{K^*}(c),$ it holds that 
    $$\| \ThTh- {\ThTh^*} S_\AAA   \|_{F}/\sqrt{N} +  \| \UU_t- \UU^*_t S_\UU \|_{F}/\sqrt{J}  \leq \delta_7c, t=1, \dots, T,  $$
    where $S_{\AAA} = sgn\left(\AAA_1^{\top}\AAA_1^*/J\right)$ \text{ and } $S_\UU$ is a $(1+p + p_z+K^*)\times (1+p +p_z+K^*)$ diagonal matrix whose diagonal elements are 1, except for the last $K^*$ diagonal elements which are equal to $S_{\AAA}$.
\end{lemma}
Proof: Without loss of generality, assume that Assumption \ref{assump: 11} holds with $t_1 = 1$ and $t_2=2$. Recall that we have
\begin{align*}
    \frac{1}{NJ} \|\ThTh \AAA_t^{\top} - \ThTh^* {\AAA_t^*}^{\top} +\XX (\BB_t- \BB_t^*)^{\top} + \mathbf{1_N} (\GG_t- \GG_t^*)^{\top} +  \ZZ_t (\VV - \VV^*)^{\top}  \|_{F}^2 \lesssim d^2(\XiXi, \XiXi^*) \text{ for } t = 1, \dots, T.
\end{align*}
For $t = 1, 2,$ project $\ThTh$ on $(\ThTh^*, \XX, \mathbf{1_N},\ZZ_t)$ and express it as  $\ThTh =  \ThTh^* M_{1t} + \XX M_{2t} + \mathbf{1_N}M_{3t} + \ZZ_t M_{4t} +\RR_{1t}$, where \( M_{1t} \), \( M_{2t} \), $M_{3t}$ and $M_{4t}$ are projection matrices of dimensions \( K^* \times K^*\), \(p \times K^*\), \(1 \times K^*\) and \(  p_z \times K^* \), respectively. The term \( \RR_{1t} \) is an \( N \times K \) matrix that is orthogonal to \( \ThTh^* \), $\XX$, \( \ZZ_t \) and $\mathbf{1_N}$. We have
\begin{align}\label{eq: ext lm 2 main eq}
    &\frac{1}{NJ} \|( \ThTh -\RR_{1t} ) \AAA_t^{\top} - \ThTh^* {\AAA_t^*}^{\top} + \XX (\BB_t - \BB_t^*)^{\top} +\mathbf{1_N} (\GG_t- \GG_t^*)^{\top}+ \ZZ_t (\VV - \VV^*)^{\top}  \|_{F}^2  
    +  \frac{1}{NJ} \|\RR_{1t}  \AAA_t^{\top} \|_{F}^2\nonumber\\
    &\lesssim d^2(\XiXi, \XiXi^*). 
\end{align}
Since $(J^{-1/2}\|\AAA_t\|_{F}) N^{-1/2}\|\RR_{1t}\|_{F} \leq (NJ)^{-1/2}\|\RR_{1t}  \AAA_t^{\top} \|_{F}$, we can derive that
\begin{align}\label{eq: ext lm 2 remainder}
    \frac{1}{\sqrt{N}}\|\RR_{1t}\|_{F}  \lesssim d(\XiXi, \XiXi^*). 
\end{align}
Combining \eqref{eq: ext lm 2 main eq} and \eqref{eq: ext lm 2 remainder}, we obtain
\begin{align*}
     \frac{1}{\sqrt{NJ}} \|& \ThTh^*(M_{1t}\AAA_t^{\top}-  {\AAA_t^*}^{\top}) +\XX\{M_{2t}\AAA_t^{\top}+(\BB_t - \BB_t^*)^{\top} \} \\
     &  +\mathbf{1_N}\{M_{3t}\AAA_t^{\top}+  (\GG_t- \GG_t^*)^{\top} \}+\ZZ_t \{ M_{4t} \AAA_t^{\top} + (\VV - \VV^*)^{\top} \} \|_{F} \lesssim d(\XiXi, \XiXi^*). 
\end{align*}
By Assumption \ref{assump: 11}, we have $\pi_{\min}((\ThTh^*,\XX, \mathbf{1_N},\ZZ_{t})^{\top}(\ThTh^*,\XX, \mathbf{1_N},\ZZ_{t}))/N \geq \kappa_7$ for $t = 1, 2$. Hence we have 
\begin{align}\label{eq: ext lm2 M and A eq}
     \frac{1}{\sqrt{J}}\|M_{1t}\AAA_t^{\top}-  {\AAA_t^*}^{\top}\|_{F} \lesssim d(\XiXi, \XiXi^*)  \text{ and } \frac{1}{\sqrt{J}}\| M_{4t}\AAA_t^{\top} +  (\VV - \VV^*)^{\top}  \|_{F} \lesssim d(\XiXi, \XiXi^*) 
\end{align}
for $t = 1, 2$. On the other hand, we have 
\begin{align*}
    &N^{-1/2}\|\ThTh^*(M_{11} - M_{12}) + \XX(M_{21} - M_{22}) + \mathbf{1_N}(M_{31} - M_{32})+ \ZZ_1 M_{41} -\ZZ_2 M_{42} \| \\
    = & N^{-1/2}\|\RR_{11} - \RR_{12}\| \lesssim d(\XiXi, \XiXi^*). 
\end{align*}
Under Assumption \ref{assump: 11}, we have
\begin{align}\label{eq: ext lm2 M convergence}
    \|M_{4t}\| \lesssim d(\XiXi, \XiXi^*) \text{ for } t = 1,2 \text{ and }\nonumber\\
    \|M_{j1} - M_{j2}\| \lesssim d(\XiXi, \XiXi^*)  \text{ for } j = 1,2, 3. 
\end{align}
Combining \eqref{eq: ext lm2 M and A eq} and \eqref{eq: ext lm2 M convergence}, we can show that
\begin{align}\label{eq: ext V convergence}
    \frac{1}{\sqrt{J}}\|\VV - \VV^* \|\lesssim d(\XiXi, \XiXi^*).
\end{align}

Therefore, by \eqref{eq: ext V convergence}, we have
\begin{align} \label{eq: ext ThTh A + 1N g convergence}
  \frac{1}{\sqrt{NJ}}\| \ThTh \AAA_t^{\top} - \ThTh^* {\AAA_t^*}^{\top}+ \mathbf{1_N} (\GG_t - \GG_t^*)^{\top} + \XX (\BB_t - \BB_t^*)^{\top}\|_{F} \lesssim d(\XiXi, \XiXi^*) \text{ for } t = 1, \dots, T.
\end{align}
Since $\ThTh^{\top}(\mathbf{1_N},X) = 0_{K^*,1+p}$ by the normalization condition, the rest of the proof follows from similar argument in the proof of Lemma \ref{lemma 2}.  

\begin{remark}
    The above proof works for general $\GG^*_t$. When $\gamma_{jt} = \gamma_j$ such that $\GG_t^* = \GG^* = (\gamma_1^*, \dots, \gamma^*_{J})^{\top}$, we can still arrive at \eqref{eq: ext ThTh A + 1N g convergence}. However,  we no longer impose the normalization constraint $\ThTh^{\top}\mathbf{1_N} = \mathbf{0}_{K^*}$. To proceed, we express $\ThTh$ as $\ThTh =  \ThTh^* M_{1} + \XX M_{2} + \mathbf{1_N}M_{3}+\RR_{1}$ using \eqref{eq: ext lm2 M convergence}, where $N^{-1/2}\|\RR_{1}\| \lesssim d(\XiXi, \XiXi^*)$. Similar to the derivation of \eqref{eq: ext lm2 M and A eq}, we can show that
\begin{align*}
    \frac{1}{\sqrt{J}}\|M_{1}\AAA_t^{\top}-{\AAA_t^*}^{\top}\| \lesssim d(\XiXi, \XiXi^*) \text{ for } t= 1, 2.
\end{align*}
Since $\AAA_t^*$ has rank $K^*$, $M_1$ must be invertible and thus we can show that 
    \begin{align*}
        (NJ)^{-1/2}\| \mathbf{1_N} \{M_3M_{1}^{-1} {\AAA_t^*}^{\top} +t(\GG - \GG^*)^{\top} \} \|_{F} \lesssim d(\XiXi, \XiXi^*) \\
        (J)^{-1/2}\|  M_3M_{1}^{-1} {\AAA_t^*}^{\top} +t(\GG - \GG^*)^{\top}  \|_{F}\lesssim d(\XiXi, \XiXi^*),
    \end{align*}
where the second line follows from the additional condition in Assumption \ref{assump: 11}. We write $\GG = m_{5t} \GG^* + \AAA_t^* M_{6t} + r_{3t}$ for $t=1,2$,  such that $r_{3t}$ is orthogonal to $(\GG^* ,\AAA_t^*)$. We have 
\begin{align*}
    (J)^{-1/2}\|  M_3M_{1}^{-1} {\AAA_t^*}^{\top} +t(m_{5t} \GG^* + \AAA_t^* M_{6t} + r_{3t} - \GG^*)^{\top}  \|_{F}\lesssim d(\XiXi, \XiXi^*),\\
      (J)^{-1/2}\|  (M_3M_{1}^{-1} +M_{6t}^{\top})  {\AAA_t^*}^{\top} +t(m_{5t}-1) {\GG^*}^{\top} \|_{F}\lesssim d(\XiXi, \XiXi^*).
\end{align*}
Hence we have $|m_{51} - 1| \lesssim d(\XiXi, \XiXi^*)$ by Assumption \ref{assump: 11}. On the other hand, we have 
\begin{align*}
   J^{-1/2} \|(m_{52}-m_{51})\GG^*  + \AAA_1^* M_{61} +  \AAA_2^* M_{62} \|_{F} \lesssim d(\XiXi, \XiXi^*).
\end{align*}
Hence we have $\|M_{61}\|\lesssim d(\XiXi, \XiXi^*)$ and thus $J^{-1/2} \|\GG -\GG^* \| \lesssim d(\XiXi, \XiXi^*)$. We can then obtain 
\begin{align*}
   \frac{1}{\sqrt{NJ}}\| \ThTh \AAA_t^{\top} - \ThTh^* {\AAA_t^*}^{\top}+ \XX (\BB_t - \BB_t^*)^{\top}\|_{F} \lesssim d(\XiXi, \XiXi^*) \text{ for } t = 1, \dots, T, 
\end{align*}
and the rest follows from the argument in the proof of Lemma 2 in \cite{Chen_etal-2021-Econometrica}. 
\end{remark}

\begin{lemma}\label{lemma: ext 3}
    Under Assumptions \ref{assp:1}, \ref{assp: 1.5}, \ref{extassump: 3}, \ref{assp:3},  \ref{assp:7} and \ref{assump: 11}, for sufficiently small $c$ and sufficiently large $N$ and $J$, it holds that 
    $$E\left( \sup_{\XiXi \in \mathcal{H}^{K^*}(c)} \left|\mathbb{W}_{NJ}(\XiXi)\right| \right) \lesssim \frac{c}{\min\left\{\sqrt{N}, \sqrt{J}\right\}}.$$
\end{lemma}
The proof of Lemma \ref{lemma: ext 3} follows directly from the argument in the proof of Lemma \ref{lemma 3}. We omit the detailed proof here. 

\subsection{Proof of Theorem \ref{Thm: ext asymptotic normality of factor}}\label{subsect: Proof of ext asymptotic normality of factor}
The proof of this theorem largely follows from the arguments in the proof of Theorem \ref{Thm: asymptotic normality of factor}. Therefore, we only present the proof of a key lemma essential to the derivation of the stochastic expansion of $\hat{\thth}_i$.  
In the context of this extension, define
\begin{align*}
   &\PhPh_{N,j} = \frac{1}{N}\sum_{i=1}^{N}\sum_{t=1}^{T} E\left(\varrho^{''}_{ijt}\right)\ee^*_{it}{\ee^*_{it}}^{\top},\quad
    \PsPs_{J,i} = \frac{1}{J}\sum_{j=1}^{J}\sum_{t=1}^{T} E\left(\varrho^{''}_{ijt}\right)\aa^*_{jt}{\aa_{jt}^*}^{\top} \text{ and } \\
      &\mathbb{P}_{NJ}(\XiXi) =  T \delta \Biggl\{\frac{1}{2J}\sum_{l=1}^{K^*} \sum_{q>l}^{K^*}\biggl( \sum_{j=1}^{J} a_{jl1}a_{jq1} \biggr)^2 + \frac{1}{2N} \sum_{l=1}^{K^*} \sum_{q>l}^{K^*} \biggl( \sum_{i=1}^{N} \theta_{il}\theta_{iq} \biggr)^2    + \frac{1}{8J}\sum_{t=1}^{T} \sum_{k=1}^{K^*} \biggl(\sum_{j=1}^{J} a_{jk1}^2 - J \biggr)^2 \\
    &\quad \quad \quad \quad \quad \quad + \frac{1}{2N}\sum_{l=1}^{K^*} \biggl(\sum_{i=1}^{N} \theta_{il} \biggr)^2   + \frac{1}{2N}\sum_{k=1}^{K^*}\sum_{l=1}^{p} \biggl( \sum_{i=1}^{N}\theta_{ik}x_{il}  \biggr) \Biggr\}
\end{align*} 
for some $\delta>0$. We further define
\begin{align*}
    &S^*(\XiXi) \\
=&\left(\underbrace{\dots, \frac{1}{\sqrt{NJ}}  \sum_{i=1}^{N}\sum_{t=1}^{T} E\left(\varrho^{'}_{ijt}(\uu_j^{\top}\ee_{it} )\right)\ee_{it}^{\top}  ,\dots}_{1\times JP}, \underbrace{\dots,\frac{1}{\sqrt{NJ}}  \sum_{j=1}^{J}\sum_{t=1}^{T} E\left(\varrho^{'}_{ijt}(\uu_j^{\top}\ee_{it})\right)\aa_{jt}^{\top} , \dots}_{1 \times NK^*}   \right)^{\top} , 
\end{align*} 
\begin{align*}
    &S(\XiXi) = S^*(\XiXi) + \partial \mathbb{P}_{NJ}(\XiXi)/\partial \XiXi,& \HHH(\XiXi) =  \partial S^*(\XiXi)/ \partial \XiXi^{\top}  + \partial \mathbb{P}_{NJ}(\XiXi)/\partial \XiXi \partial \XiXi^{\top}
\end{align*}
and let $\HHH = \HHH(\XiXi^*).$
Further, define 
\begin{align*}
    \HHH_d = 
    \begin{pmatrix}
      \HHH_d^\mathcal{U}  & 0\\
      0         & \HHH_d^\ThTh
    \end{pmatrix} 
    ,  \HHH_d^\mathcal{U}  = \frac{\sqrt{N}}{\sqrt{J}} \text{diag}\left(\PhPh_{N,1}, \dots,   \PhPh_{N,J}  \right) ,  \HHH_d^\ThTh =\frac{\sqrt{J}}{\sqrt{N}} \text{diag}\left(\PsPs_{J,1}, \dots,   \PsPs_{J,N}  \right).
\end{align*}
We have the following lemma: 
\begin{lemma}\label{ext lemma 6}
    Under Assumptions \ref{assp:1}, \ref{assp: 1.5}, \ref{extassump: 3}, \ref{assp:3}, \ref{assp:5} to \ref{assp:7}, \ref{extassump: 10}, \ref{assump: 11} and the additional condition that $\sum_{i=1}^{N}\xx_i ~= ~\boldsymbol{0}_{p}$ and $\sum_{i=1}^{N} x_{ik}x_{il} =0$ for $l,k \in \{1, \dots, p\}, l\neq k$, the matrix $\HHH$ is invertible and $\left\| \HHH^{-1} - \HHH_d^{-1} \right\|_{\max} = O(1/N)$.
\end{lemma}

Proof: We assume $p=T=K^*=2$ for simplicity, which can be generalized easily. Consider
  \begin{align*}
    \mathbb{P}_{NJ}(\XiXi) = 2 \delta \Biggl\{&   \frac{1}{2N}  \biggl(\sum_{i=1}^{N} \theta_{i1}\theta_{i2}  \biggr)^2 +  \frac{1}{2J}  \biggl(\sum_{j=1}^{J} a_{j11}a_{j21}  \biggr)^2 + \frac{1}{8J} \biggl( \sum_{j=1}^{J} a_{j11}^2-J \biggr)^2+ \frac{1}{8J} \biggl( \sum_{j=1}^{J} a_{j21}^2-J \biggr)^2  \\
    & +  \frac{1}{2N} \biggl(\sum_{i=1}^{N} \theta_{i1} \biggr)^2 +    \frac{1}{2N} \biggl(\sum_{i=1}^{N} \theta_{i2} \biggr)^2+  \frac{1}{2N}\sum_{k=1}^{2}\sum_{p=1}^{2} \biggl(\sum_{i=1}^{N} \theta_{ik}x_{ip} \biggr)^2\Biggr\}.
\end{align*}
We further define
\begin{align*}
   \mumu_1 &=(( \boldsymbol{0}_{6+p_z}^{\top},a^*_{111},0, \boldsymbol{0}_{2}^{\top}), \dots,(\boldsymbol{0}_{6+p_z}^{\top},a^*_{J11},0,\boldsymbol{0}_{2}^{\top}),\boldsymbol{0}_{2N}^{\top}   )^{\top} /\sqrt{J}, \\
    \mumu_2 &=((\boldsymbol{0}_{6+p_z}^{\top},0,a^*_{121},\boldsymbol{0}_{2}^{\top}), \dots,(\boldsymbol{0}_{6+p_z}^{\top},0,a^*_{J21},\boldsymbol{0}_{2}^{\top}),\boldsymbol{0}_{2N}^{\top}    )^{\top}/\sqrt{J}, \\
    \mumu_3 &=((\boldsymbol{0}_{6+p_z}^{\top},a^*_{121},a^*_{111},\boldsymbol{0}_{2}^{\top}), \dots,( \boldsymbol{0}_{6+p_z}^{\top},a^*_{J21},a^*_{J11},\boldsymbol{0}_{2}^{\top}),\boldsymbol{0}_{2N}^{\top}    )^{\top}/\sqrt{J}, \\
    \mumu_4 &=(\boldsymbol{0}_{PJ}^{\top},(\theta^*_{12},\theta^*_{11}), \dots,(\theta^*_{N2},\theta^*_{N1})   )^{\top}/\sqrt{N}\\
     \mumu_5 &=(\boldsymbol{0}_{PJ}^{\top},(1,0), \dots,(1,0)   )^{\top}/\sqrt{N},\\
    \mumu_6 &=(\boldsymbol{0}_{PJ}^{\top},(0,1), \dots,(0,1)   )^{\top}/\sqrt{N},\\
     \mumu_7 &=(\boldsymbol{0}_{PJ}^{\top},(x_{11},0), \dots,(x_{N1},0)   )^{\top}/\sqrt{N},\\
    \mumu_8 &=(\boldsymbol{0}_{PJ}^{\top},(0,x_{11}), \dots,(0,x_{N1})   )^{\top}/\sqrt{N},\\
     \mumu_9 &=(\boldsymbol{0}_{PJ}^{\top},(x_{12},0), \dots,(x_{N2},0)   )^{\top}/\sqrt{N},\\
    \mumu_{10} &=(\boldsymbol{0}_{PJ}^{\top},(0,x_{12}), \dots,(0,x_{N2})   )^{\top}/\sqrt{N},
\end{align*}
such that  $\partial \mathbb{P}_{NJ}(\XiXi^*) / \partial \XiXi \partial \XiXi^{\top} = 2 \delta \left(\sum_{m=1}^{10}\mumu_m\mumu_{m}^{\top} \right)$. At $t = 2$, there exists a $K^* \times K^*$ matrix $Q_2 = (q_{ij2})$ such that $\ThTh \AAA_2^{\top} = \ThTh Q_2^{-1} Q_2 \AAA_2^{\top}$, where
\begin{align*}
    (\ThTh Q_2^{-1})^{\top}\ThTh Q_2^{-1}/N \text{ is diagonal and } (\AAA_2 Q_2^{\top})^{\top}(\AAA_2 Q_2^{\top})/J = I_{K^*}. 
\end{align*}
Define $\breve{\ThTh}_2 =\ThTh Q_2^{-1} $ and $\breve{\AAA}_2 = \AAA_2 Q_2^{\top}$. At $t=1$, define
\begin{align*}
    \omom_{1,1}& =  \left(  \underbrace{(\boldsymbol{0}_{6+p_z}^{\top} ,a^*_{111}/\sqrt{J},0,\boldsymbol{0}_{2}^{\top} ), \dots,(\boldsymbol{0}_{6+p_z}^{\top} ,a^*_{J11}/\sqrt{J},0,\boldsymbol{0}_{2}^{\top} ) }_{\omom_{1\mathcal{U},1}^{\top}}, \underbrace{(-\theta^*_{11}/\sqrt{N}, 0), \dots,(-\theta^*_{N1}/\sqrt{N}, 0 )}_{\omom_{1\ThTh,1}^{\top}}  \right)^{\top}, \\
     \omom_{2,1}& =  \left( \underbrace{(\boldsymbol{0}_{6+p_z}^{\top},0,a^*_{121}/\sqrt{J},\boldsymbol{0}_{2}^{\top}), \dots, (\boldsymbol{0}_{6+p_z}^{\top} ,0,a^*_{J21}/\sqrt{J},\boldsymbol{0}_{2}^{\top})}_{\omom_{2\mathcal{U},1}^{\top}}, \underbrace{(0,-\theta^*_{12}/\sqrt{N}), \dots,(0,-\theta^*_{N2}/\sqrt{N})}_{\omom_{2\ThTh,1}^{\top}}  \right)^{\top}, \\
      \omom_{3,1}& =  \left(  \underbrace{(\boldsymbol{0}_{6+p_z}^{\top},a^*_{121}/\sqrt{J},0,\boldsymbol{0}_{2}^{\top}), \dots, (\boldsymbol{0}_{6+p_z}^{\top},a^*_{J21}/\sqrt{J},0,\boldsymbol{0}_{2}^{\top}) }_{\omom_{3\mathcal{U},1}^{\top}}, \underbrace{(0,-\theta^*_{11}/\sqrt{N}), \dots,(0,-\theta^*_{N1}/\sqrt{N} )}_{\omom_{3\ThTh,1}^{\top}}  \right)^{\top}, \\
       \omom_{4,1}& =  \left( \underbrace{(\boldsymbol{0}_{6+p_z}^{\top} ,0,a^*_{111}/\sqrt{J},\boldsymbol{0}_{2}^{\top}), \dots, (\boldsymbol{0}_{6+p_z}^{\top} ,0,a^*_{J11}/\sqrt{J},\boldsymbol{0}_{2}^{\top})}_{\omom_{4\mathcal{U},1}^{\top}}, \underbrace{(-\theta^*_{12}/\sqrt{N}, 0), \dots,(-\theta^*_{N2}/\sqrt{N}, 0 )}_{\omom_{4\ThTh,1}^{\top}}  \right)^{\top}, \\
       \omom_{5,1}&=\left( \underbrace{(a^*_{111}/\sqrt{J},\boldsymbol{0}_{9+p_z}^{\top}), \dots, (a^*_{J11}/\sqrt{J},\boldsymbol{0}_{9+p_z}^{\top})}_{\omom_{5\mathcal{U},1}^{\top}}, \underbrace{(-1/\sqrt{N}, 0), \dots,(-1/\sqrt{N}, 0 )}_{\omom_{5\ThTh,1}^{\top}}  \right)^{\top}, \\
         \omom_{6,1}&=\left( \underbrace{(a^*_{121}/\sqrt{J},\boldsymbol{0}_{9+p_z}^{\top}), \dots, (a^*_{J21}/\sqrt{J},\boldsymbol{0}_{9+p_z}^{\top})}_{\omom_{6\mathcal{U},1}^{\top}}, \underbrace{(0,-1/\sqrt{N}), \dots,(0,-1/\sqrt{N} )}_{\omom_{6\ThTh,1}^{\top}}  \right)^{\top},\\
       \omom_{7,1}&=\left( \underbrace{(0,0,a^*_{111}/\sqrt{J},\boldsymbol{0}_{7+p_z}^{\top}), \dots, (0,0,a^*_{J11}/\sqrt{J},\boldsymbol{0}_{7+p_z}^{\top})}_{\omom_{7\mathcal{U},1}^{\top}}, \underbrace{(-x_{11}/\sqrt{N}, 0), \dots,(-x_{N1}/\sqrt{N}, 0 )}_{\omom_{7\ThTh,1}^{\top}}  \right)^{\top}, \\
      \omom_{8,1}&=\left( \underbrace{(0,0,a^*_{121}/\sqrt{J},\boldsymbol{0}_{7+p_z}^{\top}), \dots, (0,0,a^*_{J21}/\sqrt{J},\boldsymbol{0}_{7+p_z}^{\top})}_{\omom_{8\mathcal{U},1}^{\top}}, \underbrace{(0,-x_{11}/\sqrt{N}), \dots,(0,-x_{N1}/\sqrt{N} )}_{\omom_{8\ThTh,1}^{\top}}  \right)^{\top},\\
       \omom_{9,1}&=\left( \underbrace{(0,0,0,a^*_{111}/\sqrt{J},\boldsymbol{0}_{6+p_z}^{\top}), \dots, (0,0,0,a^*_{J11}/\sqrt{J},\boldsymbol{0}_{6+p_z}^{\top})}_{\omom_{9\mathcal{U},1}^{\top}}, \underbrace{(-x_{12}/\sqrt{N}, 0), \dots,(-x_{N2}/\sqrt{N}, 0 )}_{\omom_{9\ThTh,1}^{\top}}  \right)^{\top}, \\
      \omom_{10,1}&=\left( \underbrace{(0,0,0,a^*_{121}/\sqrt{J},\boldsymbol{0}_{6+p_z}^{\top}), \dots, (0,0,0,a^*_{J21}/\sqrt{J},\boldsymbol{0}_{6+p_z}^{\top})}_{\omom_{10\mathcal{U},1}^{\top}}, \underbrace{(0,-x_{12}/\sqrt{N}), \dots,(0,-x_{N2}/\sqrt{N} )}_{\omom_{10\ThTh,1}^{\top}}  \right)^{\top}.
\end{align*}
On the other hand, at $t=2$, we define
\begin{align*}
       \omom_{5,2}&=\left( \underbrace{(0,\breve{a}^*_{112}/\sqrt{J},\boldsymbol{0}_{8+p_z}^{\top}), \dots, (0,\breve{a}^*_{J12}/\sqrt{J},\boldsymbol{0}_{8+p_z}^{\top})}_{\omom_{5\mathcal{U},2}^{\top}}, \underbrace{(-1/\sqrt{N}, 0), \dots,(-1/\sqrt{N}, 0 )}_{\omom_{5\ThTh,2}^{\top}}  \right)^{\top}, \\
           \omom_{6,2}&=\left( \underbrace{(0,\breve{a}^*_{122}/\sqrt{J},\boldsymbol{0}_{8+p_z}^{\top}), \dots, (0,\breve{a}^*_{J22}/\sqrt{J},\boldsymbol{0}_{8+p_z}^{\top})}_{\omom_{6\mathcal{U},2}^{\top}}, \underbrace{(0,-1/\sqrt{N}), \dots,(0,-1/\sqrt{N} )}_{\omom_{6\ThTh,2}^{\top}}  \right)^{\top},\\
       \omom_{7,2}&=\left( \underbrace{(\boldsymbol{0}_{4}^{\top},\breve{a}^*_{112}/\sqrt{J},\boldsymbol{0}_{5+p_z}^{\top}), \dots, (\boldsymbol{0}_{4}^{\top},\breve{a}^*_{J12}/\sqrt{J},\boldsymbol{0}_{5+p_z}^{\top})}_{\omom_{7\mathcal{U},2}^{\top}}, \underbrace{(-x_{11}/\sqrt{N}, 0), \dots,(-x_{N1}/\sqrt{N}, 0 )}_{\omom_{7\ThTh,2}^{\top}}  \right)^{\top}, \\
      \omom_{8,2}&=\left( \underbrace{(\boldsymbol{0}_{4}^{\top},\breve{a}^*_{122}/\sqrt{J},\boldsymbol{0}_{5+p_z}^{\top}), \dots, (\boldsymbol{0}_{4}^{\top},\breve{a}^*_{J22}/\sqrt{J},\boldsymbol{0}_{5+p_z}^{\top})}_{\omom_{8\mathcal{U},2}^{\top}}, \underbrace{(0,-x_{11}/\sqrt{N}), \dots,(0,-x_{N1}/\sqrt{N} )}_{\omom_{8\ThTh,2}^{\top}}  \right)^{\top},\\
       \omom_{9,2}&=\left( \underbrace{(\boldsymbol{0}_{5}^{\top},\breve{a}^*_{112}/\sqrt{J},\boldsymbol{0}_{4+p_z}^{\top}), \dots, (\boldsymbol{0}_{5}^{\top},\breve{a}^*_{J12}/\sqrt{J},\boldsymbol{0}_{4+p_z}^{\top})}_{\omom_{9\mathcal{U},2}^{\top}}, \underbrace{(-x_{12}/\sqrt{N}, 0), \dots,(-x_{N2}/\sqrt{N}, 0 )}_{\omom_{9\ThTh,2}^{\top}}  \right)^{\top}, \\
      \omom_{10,2}&=\left( \underbrace{(\boldsymbol{0}_{5}^{\top},\breve{a}^*_{122}/\sqrt{J},\boldsymbol{0}_{4+p_z}^{\top}), \dots, (\boldsymbol{0}_{5}^{\top},\breve{a}^*_{J22}/\sqrt{J},\boldsymbol{0}_{4+p_z}^{\top})}_{\omom_{10\mathcal{U},2}^{\top}}, \underbrace{(0,-x_{12}/\sqrt{N}), \dots,(0,-x_{N2}/\sqrt{N} )}_{\omom_{10\ThTh,2}^{\top}}  \right)^{\top}.
\end{align*}
We define $W_1= (\omom_{1,1}, \omom_{2,1}, \dots, \omom_{10,1})$ and $W_2 = (\omom_{5,2}, \omom_{6,2}, \dots, \omom_{10,2})$. It is easy to check that $\omom^{\top}_{p,t}\omom_{q,t}=0$ for $p\neq q$, $t= 1,2$. We can verify that 
\begin{align}\label{eq: ext W1W1 top}
    &W_1 W_1^{\top} = \sum_{k=1}^{10} \omom_{k,1}\omom_{k,1}^{\top}\nonumber \\
     &=  \left(
\begin{array}{cc}
    \sum_{k=1}^{10}\omom_{k\mathcal{U},1}\omom_{k\mathcal{U},1}^{\top}  & -(NJ)^{-1/2} \left\{ \left(\begin{array}{c}
        \DD_{i1}\\
        \xx_i\\
        \boldsymbol{0}_{2+p_z}\\
        \thth^*_{i}\\
        \boldsymbol{0}_2
    \end{array}  \right){\aa_{j1}^{*}}^{\top}  \right\}_{j \leq J, i \leq N} \\
    -(NJ)^{-1/2}\left\{\aa_{j1}^{*}\left(\DD_{i1}^{\top},\xx_i^{\top},\boldsymbol{0}_{2+p_z}^{\top},{\thth^*_{i}}^{\top},\boldsymbol{0}_{2}^{\top} \right) \right\}_{i \leq N, j \leq J}  & \sum_{k=1}^{10}\omom_{k\ThTh,1} \omom_{k\ThTh,1}^{\top}
\end{array}
\right),
\end{align}
\begin{align}\label{eq: ext W2W2 top}
    &W_2 W_2^{\top} = \sum_{k=5}^{10} \omom_{k,2}\omom_{k,2}^{\top}\nonumber \\
     &=  \left(
\begin{array}{cc}
    \sum_{k=5}^{10}\omom_{k\mathcal{U},2}\omom_{k\mathcal{U},2}^{\top}  & -(NJ)^{-1/2} \left\{ \left(\begin{array}{c}
       \DD_{i2}\\
       \boldsymbol{0}_{2}\\
        \xx_i\\
        \boldsymbol{0}_{p_z+4}\\
    \end{array}  \right){\breve{\aa}_{j2}^{*\top}}  \right\}_{j \leq J, i \leq N} \\
    -(NJ)^{-1/2}\left\{\breve{\aa}_{j2}^{*}\left(\DD_{i2}^{\top},\boldsymbol{0}_{2}^{\top}, \xx_i^{\top},\boldsymbol{0}_{p_z+4}^{\top}\right) \right\}_{i \leq N, j \leq J}  & \sum_{k=5}^{10}\omom_{k\ThTh,2} \omom_{k\ThTh,2}^{\top}
\end{array}
\right)\nonumber\\
&=  \left(
\begin{array}{cc}
    \sum_{k=5}^{10}\omom_{k\mathcal{U},2}\omom_{k\mathcal{U},2}^{\top}  & -(NJ)^{-1/2} \left\{ \left(\begin{array}{c}
        \DD_{i2}\\
         \boldsymbol{0}_{2}\\
        \xx_i\\
        \boldsymbol{0}_{p_z+4}
    \end{array}  \right){\aa_{j2}^{*}}^{\top} Q_{2}^{\top}  \right\}_{j \leq J, i \leq N} \\
    -(NJ)^{-1/2}\left\{Q_2\aa_{j2}^{*}\left(\DD_{i2}^{\top},\boldsymbol{0}_{2}^{\top}, \xx_i^{\top},\boldsymbol{0}_{p_z+4}^{\top}\right) \right\}_{i \leq N, j \leq J}  & \sum_{k=5}^{10}\omom_{k\ThTh,2} \omom_{k\ThTh,2}^{\top}
\end{array}
\right).
\end{align}
By writing
\begin{align*}
    \breve{W}_2 =  \begin{pmatrix}
    I_{PJ}&0\\
    0& \text{diag}(\{Q_2^{-1}\}_{i \leq N})
\end{pmatrix}W_2, \text{ and }\\
\breve{\omom}_{k\ThTh,2} =  \text{diag}(\{Q_2^{-1}\}_{i \leq N})\omom_{k\ThTh,2} \text{ for } k = 5, 6, \dots, 10,   
\end{align*}
we can express $\breve{W}_2 \breve{W}_2^{\top}$ as
\begin{align*}
&\breve{W}_2 \breve{W}_2^{\top}  \\
=&\left(
\begin{array}{cc}
    \sum_{k=5}^{10}\omom_{k\mathcal{U},2}\omom_{k\mathcal{U},2}^{\top}  & -(NJ)^{-1/2} \left\{ \left(\begin{array}{c}
        \DD_{i2}\\
         \boldsymbol{0}_{2}\\
        \xx_i\\
        \boldsymbol{0}_{p_z+4}\\
    \end{array}  \right){\aa_{j2}^{*}}^{\top}   \right\}_{j \leq J, i \leq N} \\
    -(NJ)^{-1/2}\left\{\aa_{j2}^{*}\left(\DD_{i2}^{\top},\boldsymbol{0}_{2}^{\top}, \xx_i^{\top},\boldsymbol{0}_{p_z+4}^{\top}\right)\right\}_{i \leq N, j \leq J}  & \sum_{k=5}^{10}\breve{\omom}_{k\ThTh,2} \breve{\omom}_{k\ThTh,2}^{\top}  
\end{array} 
\right).
\end{align*}
Further, it is easy to see that under our normalization criteria, 
\begin{align*}
    W_1^{\top} W_1 =
    \text{diag}\bigg(&
        \sigma_{N1}+1,
        \sigma_{N2}+1,
        \sigma_{N1}+1 ,
        \sigma_{N2}+1,
        2,
        2\\
        &1 + N^{-1}\sum_{i=1}^{N}x_{i1}^2,
         1 + N^{-1}\sum_{i=1}^{N}x_{i1}^2,
         1 +N^{-1}\sum_{i=1}^{N}x_{i2}^2,
          1 +N^{-1}\sum_{i=1}^{N}x_{i2}^2
\bigg).
\end{align*}
Moreover, we have 
\begin{align*}
    W_2^{\top} W_2 = \text{diag}\bigg(
         2,
        2 ,
        1 + N^{-1}\sum_{i=1}^{N}x_{i1}^2,
         1 + N^{-1}\sum_{i=1}^{N}x_{i1}^2,
         1 +N^{-1}\sum_{i=1}^{N}x_{i2}^2,
          1 +N^{-1}\sum_{i=1}^{N}x_{i2}^2
\bigg).
\end{align*}
At $t=1$, for $k = 1, \dots, 10$,  we project $\mumu_{k}$ onto $W_1$ and write $\mumu_k = W_1 \sss_{k,1} + \zeze_{k,1}$ for $k = 1, \dots 10$, where $\sss_{k,1} = (W_1^{\top}W_1)^{-1}W_1^{\top}\mumu_k$. At $t=2$, we write $\mumu_k = W_2 \sss_{k,2} + \zeze_{k,2}$ for $k = 5, 6,\dots,10$, where $\sss_{k,2} = (W_2^{\top}W_2)^{-1}W_2^{\top}\mumu_k$. For example, at $t=2$, we have 
\begin{align*}
     &\sss_{5,2} = \begin{pmatrix}
       -0.5\\
        0\\
        0\\
        0\\
        0\\
        0
    \end{pmatrix},
     \sss_{6,2} = \begin{pmatrix}
        0\\
        -0.5\\
        0\\
        0\\
        0\\
        0
    \end{pmatrix}
    \sss_{7,2} = \begin{pmatrix}
        0\\
        0\\
        \frac{-N^{-1}\sum_{i=1}^{N}x_{i1}^2}{1 + N^{-1}\sum_{i=1}^{N}x_{i1}^2}\\
        0\\
        0\\
        0
    \end{pmatrix}, \\
     &\sss_{8,2} = \begin{pmatrix}
        0\\
        0\\
        0\\
         \frac{-N^{-1}\sum_{i=1}^{N}x_{i1}^2}{1 + N^{-1}\sum_{i=1}^{N}x_{i1}^2}\\
        0\\
        0
    \end{pmatrix}, 
     \sss_{9,2} = \begin{pmatrix}
        0\\
        0\\
        0\\
        0\\
         \frac{-N^{-1}\sum_{i=1}^{N}x_{i2}^2}{1 + N^{-1}\sum_{i=1}^{N}x_{i2}^2}\\
        0
    \end{pmatrix}, 
     \sss_{10,2} = \begin{pmatrix}
        0\\
        0\\
        0\\
        0\\
        0\\
         \frac{-N^{-1}\sum_{i=1}^{N}x_{i2}^2}{1 + N^{-1}\sum_{i=1}^{N}x_{i2}^2}
    \end{pmatrix}. 
\end{align*}
We now define $ \mathcal{S}_{N,1} = \sum_{k=1}^{10}\sss_{k,1} \sss_{k,1}^{\top}$ and $\mathcal{S}_{N,2} = \sum_{k=5}^{10}\sss_{k,2} \sss_{k,2}^{\top}$. We set $\zeze_{k,2} = \mumu_k $ for $k = 1, \dots, 4$. Recall that we have shown in the proof of Lemma \ref{lemma 6} that there exists $\underline{\pi}$ such that $\pi_{\min}(\mathcal{S}_{N,1}) > \underline{\pi}$ for all large $N$. It is easy to see that $\pi_{\min}(\mathcal{S}_{N,2}) > \underline{\pi}$. We further Let $$\bar{\pi} = \pi_{\max}\left(\begin{pmatrix}
    I_{PJ}&0\\
    0& \text{diag}(\{Q_2^{-1}\}_{i \leq N})
\end{pmatrix}^{\top}\begin{pmatrix}
    I_{PJ}&0\\
    0& \text{diag}(\{Q_2^{-1}\}_{i \leq N})
\end{pmatrix} \right) = \max(1, \pi_{\max}({Q_2^{-1}}^{\top}Q_2^{-1})). $$
We can verify that $\bar{\pi} W_2 W_2^{\top} - \breve{W}_2 \breve{W}_2^{\top} \geq 0_{PJ + 2N, PJ + 2N}$
and thus we have 
\begin{align*}
    \delta\left\{ W_{2} \mathcal{S}_{N,2} W_{2}^{\top} - \bar{\pi}^{-1} \underline{\pi}  \breve{W}_2 \breve{W}_2^{\top}  \right\} 
    =&\delta\{ W_{2} \mathcal{S}_{N,2} W_{2}^{\top} - \underline{\pi}W_{2}  W_{2}^{\top}  \} + \delta \underline{\pi}  \{W_{2}  W_{2}^{\top} - \bar{\pi}^{-1}\breve{W}_2 \breve{W}_2^{\top}  \}\\
    \geq & 0_{PJ + 2N, PJ + 2N}.
\end{align*}
It then follows that 
\begin{align}\label{eq: ext P_NT decomposition}
    \partial \mathbb{P}_{NJ}(\XiXi^*) / \partial \XiXi \partial \XiXi^{\top} 
    &= 2 \delta \left(\sum_{k=1}^{10}\mumu_k\mumu_{k}^{\top} \right)\nonumber\\
    &= \delta \left\{W_{1} \left(\sum_{k=1}^{10}\sss_{k} \sss_{k}^{\top}\right) W_{1}^{\top}  + W_{2} \left(\sum_{k=5}^{10}\sss_{k} \sss_{k}^{\top}\right) W_{2}^{\top}\right\}+ \delta\sum_{l=1}^{2} \left(\sum_{k=1}^{10}  \zeze_{k,l} \zeze_{k,l}^{\top} \right)\nonumber\\
    &= \delta \underline{\pi} W_{1} W_{1}^{\top} + \delta \bar{\pi}^{-1} \underline{\pi}\breve{W}_2 \breve{W}_2^{\top}  + \delta \left\{ W_{1} \left( \mathcal{S}_{N,1} -  \underline{\pi} I_{10}  \right) W_{1}^{\top}\right\}\nonumber\\
    &+ \delta\left\{ W_{2} \mathcal{S}_{N,2} W_{2}^{\top} -  \bar{\pi}^{-1} \underline{\pi}  \breve{W}_2 \breve{W}_2^{\top}  \right\} + \delta\sum_{l=1}^{2} \left(\sum_{k=1}^{10}  \zeze_{k,l} \zeze_{k,l}^{\top} \right).
\end{align}
Recall that there exists $\kappa_9>0 $ such that $E(\varrho^{''}_{ijt}(\uu_{j}^{\top}\ee_{it})) > \kappa_9$ by Assumptions \ref{assp:1}, \ref{assp: 1.5} and \ref{assp:3}. Now let $\underline{\delta} = \min\{\kappa_9, \delta \bar{\pi}^{-1} \underline{\pi}  \} $. Then it follows from  \eqref{eq: ext P_NT decomposition} that 
\begin{align}\label{eq: ext H decomp}
    \mathbb{H} &= \partial S^*(\XiXi^*)/ \partial \XiXi^{\top}  + \partial \mathbb{P}_{NJ}(\XiXi^*)/\partial \XiXi \partial \XiXi^{\top}\nonumber\\
               &\geq \partial S^*(\XiXi^*)/ \partial \XiXi^{\top}  +  \underline{\delta}\sum_{l=1}^{2} W_{l} W_{l}^{\top}. 
\end{align}
Moreover, we can write 
\begin{align*}
    &\partial S^*(\XiXi^*)/ \partial \XiXi^{\top}\\
    =&(NJ)^{-1/2}\begin{pmatrix}
      \text{diag} \left(  \left\{  \sum_{i=1}^{N} \sum_{t=1}^{T} E\left(\varrho^{''}_{ijt}\right)\ee^*_{it}{\ee^*_{it}}^{\top}   \right\}_{j \leq J}  \right)&  \left\{  \sum_{t=1}^{T}E\left(\varrho^{''}_{ijt}\right)\ee^*_{it}{\aa_{jt}^*}^{\top} \right\}_{j\leq J, i \leq N}\\
       \left\{ \sum_{t=1}^{T} E\left(\varrho^{''}_{ijt}\right)\aa^*_{jt}{\ee_{it}^*}^{\top} \right\}_{i \leq N,j\leq J } &\text{diag} \left(  \left\{  \sum_{j=1}^{J} \sum_{t=1}^{T}E\left(\varrho^{''}_{ijt}\right)\aa^*_{jt}{\aa^*_{jt}}^{\top}   \right\}_{i \leq N}\right)
    \end{pmatrix}\\
    =& \underbrace{\underline{\delta} \begin{pmatrix}
        \text{diag} \left(  \left\{(NJ)^{-1/2}  \sum_{i=1}^{N}\sum_{t=1}^{T} \ee^*_{it}{\ee^*_{it}}^{\top}   \right\}_{j \leq J}  \right)& \left\{ (NJ)^{-1/2}\sum_{t=1}^{T}\left(\begin{array}{c}
        \boldsymbol{0}_{6}\\
        \zz_{it}   \\
        \boldsymbol{0}_{2} \\
        D_{it2}\thth^*_{i}
    \end{array} \right){\aa_{jt}^{*}}^{\top} \right\}_{j \leq J, i \leq N}\\
          (NJ)^{-1/2}\left\{ \sum_{t=1}^{T}\aa_{jt}^{*}\left(\boldsymbol{0}_{6}^{\top},\zz_{it}^{\top} ,\boldsymbol{0}_{2}^{\top},D_{it2}{\thth^*_{i}}^{\top} \right) \right\}_{i \leq N, j \leq J}&\text{diag} \left(  \left\{(NJ)^{-1/2}  \sum_{j=1}^{J} \sum_{t=1}^{T} \aa^*_{jt}{\aa^*_{jt}}^{\top}   \right\}_{i \leq N}  \right)
    \end{pmatrix} }_{I} \\
    &+(NJ)^{-1/2}\sum_{t=1}^{T} \underbrace{\underline{\delta} \begin{pmatrix}
         \mathit{0}_{PJ \times PJ}&  \left\{ \left(\begin{array}{c}
        \DD_{it} \\
         D_{it1}\xx_{i}\\
          D_{it2}\xx_{i}\\
        \boldsymbol{0}_{p_z}\\
        D_{it1}\thth^*_{i}\\
        \boldsymbol{0}_{2}
    \end{array} \right) {\aa_{jt}^{*}}^{\top}  \right\}_{j \leq J, i \leq N} \\
           \left\{\aa_{jt}^{*}\left( \DD_{it}^{\top} ,D_{it1}{\xx_{i}}^{\top},D_{it2}{\xx_{i}}^{\top},\boldsymbol{0}_{p_z}^{\top}, D_{it1}{\thth^*_{i}}^{\top},\boldsymbol{0}_{2}\right) \right\}_{i \leq N, j \leq J}& \mathit{0}_{2N \times 2N}
    \end{pmatrix} }_{II}\\
    &+(NJ)^{-1/2}\underbrace{\sum_{t=1}^{T}\begin{pmatrix}
      \text{diag} \left(  \left\{  \sum_{i=1}^{N} \left(E\left(\varrho^{''}_{ijt}\right) - \underline{\delta}\right)\ee^*_{it}{\ee^*_{it}}^{\top}   \right\}_{j \leq J}  \right)&  \left\{ \left(E\left(\varrho^{''}_{ijt}\right)- \underline{\delta}\right)\ee^*_{it}{\aa_{jt}^*}^{\top} \right\}_{j\leq J, i \leq N}\\
       \left\{ \left( E\left(\varrho^{''}_{ijt}\right)- \underline{\delta} \right)\aa^*_{jt}{\ee_{it}^*}^{\top} \right\}_{i \leq N,j\leq J } &\text{diag} \left(  \left\{  \sum_{j=1}^{J} \left(E\left(\varrho^{''}_{ijt}\right)- \underline{\delta}\right)\aa^*_{jt}{\aa^*_{jt}}^{\top}   \right\}_{i \leq N}\right)
    \end{pmatrix} }_{III}.
\end{align*}
The rest of the proof the follows from Assumption \ref{extassump: 10}, \eqref{eq: ext W1W1 top}, \eqref{eq: ext W2W2 top}, \eqref{eq: ext H decomp} and the arguments from the proof of Lemma \ref{lemma 6}. 
\begin{remark}
When the restriction \(\gamma_{jt} = t\gamma_j\) is imposed, the above derivations remain valid with minor modifications. Specifically, the normalization constraint \(\ThTh^{\top} \mathbf{1}_N = \boldsymbol{0}_{K^*}\) no longer applies. Therefore, in the case where \(p = T = K^* = 2\), we define
\begin{align*}
    \mathbb{P}_{NJ}(\XiXi) = 2 \delta \Biggl\{ & \frac{1}{2N} \left( \sum_{i=1}^{N} \theta_{i1}\theta_{i2} \right)^2 + \frac{1}{2J} \left( \sum_{j=1}^{J} a_{j11}a_{j21} \right)^2 + \frac{1}{8J} \left( \sum_{j=1}^{J} a_{j11}^2 - J \right)^2 + \frac{1}{8J} \left( \sum_{j=1}^{J} a_{j21}^2 - J \right)^2 \\
    & + \frac{1}{2N} \sum_{k=1}^{2} \sum_{p=1}^{2} \left( \sum_{i=1}^{N} \theta_{ik} x_{ip} \right)^2 \Biggr\}.
\end{align*}
The definitions of \(\mumu_k\) and \(\omom_{k,t}\) remain the same, except that \(\mumu_5\), \(\mumu_6\), \(\omom_{5,t}\), and \(\omom_{6,t}\) are omitted to adjust for the normalization constraints. The rest of the proof then proceeds accordingly. 
\end{remark}

\subsection{Proof of consistency results for asymptotic variance}\label{subsect: Proof of consistency results for asymptotic variance } 
It suffices to show that for $j =1, \dots, J$, $\| \hat{\PhPh}_{j}  - \PhPh_j\|_{F} = o_P(1)$. Assume that Assumptions \ref{assp:1} to \ref{assp:9} hold.
Note that 
\begin{align*}
   \hat{\PhPh}_{j} =& \frac{1}{N}\sum_{i=1}^{N}\sum_{t=1}^{T} r_{it}\rho^{''}_{ijt}(\hat{\uu}_j^{\top}\hat{\ee}_{it})  \hat{\ee}_{it}{\hat{\ee}_{it}}^{\top}   \\
                  =& \frac{1}{N}\sum_{i=1}^{N}\sum_{t=1}^{T} r_{it}\rho^{''}_{ijt}(\hat{\uu}_j^{\top}\hat{\ee}_{it})  (\hat{\ee}_{it}{\hat{\ee}_{it}}^{\top} - \ee^*_{it}{\ee^*_{it}}^{\top})\\
                  &+ \frac{1}{N}\sum_{i=1}^{N}\sum_{t=1}^{T} r_{it}\left( \rho^{''}_{ijt}(\hat{\uu}_j^{\top}\hat{\ee}_{it})  - \rho^{''}_{ijt}\right)   \ee^*_{it}{\ee^*_{it}}^{\top}\\
                  &+ \frac{1}{N}\sum_{i=1}^{N}\sum_{t=1}^{T} r_{it}   \rho^{''}_{ijt}   \ee^*_{it}{\ee^*_{it}}^{\top}.
\end{align*}
For the first term, 
\begin{align*}
\left\|\frac{1}{N}\sum_{i=1}^{N}\sum_{t=1}^{T} r_{it}\rho^{''}_{ijt}(\hat{\uu}_j^{\top}\hat{\ee}_{it})  (\hat{\ee}_{it}{\hat{\ee}_{it}}^{\top} - \ee^*_{it}{\ee^*_{it}}^{\top})\right\|_{F} &\lesssim \frac{1}{N}\sum_{i=1}^{N}\sum_{t=1}^{T} \|\hat{\ee}_{it}{\hat{\ee}_{it}}^{\top} - \ee^*_{it}{\ee^*_{it}}^{\top}\|_{F}\\
 &\lesssim \frac{1}{N}\sum_{i=1}^{N}  \| \hat{\thth}_{i} - \thth^*_{i}\|\\
&=O_P(N^{-1/2} )
\end{align*}
by Theorem \ref{Thm: asymptotic normality of factor}.
For the second term, we can show that  
\begin{align*}
   \frac{1}{N}\left\| \sum_{i=1}^{N}\sum_{t=1}^{T} r_{it}\left( \rho^{''}_{ijt}(\hat{\uu}_j^{\top}\hat{\ee}_{it})  - \rho^{''}_{ijt}\right)   \ee^*_{it}{\ee^*_{it}}^{\top}\right\|_{F}
   &\lesssim\frac{1}{N}  \sum_{i=1}^{N} \sum_{t=1}^{T} |\hat{\uu}_j^{\top}\hat{\ee}_{it}  -  {\uu^*_j}^{\top}\ee^*_{it} |\\
   &\lesssim \frac{1}{N}  \sum_{i=1}^{N} \sum_{t=1}^{T}  (\| \hat{\ee}_{it}  - \ee^*_{it} \| +   \|\uu^*_j- \hat{\uu}_{j} \|)
   & = O_P(N^{-1/2})
\end{align*}
by Theorem \ref{Thm: asymptotic normality of factor}. Therefore, we have 
\begin{align*}
    \hat{\PhPh}_{j}  - \PhPh_{N,j} = O_P(N^{-1/2}) + \frac{1}{N}\sum_{i=1}^{N}\sum_{t=1}^{T} (r_{it} - E(r_{it}))   \rho^{''}_{ijt}   \ee^*_{it}{\ee^*_{it}}^{\top} = O_P(N^{-1/2})
\end{align*}
and consequently $\| \hat{\PhPh}_{j}  - \PhPh_j\|_{F} = o_P(1)$ since $\PhPh_j = \lim_{N\to \infty}\PhPh_{N,j}. $

\section{Additional Simulation and Real Data Analysis results}\label{app: Additional Simulation Results}

This section provides additional numerical results. Section~\ref{app: Sensitivity Analysis} reports additional simulation results for the main model. Section~\ref{app:result for ext} presents results for the model extensions introduced in Sections~\ref{subsect: ext 2} and~\ref{subsect: gamma_structure}. Finally, Section~\ref{app: subsect Additional Results in Real Data Analysis} reports additional results in the real data analysis under both the main model and its extensions.

\subsection{Additional Simulation results for the Main Model} \label{app: Sensitivity Analysis}
We present three additional sets of simulation results for the main model. Table~\ref{tab: tuning_sensitivity_combined} reports the sensitivity analysis results for the constraint parameter $c$, as discussed in Section~\ref{sect: Simulation Study}. 

We also compare the computation time of the proposed method using the candidate sets $\mathcal{K} = \{K^*\}$ and $\mathcal{K} = \{1, \dots, 10\}$ with that of standard logistic regression (LR) method and logistic regression with a random intercept (LRRI). Specifically, Table~\ref{tab: computational time} reports the total time required to complete five replications across the settings described in Section~\ref{sect: Simulation Study}. When the candidate set $\mathcal{K}$ equals the true latent dimension $K^*$, the proposed method is significantly faster than LRRI, one of the simplest traditional latent variable models that does not account for dependence between items. This demonstrates the superior computational efficiency of our method over existing latent variable approaches. Even with a broader candidate set $\mathcal{K} = \{1, \dots, 10\}$, the proposed method remains comparable in speed to LRRI and is faster in some cases, highlighting its strong scalability for larger datasets even when the number of latent factors needs to be estimated.

Finally, we evaluate the performance of the proposed estimator in estimating the latent variables. As noted in Remark~\ref{rmk: eigengap}, having distinct eigenvalues in $(\ThTh^*)^\top \ThTh^*/N$ is essential for identifying the latent factors. To ensure this, we generate each $\theta_{ik}$ from a truncated standard normal distribution on $[-1, 1]$, multiplied by a constant factor $k/2$. All other parameters are generated as in the procedures described in Section~\ref{sect: Simulation Study}.

Following the evaluation criteria outlined in Section~\ref{subsect: Evaluation Criteria}, we compute the Frobenius losses for $\hat{\AAA}$ and $\hat{\ThTh}$, denoted by ``Aloss'' and ``Tloss'', respectively:
\begin{align*}
\text{ALoss}  = \frac{1}{\sqrt{J}} \left\| \hat{\AAA} - \AAA^* \hat{S}_{\AAA} \right\|_{F}, 
\quad \text{TLoss}  = \frac{1}{\sqrt{N}} \left\| \hat{\ThTh} - \ThTh^* \hat{S}_{\AAA} \right\|_{F},
\end{align*}
where $\hat{S}_{\AAA}$ is a diagonal matrix correcting for sign indeterminacy, as defined in Theorem~\ref{Thm: convergence of factors }.

To further assess the estimator’s performance at the individual parameter level, we compute the mean squared errors for each $a_{jk}$ and $\theta_{ik}$, where $j = 1, \dots, J$, $i = 1, \dots, N$, and $k = 1, \dots, K^*$. The maximum MSE across all simulation trials for the entries in $\AAA$ and $\ThTh$ are reported as ``MAMSE'' and ``MTMSE'', repsectively.

In addition, we construct 95\% confidence intervals for $a_{jk}$ and $\theta_{ik}$ using the asymptotic variance estimation method described in Remark~\ref{rmk: Asymptotic variance estimation}. The empirical coverage probabilities (AECP for $\AAA$ and TECP for $\ThTh$) are computed by aggregating the coverage rates across all parameters and simulation repetitions. The results are summarized in Table~\ref{tab:proposed_metrics}. 

The results validate Theorem~\ref{Thm: convergence of factors }, indicated by decreasing trends in ``ALoss'', ``TLoss'', ``MAMSE'', and ``MTMSE'' with larger $N$ and $J$. Empirical coverage probabilities (AECP and TECP) approach the nominal 95\% level as $N$ increases, supporting Theorem~\ref{Thm: asymptotic normality of factor}. We note that coverage rates worsen as $K^*$ increases from $3$ to $8$, especially when $J = 100$. Therefore, while the theoretical properties are supported in large samples, caution is required when conducting inference for latent variables in practical settings, especially when the estimated number of factors is large relative to the sample size.

\begin{table}[ht]
\scriptsize
\centering
\caption{Summary statistics for the constraint parameter sensitivity analysis. The results for the proposed method under different choices of the constraint parameter $c$ across various combinations of $N$ and $J$ are reported, for $K^* = 3$ and $K^* = 8$, respectively. Definitions of Loss, Bloss, and MMSE are provided in Table~\ref{tab: result under proposed}.}
\label{tab: tuning_sensitivity_combined}
\begin{tabular}{llcccccccc}
\toprule
& & \multicolumn{4}{c}{$N = 5J$} & \multicolumn{4}{c}{$N = 10J$} \\
\cmidrule(lr){3-6} \cmidrule(lr){7-10}
$K^*$ & Metric ($c$) & $J=100$ & $J=200$ & $J=300$ & $J=400$ & $J=100$ & $J=200$ & $J=300$ & $J=400$ \\
\midrule
\multirow{15}{*}{3} 
&Loss ($c=3$)             & 0.54  & 0.36 & 0.28 & 0.24 & 0.48  & 0.32  & 0.26 & 0.22 \\
&Loss ($c=4$)             & 0.55  & 0.36 & 0.28 & 0.25 & 0.48  & 0.32  & 0.26 & 0.22 \\
&Loss ($c=5$)            & 0.55  & 0.36 & 0.29 & 0.24 & 0.48  & 0.32  & 0.26 & 0.22 \\
&Loss ($c=6$)            & 0.55  & 0.36 & 0.29 & 0.24 & 0.48  & 0.32  & 0.26 & 0.22 \\
&Loss ($c=7$)           & 0.55  & 0.36 & 0.28 & 0.24 & 0.48  & 0.32  & 0.26 & 0.22 \\
\addlinespace
&Bloss ($c=3$)            & 0.48  & 0.32 & 0.25 & 0.22 & 0.33  & 0.22  & 0.18 & 0.16 \\
&Bloss ($c=4$)            & 0.48  & 0.32 & 0.25 & 0.22 & 0.33  & 0.22  & 0.18 & 0.16 \\
&Bloss ($c=5$)            & 0.49  & 0.32 & 0.25 & 0.22 & 0.33  & 0.22  & 0.18 & 0.16 \\
&Bloss ($c=6$)            & 0.48  & 0.32 & 0.25 & 0.22 & 0.33  & 0.22  & 0.18 & 0.16 \\
&Bloss ($c=7$)            & 0.49  & 0.32 & 0.25 & 0.22 & 0.33  & 0.22  & 0.18 & 0.16 \\
\addlinespace
&MMSE ($c=3$)          & 0.11  & 0.05 & 0.03 & 0.03 & 0.06  & 0.02  & 0.02 & 0.01 \\
&MMSE ($c=4$)          & 0.15  & 0.05 & 0.04 & 0.03 & 0.06  & 0.03  & 0.02 & 0.01 \\
&MMSE ($c=5$)       & 0.12  & 0.06 & 0.03 & 0.03 & 0.06  & 0.03  & 0.02 & 0.02 \\
&MMSE ($c=6$)           & 0.13  & 0.06 & 0.03 & 0.03 & 0.06  & 0.03  & 0.02 & 0.01 \\
&MMSE ($c=7$)           & 0.13  & 0.05 & 0.03 & 0.03 & 0.06  & 0.02  & 0.02 & 0.01 \\
\midrule
\multirow{15}{*}{8} 
&Loss ($c=3$)           & 1.28  & 0.69 & 0.52 & 0.44 & 1.06  & 0.62  & 0.48 & 0.41 \\
&Loss ($c=4$)               & 1.31  & 0.69 & 0.52 & 0.44 & 1.08  & 0.63  & 0.48 & 0.41 \\
&Loss ($c=5$)              & 1.33  & 0.68 & 0.52 & 0.44 & 1.09  & 0.62  & 0.48 & 0.41 \\
&Loss ($c=6$)              & 1.38  & 0.68 & 0.52 & 0.44 & 1.10  & 0.62  & 0.48 & 0.41 \\
&Loss ($c=7$)            & 1.35  & 0.69 & 0.52 & 0.44 & 1.10  & 0.63  & 0.48 & 0.41 \\
\addlinespace
&Bloss ($c=3$)           & 0.66  & 0.39 & 0.31 & 0.26 & 0.43  & 0.27  & 0.22 & 0.19 \\
&Bloss ($c=4$)            & 0.67  & 0.39 & 0.31 & 0.26 & 0.44  & 0.27  & 0.21 & 0.19 \\
&Bloss ($c=5$)           & 0.67  & 0.39 & 0.31 & 0.26 & 0.44  & 0.27  & 0.22 & 0.19 \\
&Bloss ($c=6$)           & 0.69  & 0.39 & 0.31 & 0.26 & 0.44  & 0.27  & 0.22 & 0.19 \\
&Bloss ($c=7$)           & 0.68  & 0.39 & 0.31 & 0.26 & 0.44  & 0.27  & 0.21 & 0.19 \\
\addlinespace
&MMSE ($c=3$)          & 1.21  & 0.09 & 0.06 & 0.04 & 0.15  & 0.04  & 0.03 & 0.02 \\
&MMSE ($c=4$)          & 1.33  & 0.08 & 0.05 & 0.04 & 0.13  & 0.05  & 0.02 & 0.02 \\
&MMSE ($c=5$)           & 1.10  & 0.08 & 0.05 & 0.04 & 0.13  & 0.04  & 0.03 & 0.02 \\
&MMSE ($c=6$)      & 2.21  & 0.09 & 0.07 & 0.04 & 0.13  & 0.05  & 0.02 & 0.02 \\
&MMSE ($c=7$)          & 1.03  & 0.09 & 0.06 & 0.03 & 0.15  & 0.05  & 0.03 & 0.02\\
\bottomrule
\end{tabular}
\end{table}

\begin{table}[]
\centering
\caption{Computation time (in minutes) over 5 simulations under various settings.}
\begin{tabular}{llllll}
\toprule
N   &  J   & Proposed($\mathcal K = \{K^*\}$) & Proposed($\mathcal K = \{1,\dots, 10\}$) & LR   & LRRI   \\
\midrule
\multicolumn{6}{l}{\textbf{$K^* = 3$}} \\
5J                   & 100 & 0.09            & 22.72          & 0.01 & 9.54   \\
5J                    & 200 & 0.28            & 78.91          & 0.05 & 37.40  \\
5J                  & 300 & 0.63            & 163.12         & 0.10 & 82.80  \\
5J                 & 400 & 1.15            & 386.44         & 0.18 & 152.36 \\
10J                 & 100 & 0.15            & 45.14          & 0.02 & 17.46  \\
10J                   & 200 & 0.55            & 202.89         & 0.09 & 77.52  \\
10J                  & 300 & 1.23            & 584.78         & 0.19 & 180.16 \\
10J                   & 400 & 2.25            & 1132.89        & 0.35 & 332.14 \\
\midrule
\multicolumn{6}{l}{\textbf{$K^* = 8$}} \\
5J                  & 100 & 0.24            & 11.46          & 0.01 & 12.12  \\
5J                   & 200 & 0.70            & 40.65          & 0.04 & 44.56  \\
5J                   & 300 & 1.54            & 74.08          & 0.10 & 100.88 \\
5J                   & 400 & 2.51            & 119.48         & 0.17 & 185.35 \\
10J                  & 100 & 0.42            & 21.09          & 0.02 & 21.92  \\
10J                  & 200 & 1.38            & 73.46          & 0.09 & 91.71  \\
10J                   & 300 & 2.57            & 174.58         & 0.19 & 213.08 \\
10J                  & 400 & 4.59            & 229.82         & 0.37 & 389.39\\
\bottomrule
\end{tabular}
\label{tab: computational time}
\end{table}

\begin{table}[!t]
\small
\centering
\caption{Summary statistics on latent variables estimations for the proposed method across different values of $N$, $K^*$ and $J$.}
\begin{tabular}{llrrrrrrrr}
\toprule
$N$ & $J$ & Aloss & Tloss & MAMSE & MTMSE & AECP & TECP \\
\midrule
\multicolumn{8}{l}{\textbf{$K^* = 3$}} \\
5J  & 100 & 0.39 & 0.36 & 0.28 & 0.11  & 0.88 & 0.94 \\
5J  & 200 & 0.23 & 0.24 & 0.06 & 0.06  & 0.92 & 0.94 \\
5J  & 300 & 0.20 & 0.19 & 0.05 & 0.03  & 0.93 & 0.95 \\
5J  & 400 & 0.17 & 0.17 & 0.04 & 0.02  & 0.94 & 0.95 \\
10J & 100 & 0.25 & 0.35 & 0.08 & 0.12  & 0.89 & 0.94 \\
10J & 200 & 0.18 & 0.23 & 0.07 & 0.05  & 0.91 & 0.95 \\
10J & 300 & 0.14 & 0.19 & 0.03 & 0.04  & 0.93 & 0.95 \\
10J & 400 & 0.12 & 0.17 & 0.02 & 0.03  & 0.94 & 0.95 \\
\midrule
\multicolumn{8}{l}{\textbf{$K^* = 8$}} \\
5J  & 100 & 1.86 & 3.69 & 3.74 & 31.94 & 0.36 & 0.65 \\
5J  & 200 & 0.46 & 0.89 & 0.29 & 1.38  & 0.81 & 0.91 \\
5J  & 300 & 0.32 & 0.61 & 0.14 & 0.49  & 0.87 & 0.93 \\
5J  & 400 & 0.27 & 0.50 & 0.09 & 0.36  & 0.89 & 0.93 \\
10J & 100 & 0.82 & 2.16 & 0.81 & 14.85 & 0.55 & 0.85 \\
10J & 200 & 0.32 & 0.82 & 0.11 & 1.67  & 0.81 & 0.93 \\
10J & 300 & 0.23 & 0.58 & 0.06 & 0.56  & 0.87 & 0.94 \\
10J & 400 & 0.19 & 0.48 & 0.04 & 0.42  & 0.90 & 0.94 \\
\bottomrule
\end{tabular}
\begin{tablenotes}
    \item \textbf{Aloss / Tloss:} Frobenius loss measuring convergence of $\hat{\AAA}/\hat{\ThTh}$.
    \item \textbf{MAMSE / MTMSE:}   Maximum mean squared error across all estimated $a_{jk}/\theta_{ik}$s.
    \item \textbf{AECP / TECP:} Empirical coverage probability of the confidence intervals across all estimated $a_{jk}/\theta_{ik}$s.
\end{tablenotes}
\label{tab:proposed_metrics}
\end{table}

\subsection{Additional Simulation Results for Models in Section~\ref{subsect: ext 2} and~\ref{subsect: gamma_structure}}\label{app:result for ext}
To assess the performance of the model variants introduced in Sections~\ref{subsect: ext 2} and~\ref{subsect: gamma_structure}, we conduct additional simulations using the same combinations of $N$, $J$, and $K^*$ as in Section~\ref{sect: Simulation Study}. For each setting, we consider the candidate set $\mathcal{K} = \{1, 2, \ldots, 10\}$ and generate 100 independent replications.

For the extended model in Section~\ref{subsect: ext 2}, data are generated from the logistic model:
\begin{equation}\label{eq: ext sim logistic}
 P(y_{ijt}=1 \mid \gamma_{jt}, \aa_{jt}, \thth_i, \bbb_{jt}, \xx_i) = \frac{\exp( \gamma_{jt} + \sum_{k=1}^{K^*}a_{jkt}\theta_{ik} + \sum_{l=1}^{5} \beta_{jlt}x_{il} )}{1+\exp(\gamma_{jt} + \sum_{k=1}^{K^*}a_{jkt}\theta_{ik} + \sum_{l=1}^{5} \beta_{jlt}x_{il} )}.
\end{equation}
The variables are generated following a similar procedure to that described in Section~\ref{sect: Simulation Study} for the main model. As before, we slightly abuse notation by using the same symbols before and after normalization. Specifically, the covariates $\xx_i$, intercepts $\gamma_{jt}$, preliminary latent factors $\theta_{ik}$, and missingness indicators $\rr_i$ are generated as in the main setting. The preliminary time-varying regression coefficients $\beta_{jlt}$ are independently sampled from a uniform distribution $U[0.5, 1]$, and the time-varying factor loadings $a_{jkt}$ are sampled from truncated standard normal distributions on $[-3, 3]$. We then apply the normalization procedure described in Section~\ref{app: Normalization algorithm} to ensure identifiability conditions in this model are satisfied. Following normalization, we set half of the coefficient pairs $(\beta_{j1t}, \beta_{j2t})$ to zero. The same procedure is applied independently to the pairs $(\beta_{j3t}, \beta_{j4t})$, and half of the individual coefficients $\beta_{j5t}$ are also set to zero.

For simulations under the constraint $\gamma_{jt} = t \gamma_j$ (Section~\ref{subsect: gamma_structure}), we substitute $\gamma_{jt}$ accordingly in both \eqref{eq: sim logistic} and \eqref{eq: ext sim logistic}, where $\gamma_{j}$ is drawn from $U[-0.25,0.25]$.

The models described above are evaluated using the performance metrics analogous to those introduced in Section \ref{subsect: Evaluation Criteria}, focusing on the convergence of parameters and accuracy in determining the number of factors.  In particular, for the extension in  Section~\ref{subsect: ext 2}, the convergence of the estimated time-varying regression coefficients $\hat{B} = (\hat{B}_1, \dots, \hat{B}_T)$ is assessed by 
\begin{align*}
    \text{Bloss} = \frac{1}{\sqrt{J}} \max_{t = 1, \dots, T}\left\| \hat{B}_t - {B}^*_t \right\|_{F},
\end{align*}
and the convergence of $\hat{\XiXi}$ is evaluated by 
\begin{align*}
    \text{Loss} =\max_{t = 1, \dots, T} \frac{\left\| \hat{\Theta}\hat{A}_t^{\top} - \Theta^* {A^*_t}^{\top} + X(\hat{B}_t -B_t^*) ^{\top} + \mathbf{1_{N}}(\hat{\GG}_{t} -\GG^*_{t}) ^{\top} \right\|_{F}}{\sqrt{NJ}} .
\end{align*}

We replace $\hat{\GG}_{t}$ and $\GG^*_{t}$ by $t\hat{\GG}$ and $t\GG^*$, respectively, in the ``Loss'' metric when the model in Section~\ref{subsect: gamma_structure} is evaluated.
Table~\ref{tab: ext results} presents the simulation results for the model variants described in Sections~\ref{subsect: ext 2} and~\ref{subsect: gamma_structure}, as well as the combined model incorporating both specifications. The results are consistent with the theory of these methods. In particular, the performance metrics ``Loss'', ``Bloss'', and ``MMSE'' decrease as $N$ and $J$ increase, regardless of the number of factors or the ratio between $N$ and $J$. Furthermore, the proposed information criterion is effective, with $P(\hat{K} = K^*)$ consistently equals 1 across all settings and model variants.

\begin{table}[]
\caption{Simulation results for the model variants introduced in Sections~\ref{subsect: ext 2} and~\ref{subsect: gamma_structure} across different combinations of $N$, $J$, and $K^*$.}
\centering
\tiny
\begin{tabular}{llrrrrrrrrrrrrrr}
\toprule
\multicolumn{2}{l}{} & \multicolumn{4}{c}{\textbf{Section 2.3.2}} && \multicolumn{4}{c}{\textbf{Section 2.3.3}} && \multicolumn{4}{c}{\textbf{Sections 2.3.2 and 2.3.3}} \\
 \cmidrule(lr){3-6} \cmidrule(lr){8-11} \cmidrule(lr){13-16}
N                        & J   & Loss & $P(\hat{K}=K^*)$ & Bloss & MMSE  && Loss & $P(\hat{K}=K^*)$ & Bloss & MMSE && Loss & $P(\hat{K}=K^*)$ & Bloss & MMSE  \\
 \midrule
$K^*=3$ &     &           &                  &       &       &           &&                  &       &      &           &&                  &       &       \\
5J                       & 100 & 0.88      & 1                & 1.06  & 0.85  && 0.51      & 1                & 0.42  & 0.11 && 0.83      & 1                & 0.94  & 1.02  \\
5J                       & 200 & 0.51      & 1                & 0.67  & 0.29  && 0.34      & 1                & 0.29  & 0.04 && 0.50      & 1                & 0.60  & 0.23  \\
5J                       & 300 & 0.40      & 1                & 0.55  & 0.16  && 0.28      & 1                & 0.23  & 0.03 && 0.38      & 1                & 0.48  & 0.12  \\
5J                       & 400 & 0.34      & 1                & 0.45  & 0.11  && 0.24      & 1                & 0.20  & 0.02 && 0.33      & 1                & 0.41  & 0.08  \\
10J                      & 100 & 0.66      & 1                & 0.71  & 0.35  && 0.46      & 1                & 0.29  & 0.05 && 0.63      & 1                & 0.63  & 0.33  \\
10J                      & 200 & 0.40      & 1                & 0.47  & 0.11  && 0.31      & 1                & 0.20  & 0.02 && 0.40      & 1                & 0.42  & 0.11  \\
10J                      & 300 & 0.32      & 1                & 0.38  & 0.08  && 0.25      & 1                & 0.16  & 0.02 && 0.31      & 1                & 0.34  & 0.06  \\
10J                      & 400 & 0.27      & 1                & 0.32  & 0.06  && 0.22      & 1             & 0.14  & 0.01 && 0.27      & 1                & 0.29  & 0.05  \\
 \midrule
$K^*=8$ &     &           &                  &       &       &           &&                  &       &      &           &&                  &       &       \\
5J                       & 100 & 7.63      & 1                & 3.55  & 61.55 && 1.24      & 1                & 0.58  & 0.31 && 7.68      & 1                & 3.03  & 51.17 \\
5J                       & 200 & 1.01      & 1                & 0.86  & 0.53  && 0.67      & 1                & 0.35  & 0.09 && 0.99      & 1                & 0.75  & 0.35  \\
5J                       & 300 & 0.73      & 1                & 0.64  & 0.25  && 0.51      & 1                & 0.28  & 0.04 && 0.72      & 1                & 0.58  & 0.17  \\
5J                       & 400 & 0.59      & 1                & 0.55  & 0.17  && 0.43      & 1                & 0.24  & 0.03 && 0.58      & 1                & 0.49  & 0.13  \\
10J                      & 100 & 2.96      & 1                & 1.24  & 24.35 && 1.09      & 1                & 0.40  & 0.15 && 3.21      & 1                & 1.23  & 15.53 \\
10J                      & 200 & 0.79      & 1                & 0.57  & 0.21  && 0.62      & 1                & 0.25  & 0.04 && 0.78      & 1                & 0.50  & 0.19  \\
10J                      & 300 & 0.59      & 1                & 0.45  & 0.11  && 0.48      & 1                & 0.20  & 0.02 && 0.59      & 1                & 0.41  & 0.10  \\
10J                      & 400 & 0.49      & 1                & 0.38  & 0.08  && 0.40      & 1                & 0.17  & 0.01 && 0.49      & 1                & 0.34  & 0.07 \\
\bottomrule
\end{tabular}
\begin{tablenotes}
    \tiny
     \item \textbf{Loss:} Frobenius loss measuring the convergence of $\hat{\Xi}$. 
     \item \textbf{P($\hat{K} = K^*$):} Proportion of instances where the correct number of factors is identified.
     \item \textbf{Bloss:} Frobenius loss measuring convergence of $\hat{B}$.
     \item \textbf{MMSE:} Maximum mean squared error across all elements in $\hat{B}$.
\end{tablenotes}
\label{tab: ext results}
\end{table}

\subsection{Additional Results in Real Data Analysis}\label{app: subsect Additional Results in Real Data Analysis}
This section presents additional results relevant to the real data analysis presented in Section \ref{sect: Application to The Complete Journey dataset }. Table \ref{tab: data analysis - result with size } reports the items selected by the BY procedure for the covariates household size, the corresponding regression coefficients, p-value, BY-adjusted p-value (adj p-value), category, subcategory, average price, and package size, as described in Section \ref{subsect: Statistical Inference}. Table~\ref{tab: data analysis - result with income } reports the same quantities associated with the covariate household income. 

The rest of the section provides additional results for the performance of the model variants discussed in Section \ref{subsect: model extensions} in the prediction setting. In particular, following the setup in Section \ref{subsect: Prediction}, for the method proposed in Section \ref{subsect: ext 2}, we substitute $\hat{\bbb}_{j,T}$ and $\hat{\aa}_{j,T}$ for $\bbb_{j,T+1}$ and $\aa_{j,T+1}$ in the prediction model, respectively. The number of factors $\hat{K}$ selected by each method for different numbers of items $J = 100, 200, 300, 400$ is reported in Table \ref{Tab: real data number of factors}. Moreover, Table \ref{tab:residual_deviance} reports the in-sample and out-of-sample residual deviances $D^{\text{res}} = \sum_{t} \sum_{j=1}^{J} D^{\text{res}}_{jt}$, computed using $t \in \{1,\dots,24\}$ and $t = 25$, respectively. Finally, Table \ref{tab: Full Sensitivity based on Number of Recommendations}  displays the sensitivities for \(10\), \(20\), \(30\), and \(40\) recommendations across different values of \(J\) for the methods based on the model variants alongside the methods presented in Section \ref{subsect: Prediction}.

Consistent with the discussions in Section \ref{subsect: Prediction}, Prop demonstrates superior performance, achieving higher sensitivity (Table~\ref{tab: Full Sensitivity based on Number of Recommendations}) and yielding the smallest in-sample and out-of-sample residual deviances (Table~\ref{tab:residual_deviance}). The diminished effectiveness observed under the intercept constraint $\gamma_{jt} = t\gamma_j$ suggests the absence of a linear temporal trend in consumer preferences, which aligns with expectations in the context of grocery shopping data. This pattern is evident in Figures \ref{fig:res_dev}, where the fit of Prop (sect 2.3.3) compared to other methods significantly worsens as $t$ increases. Furthermore, the uncompetitiveness of the model extension incorporating time-dependent regression coefficients and factor loadings may be attributable to overparameterization. For example, when $J = 100$ with 24 time points and 4 estimated factors, the model entails 192 parameters. 

We note that when two models use the same number of latent factors, the original model and its extensions form nested pairs. For example, given the same latent dimension, Prop is a submodel of Prop (Sect.~2.3.2), and the latter would be expected to achieve a better in-sample fit. However, our comparison is based on models under different latent dimensions chosen by the proposed information criterion. Specifically, the extension with time-varying coefficients and factor loadings consistently selects a smaller number of factors than the original model (Table~\ref{Tab: real data number of factors}).

\begin{table}[p]
\centering
\tiny
\caption{Characteristics and Estimated coefficients of selected products based on household size}
\label{tab: data analysis - result with size }
\begin{tabular}{llrrrrrr}
\toprule
\textbf{Category}          & \textbf{Subcategory}             & \textbf{Price} & \textbf{Size}       & $\mathbf{\hat{\bbb}_1}$ & $\mathbf{\hat{\bbb}_2}$ & \textbf{Adj p-value}&\textbf{p-value}\\
\midrule
Beverages      & DAIRY CASE 100\% PURE JUICE - O & 1.36 & NA         & -0.29 & 0.03  & 0.000 & 0.000 \\
Beverages      & DAIRY CASE TEA WITH SUGAR OR S  & 0.99 & 1 GA       & 0.56  & 0.80  & 0.000 & 0.000 \\
Beverages      & SFT DRNK 2 LITER BTL CARB INCL  & 1.16 & 2 LTR      & 0.08  & 0.46  & 0.000 & 0.000 \\
Beverages      & SFT DRNK 2 LITER BTL CARB INCL  & 1.23 & 2 LTR      & 0.15  & 0.63  & 0.000 & 0.000 \\
Beverages      & SFT DRNK 2 LITER BTL CARB INCL  & 1.16 & 2 LTR      & 0.04  & 0.86  & 0.000 & 0.000 \\
Beverages      & SFT DRNK SNGL SRV BTL CARB (EX  & 1.07 & 20 OZ      & -0.89 & 0.19  & 0.000 & 0.000 \\
Beverages      & SOFT DRINKS 12/18\&15PK CAN CAR & 3.41 & 12 OZ      & 0.10  & 0.74  & 0.000 & 0.000 \\
Beverages      & SOFT DRINKS 12/18\&15PK CAN CAR & 3.33 & 12 OZ      & 0.38  & 0.31  & 0.000 & 0.000 \\
Beverages      & SOFT DRINKS 12/18\&15PK CAN CAR & 3.49 & 12 OZ      & -0.05 & 0.49  & 0.000 & 0.000 \\
Beverages      & SOFT DRINKS 12/18\&15PK CAN CAR & 3.41 & 12 OZ      & 0.24  & 0.65  & 0.000 & 0.000 \\
Breakfast      & HAMBURGER BUNS                  & 0.95 & 12 OZ      & 0.44  & 0.76  & 0.000 & 0.000 \\
Breakfast      & HOT DOG BUNS                    & 0.95 & 11 OZ      & 0.40  & 0.88  & 0.000 & 0.000 \\
Breakfast      & MAINSTREAM WHITE BREAD          & 0.97 & 20 OZ      & -0.08 & 0.73  & 0.000 & 0.000 \\
Breakfast      & MAINSTREAM WHITE BREAD          & 1.65 & 20 OZ      & 0.09  & 0.65  & 0.000 & 0.000 \\
Breakfast      & MAINSTREAM WHITE BREAD          & 1.47 & 24 OZ      & 0.48  & 0.85  & 0.000 & 0.000 \\
Breakfast      & MAINSTREAM WHITE BREAD          & 0.97 & 20 OZ      & -0.17 & 1.03  & 0.000 & 0.000 \\
Breakfast      & SW GDS:DONUTS                   & 0.49 & NA         & -0.17 & 0.10  & 0.023 & 0.003 \\
Dairy and Eggs & CHOCOLATE MILK                  & 1.28 & NA         & -0.18 & 0.50  & 0.000 & 0.000 \\
Dairy and Eggs & CHOCOLATE MILK                  & 2.36 & 1 GA       & 0.16  & 0.60  & 0.000 & 0.000 \\
Dairy and Eggs & CREAM CHEESE                    & 1.55 & 8 OZ       & 0.12  & 0.34  & 0.008 & 0.001 \\
Dairy and Eggs & CREAM CHEESE                    & 0.98 & 8 OZ       & -0.09 & 0.47  & 0.000 & 0.000 \\
Dairy and Eggs & EGGS - LARGE                    & 1.02 & 1 DZ       & 0.01  & 0.26  & 0.000 & 0.000 \\
Dairy and Eggs & EGGS - LARGE                    & 1.40 & 18 CT      & 0.02  & 0.35  & 0.000 & 0.000 \\
Dairy and Eggs & EGGS - MEDIUM                   & 0.71 & 1 DZ       & -0.16 & 0.28  & 0.000 & 0.000 \\
Dairy and Eggs & EGGS - X-LARGE                  & 1.06 & 1 DZ       & 0.31  & 0.29  & 0.000 & 0.000 \\
Dairy and Eggs & FLUID MILK WHITE ONLY           & 1.35 & NA         & 0.25  & -0.10 & 0.000 & 0.000 \\
Dairy and Eggs & FLUID MILK WHITE ONLY           & 1.35 & NA         & 0.03  & 0.23  & 0.001 & 0.000 \\
Dairy and Eggs & FLUID MILK WHITE ONLY           & 2.43 & 1 GA       & 0.37  & 1.00  & 0.000 & 0.000 \\
Dairy and Eggs & FLUID MILK WHITE ONLY           & 1.37 & NA         & -0.35 & -0.05 & 0.000 & 0.000 \\
Dairy and Eggs & FLUID MILK WHITE ONLY           & 2.41 & 1 GA       & 0.47  & 0.77  & 0.000 & 0.000 \\
Dairy and Eggs & FLUID MILK WHITE ONLY           & 2.43 & 1 GA       & -0.16 & 0.64  & 0.000 & 0.000 \\
Dairy and Eggs & FLUID MILK WHITE ONLY           & 2.43 & 1 GA       & 0.16  & 0.42  & 0.000 & 0.000 \\
Dairy and Eggs & FLUID MILK WHITE ONLY           & 1.32 & NA         & -0.30 & 0.27  & 0.000 & 0.000 \\
Dairy and Eggs & IWS SINGLE CHEESE               & 2.35 & 16 OZ      & 0.07  & 0.47  & 0.000 & 0.000 \\
Dairy and Eggs & IWS SINGLE CHEESE               & 1.89 & 12 OZ      & 0.26  & 0.59  & 0.000 & 0.000 \\
Dairy and Eggs & IWS SINGLE CHEESE               & 1.49 & 12 OZ      & -0.13 & 0.45  & 0.000 & 0.000 \\
Dairy and Eggs & SHREDDED CHEESE                 & 1.57 & 8 OZ       & 0.03  & 0.65  & 0.000 & 0.000 \\
Dairy and Eggs & SOUR CREAMS                     & 1.11 & 16 OZ      & 0.28  & 0.68  & 0.000 & 0.000 \\
Fruits         & APPLES GRANNY SMITH (BULK\&BAG) & 2.65 & NA         & 0.28  & 0.50  & 0.000 & 0.000 \\
Fruits         & CANTALOUPE WHOLE                & 2.13 & NA         & 0.31  & 0.15  & 0.002 & 0.000 \\
Fruits         & GRAPES RED                      & 3.19 & 18 LB      & 0.14  & 0.31  & 0.002 & 0.000 \\
Fruits         & GRAPES WHITE                    & 3.63 & 18 LB      & 0.08  & 0.25  & 0.011 & 0.002 \\
Fruits         & STRAWBERRIES                    & 2.61 & 16 OZ      & 0.29  & 0.42  & 0.000 & 0.000 \\
Vegetables     & BEANS GREEN: FS/WHL/CUT         & 0.52 & 14.5 OZ    & 0.02  & 0.48  & 0.000 & 0.000 \\
Vegetables     & BROCCOLI WHOLE\&CROWNS          & 1.65 & NA         & 0.24  & 0.39  & 0.000 & 0.000 \\
Vegetables     & CABBAGE                         & 1.27 & 14-18 CT   & 0.35  & 0.15  & 0.001 & 0.000 \\
Vegetables     & CARROTS MINI PEELED             & 1.57 & 1 LB       & 0.03  & 0.28  & 0.000 & 0.000 \\
Vegetables     & CELERY                          & 1.33 & NA         & 0.24  & 0.17  & 0.009 & 0.001 \\
Vegetables     & CORN                            & 0.51 & 15.25 OZ   & 0.03  & 0.42  & 0.000 & 0.000 \\
Vegetables     & CORN YELLOW                     & 0.36 & 48 CT      & 0.25  & 0.02  & 0.046 & 0.007 \\
Vegetables     & CUCUMBERS                       & 0.70 & 36 CT      & 0.28  & 0.30  & 0.000 & 0.000 \\
Vegetables     & GARDEN PLUS                     & 2.31 & 10 OZ      & 0.46  & 0.37  & 0.000 & 0.000 \\
Vegetables     & GARDEN PLUS                     & 2.26 & 12 OZ      & 0.26  & 0.27  & 0.048 & 0.007 \\
Vegetables     & HEAD LETTUCE                    & 0.98 & 24 CT      & 0.25  & 0.40  & 0.000 & 0.000 \\
Vegetables     & HEAD LETTUCE                    & 0.99 & 24 CT      & 0.23  & 0.43  & 0.001 & 0.000 \\
Vegetables     & MUSHROOMS WHITE SLICED PKG      & 1.86 & 8 OZ       & -0.06 & 0.32  & 0.002 & 0.000 \\
Vegetables     & ONIONS OTHER                    & 0.53 & 48 CT      & 0.19  & 0.36  & 0.000 & 0.000 \\
Vegetables     & ONIONS SWEET (BULK\&BAG)        & 1.17 & 40 LB      & 0.28  & 0.14  & 0.011 & 0.002 \\
Vegetables     & POTATOES RUSSET (BULK\&BAG)     & 3.48 & 10 LB      & 0.34  & 0.40  & 0.000 & 0.000 \\
Vegetables     & POTATOES RUSSET (BULK\&BAG)     & 2.44 & 5 LB       & 0.11  & 0.31  & 0.001 & 0.000 \\
Vegetables     & POTATOES SWEET\&YAMS            & 1.81 & 40 LB      & 0.40  & 0.20  & 0.000 & 0.000 \\
Vegetables     & ROMA TOMATOES (BULK/PKG)        & 1.93 & 25 LB      & -0.39 & -0.33 & 0.000 & 0.000 \\
Vegetables     & SALAD BAR FRESH FRUIT           & 2.37 & NA         & 0.15  & -0.12 & 0.007 & 0.001 \\
Vegetables     & TOMATOES HOTHOUSE ON THE VINE   & 2.57 & 13 LB      & 0.16  & -0.05 & 0.017 & 0.003 \\
Miscellaneous  & CANDY BARS (SINGLES)(INCLUDING  & 0.42 & 1.6 OZ     & -0.11 & 0.57  & 0.000 & 0.000 \\
Miscellaneous  & POTATO CHIPS                    & 1.91 & 11.5 OZ    & 0.52  & 1.00  & 0.000 & 0.000 \\
Miscellaneous  & SOUP CRACKERS (SALTINE/OYSTER)  & 1.05 & 16 OZ      & 0.12  & 0.37  & 0.006 & 0.001 \\
Miscellaneous  & TORTILLA/NACHO CHIPS            & 2.33 & 12.5 OZ    & 0.10  & 0.59  & 0.000 & 0.000 \\
Miscellaneous  & LEAN                            & 3.31 & NA         & 0.00  & 0.28  & 0.001 & 0.000 \\
Miscellaneous  & MEAT: LUNCHMEAT BULK            & 2.75 & NA         & 0.32  & 0.77  & 0.000 & 0.000 \\
Miscellaneous  & MEAT: SAUS DRY BULK             & 3.29 & NA         & 0.15  & 0.73  & 0.000 & 0.000 \\
Miscellaneous  & PREMIUM - MEAT                  & 2.50 & 1 LB       & 0.42  & 1.19  & 0.000 & 0.000 \\
Miscellaneous  & CIGARETTES                      & 3.56 & 974246  PK & -1.00 & -0.35 & 0.000 & 0.000 \\
Miscellaneous  & CONDENSED SOUP                  & 0.64 & 10.5 OZ     & 0.05  & 0.49  & 0.000 & 0.000 \\
Miscellaneous  & GASOLINE-REG UNLEADED           & 0.00 & NA         & -0.64 & -0.09 & 0.000 & 0.000 \\
Miscellaneous  & SUGAR                           & 2.01 & 4 LB        & -0.10 & 0.37  & 0.000 & 0.000 \\
Miscellaneous  & NA                              & NA  & NA         & 0.26  & 0.43  & 0.000 & 0.000 \\
Miscellaneous  & NA                              & NA  & NA         & 0.00  & 0.17  & 0.044 & 0.007 \\
Miscellaneous  & NEWSPAPER                       & 1.42 & NA         & -0.19 & 0.52  & 0.000 & 0.000 \\
Miscellaneous  & TOILET TISSUE                   & 1.02 & 83.5 SQ FT & -0.21 & 0.34  & 0.000 & 0.000\\
\bottomrule
\end{tabular}
\end{table}

\begin{table}[p]
\centering
\tiny
\caption{Characteristics and Estimated coefficients of selected products based on household income}
\label{tab: data analysis - result with income }
\begin{tabular}{llrrrrrr}
\toprule
\textbf{Category}          & \textbf{Subcategory}             & \textbf{Price} & \textbf{Size}       & $\mathbf{\hat{\bbb}_3}$ & $\mathbf{\hat{\bbb}_4}$& \textbf{Adj p-value} & \textbf{p-value} \\
\midrule
Beverages      & DAIRY CASE 100\% PURE JUICE - O & 1.36  & NA         & -0.33    & -0.97       & 0.000             & 0.000             \\
Beverages      & DAIRY CASE TEA WITH SUGAR OR S  & 0.99  & 1 GA       & -0.71    & -3.23       & 0.000             & 0.000             \\
Beverages      & SFT DRNK 2 LITER BTL CARB INCL  & 1.16  & 2 LTR      & -0.54    & -1.35       & 0.000             & 0.000             \\
Beverages      & SFT DRNK 2 LITER BTL CARB INCL  & 1.18  & 2 LTR      & -0.49    & -0.41       & 0.001             & 0.000             \\
Beverages      & SFT DRNK 2 LITER BTL CARB INCL  & 1.23  & 2 LTR      & -0.21    & -0.62       & 0.000             & 0.000             \\
Beverages      & SFT DRNK 2 LITER BTL CARB INCL  & 1.16  & 2 LTR      & -0.48    & -1.72       & 0.000             & 0.000             \\
Beverages      & SFT DRNK SNGL SRV BTL CARB (EX  & 1.07  & 20 OZ      & -0.54    & -1.57       & 0.000             & 0.000             \\
Beverages      & SOFT DRINKS 12/18\&15PK CAN CAR & 3.33  & 12 OZ      & 0.21     & 0.96        & 0.000             & 0.000             \\
Beverages      & SOFT DRINKS 12/18\&15PK CAN CAR & 3.49  & 12 OZ      & -0.24    & -0.95       & 0.000             & 0.000             \\
Beverages      & SOFT DRINKS 12/18\&15PK CAN CAR & 3.41  & 12 OZ      & -0.28    & -1.43       & 0.000             & 0.000             \\
Breakfast      & HAMBURGER BUNS                  & 0.95  & 12 OZ      & -0.09    & -0.48       & 0.000             & 0.000             \\
Breakfast      & HOT DOG BUNS                    & 0.95  & 11 OZ      & -0.22    & -0.49       & 0.000             & 0.000             \\
Breakfast      & MAINSTREAM WHEAT/MULTIGRAIN BR  & 0.96  & 20 OZ      & -0.24    & -0.52       & 0.000             & 0.000             \\
Breakfast      & MAINSTREAM WHITE BREAD          & 0.97  & 20 OZ      & -0.33    & -0.66       & 0.000             & 0.000             \\
Breakfast      & MAINSTREAM WHITE BREAD          & 1.65  & 20 OZ      & 0.06     & -0.45       & 0.000             & 0.000             \\
Breakfast      & MAINSTREAM WHITE BREAD          & 1.47  & 24 OZ      & -0.09    & -1.30       & 0.000             & 0.000             \\
Breakfast      & MAINSTREAM WHITE BREAD          & 0.97  & 20 OZ      & -0.82    & -1.40       & 0.000             & 0.000             \\
Dairy and Eggs & CHOCOLATE MILK                  & 1.28  & NA         & -0.05    & -0.30       & 0.001             & 0.000             \\
Dairy and Eggs & COTTAGE CHEESE                  & 2.09  & 24 OZ      & 0.36     & -0.17       & 0.000             & 0.000             \\
Dairy and Eggs & CREAM CHEESE                    & 1.55  & 8 OZ       & 0.26     & 0.35        & 0.011             & 0.002             \\
Dairy and Eggs & CREAM CHEESE                    & 0.98  & 8 OZ       & 0.55     & -0.02       & 0.000             & 0.000             \\
Dairy and Eggs & EGGS - LARGE                    & 1.02  & 1 DZ       & 0.27     & 0.44        & 0.000             & 0.000             \\
Dairy and Eggs & EGGS - LARGE                    & 1.40  & 18 CT      & -0.36    & -0.25       & 0.000             & 0.000             \\
Dairy and Eggs & EGGS - MEDIUM                   & 0.71  & 1 DZ       & -0.19    & -0.73       & 0.000             & 0.000             \\
Dairy and Eggs & FLUID MILK WHITE ONLY           & 1.35  & NA         & 0.45     & 0.71        & 0.000             & 0.000             \\
Dairy and Eggs & FLUID MILK WHITE ONLY           & 1.35  & NA         & -0.14    & -0.28       & 0.000             & 0.000             \\
Dairy and Eggs & FLUID MILK WHITE ONLY           & 2.43  & 1 GA       & -0.20    & -0.28       & 0.000             & 0.000             \\
Dairy and Eggs & FLUID MILK WHITE ONLY           & 1.37  & NA         & 0.37     & 1.00        & 0.000             & 0.000             \\
Dairy and Eggs & FLUID MILK WHITE ONLY           & 2.41  & 1 GA       & 0.06     & 0.44        & 0.000             & 0.000             \\
Dairy and Eggs & FLUID MILK WHITE ONLY           & 2.43  & 1 GA       & -0.67    & -0.98       & 0.000             & 0.000             \\
Dairy and Eggs & FLUID MILK WHITE ONLY           & 2.43  & 1 GA       & 0.30     & 1.07        & 0.000             & 0.000             \\
Dairy and Eggs & FLUID MILK WHITE ONLY           & 1.32  & NA         & -0.48    & -1.11       & 0.000             & 0.000             \\
Dairy and Eggs & IWS SINGLE CHEESE               & 2.35  & 16 OZ      & 0.17     & -0.55       & 0.000             & 0.000             \\
Dairy and Eggs & IWS SINGLE CHEESE               & 1.89  & 12 OZ      & -0.06    & -0.54       & 0.000             & 0.000             \\
Dairy and Eggs & IWS SINGLE CHEESE               & 1.49  & 12 OZ      & -0.32    & -0.72       & 0.000             & 0.000             \\
Dairy and Eggs & SHREDDED CHEESE                 & 1.57  & 8 OZ       & -0.02    & -0.39       & 0.000             & 0.000             \\
Dairy and Eggs & SOUR CREAMS                     & 1.11  & 16 OZ      & 0.32     & -0.41       & 0.000             & 0.000             \\
Fruits         & APPLES GALA (BULK\&BAG)         & 2.36  & NA         & 0.45     & 1.08        & 0.000             & 0.000             \\
Fruits         & APPLES GRANNY SMITH (BULK\&BAG) & 2.65  & NA         & 0.35     & 1.23        & 0.000             & 0.000             \\
Fruits         & BANANAS                         & 0.95  & 40 LB      & 0.43     & 1.08        & 0.000             & 0.000             \\
Fruits         & CANTALOUPE WHOLE                & 2.13  & NA         & 0.26     & 0.64        & 0.000             & 0.000             \\
Fruits         & GRAPES RED                      & 3.19  & 18 LB      & 0.22     & 0.53        & 0.000             & 0.000             \\
Fruits         & GRAPES WHITE                    & 3.63  & 18 LB      & -0.12    & 0.40        & 0.000             & 0.000             \\
Fruits         & LEMONS                          & 0.61  & NA         & 0.36     & 1.19        & 0.000             & 0.000             \\
Fruits         & ORANGES NAVELS ALL              & 0.51  & NA         & -0.06    & 0.54        & 0.000             & 0.000             \\
Fruits         & STRAWBERRIES                    & 2.61  & 16 OZ      & 0.26     & 0.95        & 0.000             & 0.000             \\
Vegetables     & BROCCOLI WHOLE\&CROWNS          & 1.65  & NA         & 0.20     & 1.09        & 0.000             & 0.000             \\
Vegetables     & CABBAGE                         & 1.27  & 14-18 CT   & -0.18    & -0.37       & 0.005             & 0.001             \\
Vegetables     & CARROTS MINI PEELED             & 1.57  & 1 LB       & 0.36     & 0.62        & 0.000             & 0.000             \\
Vegetables     & CELERY                          & 1.33  & NA         & 0.13     & 0.37        & 0.000             & 0.000             \\
Vegetables     & CORN                            & 0.51  & 15.25 OZ   & 0.22     & -0.32       & 0.000             & 0.000             \\
Vegetables     & CORN YELLOW                     & 0.36  & 48 CT      & 0.34     & 0.57        & 0.000             & 0.000             \\
Vegetables     & CUCUMBERS                       & 0.70  & 36 CT      & 0.16     & 0.35        & 0.000             & 0.000             \\
Vegetables     & GARDEN PLUS                     & 2.31  & 10 OZ      & 0.52     & 1.19        & 0.000             & 0.000             \\
Vegetables     & HEAD LETTUCE                    & 0.98  & 24 CT      & 0.03     & -0.18       & 0.020             & 0.003             \\
Vegetables     & MUSHROOMS WHITE SLICED PKG      & 1.86  & 8 OZ       & 0.01     & 0.93        & 0.000             & 0.000             \\
Vegetables     & ONIONS OTHER                    & 0.53  & 48 CT      & 0.23     & 0.58        & 0.000             & 0.000             \\
Vegetables     & ONIONS SWEET (BULK\&BAG)        & 1.17  & 40 LB      & 0.38     & 0.83        & 0.000             & 0.000             \\
Vegetables     & ONIONS SWEET (BULK\&BAG)        & 1.04  & 40 LB      & 0.34     & 0.59        & 0.000             & 0.000             \\
Vegetables     & ONIONS YELLOW (BULK\&BAG)       & 1.91  & 3 LB       & -0.30    & -0.30       & 0.001             & 0.000             \\
Vegetables     & PEPPERS GREEN BELL              & 0.72  & 48-54 CT   & 0.17     & 0.47        & 0.000             & 0.000             \\
Vegetables     & POTATOES RUSSET (BULK\&BAG)     & 3.48  & 10 LB      & -0.33    & -0.76       & 0.000             & 0.000             \\
Vegetables     & POTATOES RUSSET (BULK\&BAG)     & 2.44  & 5 LB       & -0.13    & -0.42       & 0.000             & 0.000             \\
Vegetables     & POTATOES SWEET\&YAMS            & 1.81  & 40 LB      & 0.35     & 0.80        & 0.000             & 0.000             \\
Vegetables     & REGULAR GARDEN                  & 1.47  & 1 LB       & -0.14    & -0.30       & 0.020             & 0.003             \\
Vegetables     & ROMA TOMATOES (BULK/PKG)        & 1.93  & 25 LB      & -0.07    & 0.46        & 0.000             & 0.000             \\
Vegetables     & SQUASH ZUCCHINI                 & 1.38  & 18 LB      & 0.55     & 1.47        & 0.000             & 0.000             \\
Vegetables     & TOMATOES GRAPE                  & 2.59  & PINT       & 0.11     & 0.79        & 0.000             & 0.000             \\
Vegetables     & TOMATOES HOTHOUSE ON THE VINE   & 2.57  & 13 LB      & 0.27     & 0.69        & 0.000             & 0.000             \\
Miscellaneous  & CANDY BARS (SINGLES)(INCLUDING  & 0.42  & 1.6 OZ     & -0.09    & -1.37       & 0.000             & 0.000             \\
Miscellaneous  & POTATO CHIPS                    & 1.91  & 11.5 OZ    & -0.12    & -0.35       & 0.020             & 0.003             \\
Miscellaneous  & SOUP CRACKERS (SALTINE/OYSTER)  & 1.05  & 16 OZ      & -0.52    & -0.68       & 0.000             & 0.000             \\
Miscellaneous  & TORTILLA/NACHO CHIPS            & 2.33  & 12.5 OZ    & -0.07    & -0.70       & 0.000             & 0.000             \\
Miscellaneous  & CHICKEN BREAST BONELESS         & 4.46  & NA         & 0.25     & 0.50        & 0.000             & 0.000             \\
Miscellaneous  & LEAN                            & 3.31  & NA         & -0.44    & -1.27       & 0.000             & 0.000             \\
Miscellaneous  & MEAT: LUNCHMEAT BULK            & 2.75  & NA         & 0.26     & -0.77       & 0.000             & 0.000             \\
Miscellaneous  & MEAT: SAUS DRY BULK             & 3.29  & NA         & 0.43     & 0.27        & 0.000             & 0.000             \\
Miscellaneous  & PREMIUM - MEAT                  & 2.50  & 1 LB       & -0.14    & -0.54       & 0.000             & 0.000             \\
Miscellaneous  & PRIMAL                          & 3.81  & NA         & -0.02    & -0.53       & 0.000             & 0.000             \\
Miscellaneous  & CIGARETTES                      & 3.56  & 974246  PK & 0.85     & 0.83        & 0.000             & 0.000             \\
Miscellaneous  & CONDENSED SOUP                  & 0.64  & 10.5 OZ     & 0.01     & -0.59       & 0.000             & 0.000             \\
Miscellaneous  & GARLIC WHOLE CLOVES             & 0.46  & 10 LB      & 0.18     & 1.15        & 0.000             & 0.000             \\
Miscellaneous  & GASOLINE-REG UNLEADED           & 0.00  & NA         & 0.61     & 1.08        & 0.000             & 0.000             \\
Miscellaneous  & NA                              & NA   & NA         & 0.29     & 0.40        & 0.000             & 0.000             \\
Miscellaneous  & NA                              & NA   & NA         & 0.27     & 0.43        & 0.000             & 0.000             \\
Miscellaneous  & PAPER TOWELS \& HOLDERS         & 0.55  & 57 SQ FT   & 0.09     & -0.59       & 0.000             & 0.000             \\
Miscellaneous  & TOILET TISSUE                   & 1.02  & 83.5 SQ FT & -0.29    & -1.10       & 0.000             & 0.000            \\
\bottomrule
\end{tabular}
\end{table}

\begin{table}[h]
\centering
\caption{Number of factors $\hat{K}$ selected by each method for varying values of $J$ in real data analysis.}
\label{Tab: real data number of factors}
\begin{tabular}{lcccc}
\toprule
Method & $J = 100$ & $J = 200$ & $J = 300$ & $J = 400$ \\
\midrule
Prop & 8  & 8  & 11 & 11 \\
Prop (Sect 2.3.2) & 4  & 2  & 2  & 1  \\
Prop (Sect 2.3.3) & 10 & 11 & 11 & 14 \\
Prop (Sect 2.3.2 \& 2.3.3) & 5  & 3  & 3  & 3 \\
\bottomrule
\end{tabular}
\end{table}

\begin{table}[!t]
\centering
\caption{Residual deviances $D^{\text{res}}$ for each model across different $J$.}
\label{tab:residual_deviance}
\scriptsize
\begin{tabular}{@{}llcccccccc@{}}
\toprule
& & \multicolumn{4}{c}{\textbf{In-sample residual deviances}} 
  & \multicolumn{4}{c}{\textbf{Out-of-sample residual deviances}} \\
\cmidrule(lr){3-6} \cmidrule(lr){7-10}
& & $J=100$ & $J=200$ & $J=300$ & $J=400$
  & $J=100$ & $J=200$ & $J=300$ & $J=400$ \\
\midrule
& Prop                       
& 738553 & 1146654 & 1432926 & 1677555 
& 36541  & 56207   & 70515   & 82853 \\

& Prop (sect 2.3.2)          
& 746002 & 1195027 & 1504740 & 1820744 
& 38481  & 60997   & 78649   & 94701 \\

& Prop (sect 2.3.3)          
& 858980 & 1364673 & 1757449 & 2067403 
& 50899  & 81170   & 104893  & 125275 \\

& Prop (sect 2.3.2 \& 2.3.3) 
& 775049 & 1215675 & 1542648 & 1805447 
& 39930  & 61543   & 79038   & 93154 \\
\bottomrule
\end{tabular}
\end{table}

\begin{table}[!t]
\centering
\caption{Sensitivity Based on Number of Recommendations}
\label{tab: Full Sensitivity based on Number of Recommendations}
\small
\begin{tabular}{@{}llcccccccc@{}}
\toprule
& & \multicolumn{4}{c}{10 recommendations} & \multicolumn{4}{c}{20 recommendations} \\
\cmidrule(lr){3-6} \cmidrule(lr){7-10}
& & J=100 & J=200 & J=300 & J=400 & J=100 & J=200 & J=300 & J=400 \\
\midrule
&Hist                    & 0.448    & 0.348   & 0.297   & 0.264   & 0.652    & 0.518   & 0.447   & 0.404   \\
&Prop                & 0.352    & 0.256   & 0.223   & 0.199   & 0.533    & 0.394   & 0.340   & 0.304   \\
&Prop (sect 2.3.2)         & 0.323    & 0.208   & 0.173   & 0.143   & 0.498    & 0.332   & 0.275   & 0.224   \\
&Prop (sect 2.3.3)         & 0.337    & 0.247   & 0.208   & 0.189   & 0.489    & 0.360   & 0.303   & 0.278   \\
&Prop (sect 2.3.2 \& 2.3.3) & 0.297    & 0.197   & 0.164   & 0.143   & 0.462    & 0.317   & 0.262   & 0.231   \\
&Hist-Hist               & 0.451    & 0.350   & 0.299   & 0.266   & 0.654    & 0.519   & 0.450   & 0.406   \\
&Hist-Prop               & 0.456    & 0.352   & 0.301   & 0.269   & 0.659    & 0.525   & 0.453   & 0.407   \\
\midrule
& & \multicolumn{4}{c}{30 recommendations} & \multicolumn{4}{c}{40 recommendations} \\
\cmidrule(lr){3-6} \cmidrule(lr){7-10}
& & J=100 & J=200 & J=300 & J=400 & J=100 & J=200 & J=300 & J=400 \\
\midrule
&Hist                    & 0.774    & 0.627   & 0.546   & 0.496   & 0.848    & 0.705   & 0.618   & 0.565   \\
&Prop               & 0.658    & 0.490   & 0.425   & 0.382   & 0.752    & 0.570   & 0.496   & 0.447   \\
&Prop (sect 2.3.2)         & 0.624    & 0.430   & 0.355   & 0.292   & 0.723    & 0.509   & 0.424   & 0.346   \\
&Prop (sect 2.3.3)         & 0.612    & 0.456   & 0.382   & 0.348   & 0.712    & 0.529   & 0.445   & 0.406   \\
&Prop (sect 2.3.2 \& 2.3.3) & 0.597    & 0.415   & 0.341   & 0.300   & 0.696    & 0.495   & 0.406   & 0.358   \\
&Hist-Hist               & 0.775    & 0.629   & 0.549   & 0.499   & 0.852    & 0.709   & 0.621   & 0.570   \\
&Hist-Prop               & 0.781    & 0.635   & 0.555   & 0.505   & 0.860    & 0.714   & 0.626   & 0.575  \\
\bottomrule
\end{tabular}
\end{table}

\FloatBarrier

\bibliographystyle{apalike}
\bibliography{DisEHA_FA}

@book{Van_etal-1996-weak,
  title={Weak Convergence and Empirical Processes},
  author={Van Der Vaart, Aad W and Wellner, Jon A and van der Vaart, Aad W and Wellner, Jon A},
  year={1996},
  publisher={Springer}
}

@article{stock_Watson-2002-JASA,
  title={Forecasting using principal components from a large number of predictors},
  author={Stock, James H and Watson, Mark W},
  journal={Journal of the American Statistical Association},
  volume={97},
  number={460},
  pages={1167--1179},
  year={2002},
  url = {https://doi.org/10.1198/016214502388618960},
  publisher={Taylor \& Francis}
}

@article{Hsieh_etal-2010-MBR,
  title={Using a multivariate multilevel polytomous item response theory model to study parallel processes of change: The dynamic association between adolescents' social isolation and engagement with delinquent peers in the national youth survey},
  author={Hsieh, Chueh-An and von Eye, Alexander A and Maier, Kimberly S},
  journal={Multivariate Behavioral Research},
  volume={45},
  number={3},
  pages={508--552},
  year={2010},
  url = {https://doi.org/10.1080/00273171.2010.483387},
  publisher={Taylor \& Francis}
}

@article{Galvao_Kato-2016-JE,
  title={Smoothed quantile regression for panel data},
  author={Galvao, Antonio F and Kato, Kengo},
  journal={Journal of Econometrics},
  volume={193},
  number={1},
  pages={92--112},
  url = {https://doi.org/10.1016/j.jeconom.2016.01.008},
  year={2016},
  publisher={Elsevier}
}

@article{Zhang_etal-2020-Psychometrika,
  title={A note on exploratory item factor analysis by singular value decomposition},
  author={Zhang, Haoran and Chen, Yunxiao and Li, Xiaoou},
  journal={Psychometrika},
  volume={85},
  pages={358--372},
  year={2020},
  publisher={Springer}
}

@article{Liu_etal-2023-Psychometrika,
  title={ROTATION TO SPARSE LOADINGS USING {$L^P$} LOSSES AND RELATED INFERENCE PROBLEMS},
  author={Liu, Xinyi and Wallin, Gabriel and Chen, Yunxiao and Moustaki, Irini},
  journal={Psychometrika},
  volume={88},
  pages={527--553},
  year={2023},
  url = {https://doi.org/10.1007/s11336-023-09911-y},
  publisher={Springer}
}

@article{Rohe_Zeng-2023-JRSSB,
  title={Vintage factor analysis with Varimax performs statistical inference},
  author={Rohe, Karl and Zeng, Muzhe},
  journal={Journal of the Royal Statistical Society Series B: Statistical Methodology},
  volume={85},
  pages={1037-1060},
  url = {https://doi.org/10.1093/jrsssb/qkad029},
  year={2023},
}

@article{Zhu_etal-2016-JASA,
  title={Personalized prediction and sparsity pursuit in latent factor models},
  author={Zhu, Yunzhang and Shen, Xiaotong and Ye, Changqing},
  journal={Journal of the American Statistical Association},
  volume={111},
  number={513},
  pages={241--252},
  year={2016},
  url = {https://doi.org/10.1080/01621459.2014.999158},
  publisher={Taylor \& Francis}
}

@book{Molenberghs_Verbeke-2005-Springer,
  title={Models for discrete longitudinal data},
  author={Molenberghs, Geert and Verbeke, Geert},
  publisher={Springer},
  year={2005},
  address = {New York, NY}
}

@article{Parikh_Boyd-2014-Optimization,
  title={Proximal algorithms},
  author={Parikh, Neal and Boyd, Stephen and others},
  journal={Foundations and trends{\textregistered} in Optimization},
  volume={1},
  number={3},
  pages={127--239},
  year={2014},
  publisher={Now Publishers, Inc.}
}

@article{Galvao_Kato-2016-JoE,
  title={Smoothed quantile regression for panel data},
  author={Galvao, Antonio F and Kato, Kengo},
  journal={Journal of econometrics},
  volume={193},
  number={1},
  pages={92--112},
  year={2016},
  publisher={Elsevier}
}

@article{Newey_Mcfadden-1994-HoE,
  title={Large sample estimation and hypothesis testing},
  author={Newey, Whitney K and McFadden, Daniel},
  journal={Handbook of econometrics},
  volume={4},
  pages={2111--2245},
  year={1994},
  publisher={Elsevier}
}

@book{cook_etal-2007-recurbook,
    address = {New York, NY},
	author = {Cook, Richard John and Lawless, Jerald F.},
	publisher = {Springer},
	title = {The statistical analysis of recurrent events},
	year = {2007}}

@article{bai_li-2012-aos,
	author = {Bai, Jushan and Li, Kunpeng},
	journal = {The Annals of Statistics},
	number = {1},
	pages = {436--465},
	publisher = {Institute of Mathematical Statistics},
    doi = {10.1214/11-AOS966},
url = {https://doi.org/10.1214/11-AOS966}, 
	title = {Statistical analysis of factor models of high dimension},
	volume = {40},
	year = {2012}}

@article{chen_Li_Zhang-2019-Psychometrika,
	author = {Chen, Yunxiao and Li, Xiaoou and Zhang, Siliang},
	journal = {Psychometrika},
	number = {1},
	pages = {124--146},
	publisher = {Springer},
	title = {Joint maximum likelihood estimation for high-dimensional exploratory item factor analysis},
	volume = {84},
	year = {2019}}

@article{bai-2003-Econometrica,
	author = {Bai, Jushan},
	journal = {Econometrica},
	number = {1},
	pages = {135--171},
	publisher = {Wiley Online Library},
	title = {Inferential theory for factor models of large dimensions},
	volume = {71},
	year = {2003}}

@article{Chen_Li_Zhang-2020-JASA,
	author = {Chen, Yunxiao and Li, Xiaoou and Zhang, Siliang},
	journal = {Journal of the American Statistical Association},
	number = {532},
	pages = {1756--1770},
	publisher = {Taylor \& Francis},
	title = {Structured latent factor analysis for large-scale data: Identifiability, estimability, and their implications},
    url = {https://doi.org/10.1080/01621459.2019.1635485},
	volume = {115},
	year = {2020}}

@article{chen_Li-2022-Biometrika,
  title={Determining the number of factors in high-dimensional generalized latent factor models},
  author={Chen, Yunxiao and Li, Xiaoou},
  journal={Biometrika},
  volume={109},
  number={3},
url = {https://doi.org/10.1093/biomet/asab044},
  pages={769--782},
  year={2022},
  publisher={Oxford University Press}
}

@article{Bai_Ng-2002-Econometrica,
  title={Determining the number of factors in approximate factor models},
  author={Bai, Jushan and Ng, Serena},
  journal={Econometrica},
  volume={70},
  number={1},
  pages={191--221},
  year={2002},
  publisher={Wiley Online Library}
}

@article{French_etal-2019-BMC,
  title={Nutrition quality of food purchases varies by household income: the SHoPPER study},
  author={French, Simone A and Tangney, Christy C and Crane, Melissa M and Wang, Yamin and Appelhans, Bradley M},
  journal={BMC Public Health},
  volume={19},
  pages={1--7},
  year={2019},
  publisher={Springer},
  url = {https://link.springer.com/article/10.1186/s12889-019-6546-2}
}

@article{Chen_etal-2021-Econometrica,
	author = {Chen, Liang and Dolado, Juan J and Gonzalo, Jes{\'u}s},
	journal = {Econometrica},
	number = {2},
	pages = {875--910},
	publisher = {Wiley Online Library},
	title = {Quantile factor models},
    doi = {10.3982/ECTA15746},
url = {https://onlinelibrary.wiley.com/doi/abs/10.3982/ECTA15746},
	volume = {89},
	year = {2021}}

@book{vandervaart1996_weak,
	address = {New York ; Hong Kong},
	author = {A.W. van der Vaart and Jon Wellner},
	publisher = {Springer},
	series = {Springer series in statistics},
	title = {Weak convergence and empirical processes : with applications to statistics},
	year = {1996}}

@book{bolger2013intensive,
	address = {New York, NY},
	author = {Bolger, Niall. and Laurenceau, Jean-Philippe.},
	booktitle = {Intensive longitudinal methods : an introduction to diary and experience sampling research},
	isbn = {9781462506781},
	language = {eng},
	lccn = {2012032402},
	publisher = {Guilford Press},
    title = {Intensive longitudinal methods : An introduction to diary and experience sampling research},
	year = {2013}
}

@inproceedings{zhang2020adverse,
  title={Adverse drug reaction discovery from electronic health records with deep neural networks},
  author={Zhang, Wei and Kuang, Zhaobin and Peissig, Peggy and Page, David},
  booktitle={Proceedings of the ACM Conference on Health, Inference, and Learning},
  pages={30--39},
  year={2020},
  url = {https://doi.org/10.1145/3368555.3384459}
}

@inproceedings{lian2015multitask,
  title={A multitask point process predictive model},
  author={Lian, Wenzhao and Henao, Ricardo and Rao, Vinayak and Lucas, Joseph and Carin, Lawrence},
  booktitle={International Conference on Machine Learning},
  pages={2030--2038},
  year={2015},
url = {https://proceedings.mlr.press/v37/lian15.html}
}

@article{chen2019statistical,
  title={Statistical analysis of complex problem-solving process data: An event history analysis approach},
  author={Chen, Yunxiao and Li, Xiaoou and Liu, Jingchen and Ying, Zhiliang},
  journal={Frontiers in Psychology},
  volume={10},
  pages={486},
  year={2019},
  doi = {10.3389/fpsyg.2019.00486},
url = {https://doi.org/10.3389/fpsyg.2019.00486},
  publisher={Frontiers Media SA}
}

@article{chen2020continuous,
  title={A continuous-time dynamic choice measurement model for problem-solving process data},
  author={Chen, Yunxiao},
  journal={Psychometrika},
  volume={85},
  number={4},
url = {https://link.springer.com/article/10.1007/s11336-020-09734-1},
  pages={1052--1075},
  year={2020},
  publisher={Springer}
}

@inproceedings{wan2017modeling,
  title={Modeling consumer preferences and price sensitivities from large-scale grocery shopping transaction logs},
  author={Wan, Mengting and Wang, Di and Goldman, Matt and Taddy, Matt and Rao, Justin and Liu, Jie and Lymberopoulos, Dimitrios and McAuley, Julian},
  booktitle={Proceedings of the 26th International Conference on World Wide Web},
  pages={1103--1112},
  year={2017},
  url = {https://doi.org/10.1145/3038912.3052568}
}

@inproceedings{wan2018representing,
  title={Representing and recommending shopping baskets with complementarity, compatibility and loyalty},
  author={Wan, Mengting and Wang, Di and Liu, Jie and Bennett, Paul and McAuley, Julian},
  booktitle={Proceedings of the 27th ACM International Conference on Information and Knowledge Management},
  url = {https://doi.org/10.1145/3269206.3271786},
  pages={1133--1142},
  year={2018}
}

@inproceedings{wang2014studentlife,
  title={StudentLife: assessing mental health, academic performance and behavioral trends of college students using smartphones},
  author={Wang, Rui and Chen, Fanglin and Chen, Zhenyu and Li, Tianxing and Harari, Gabriella and Tignor, Stefanie and Zhou, Xia and Ben-Zeev, Dror and Campbell, Andrew T},
  booktitle={Proceedings of the 2014 ACM International Joint Conference on Pervasive and Ubiquitous Computing},
  pages={3--14},
  year={2014}, 
  url = {https://doi.org/10.1145/2632048.2632054}
}

@article{ten1999mixed,
  title={Mixed effects models with bivariate and univariate association parameters for longitudinal bivariate binary response data},
  author={Ten Have, Thomas R and Morabia, Alfredo},
  journal={Biometrics},
  volume={55},
  number={1},
  pages={85--93},
  year={1999},
  url = {https://doi.org/10.1111/j.0006-341X.1999.00085.x},
  publisher={Wiley Online Library}
}

@article{liu2023generalized,
  title={Generalized factor model for ultra-high dimensional correlated variables with mixed types},
  author={Liu, Wei and Lin, Huazhen and Zheng, Shurong and Liu, Jin},
  journal={Journal of the American Statistical Association},
  volume={118},
  number={542},
  url = {https://doi.org/10.1080/01621459.2021.1999818},
  pages={1385--1401},
  year={2023},
  publisher={Taylor \& Francis}
}

@book{skrondal2004generalized,
  title={Generalized latent variable modeling: Multilevel, longitudinal, and structural equation models},
  author={Skrondal, Anders and Rabe-Hesketh, Sophia},
  year={2004},
  publisher={Chapman and Hall/CRC},
}

@book{bertsekas1999nonlinear,
  title={Nonlinear Programming},
  author={Bertsekas, Dimitri},
  year={1999},
  publisher={Athena Scientific}
}

@article{liang1986longitudinal,
  title={Longitudinal data analysis using generalized linear models},
  author={Liang, Kung-Yee and Zeger, Scott L},
  journal={Biometrika},
  volume={73},
  number={1},
  pages={13--22},
  year={1986},
  publisher={Oxford University Press}
}

@article{prentice1988correlated,
  title={Correlated binary regression with covariates specific to each binary observation},
  author={Prentice, Ross L},
  journal={Biometrics},
  pages={1033--1048},
  year={1988},
  publisher={JSTOR}
}

@article{carey1993modelling,
  title={Modelling multivariate binary data with alternating logistic regressions},
  author={Carey, Vincent and Zeger, Scott L and Diggle, Peter},
  journal={Biometrika},
  volume={80},
  number={3},
  pages={517--526},
  year={1993},
  publisher={Oxford University Press}
}

@article{gray2000multidimensional,
  title={Multidimensional longitudinal data: estimating a treatment effect from continuous, discrete, or time-to-event response variables},
  author={Gray, Sarah M and Brookmeyer, Ron},
  journal={Journal of the American Statistical Association},
  volume={95},
  number={450},
  pages={396--406},
  year={2000},
  publisher={Taylor \& Francis}
}

@article{zeng2007transition,
  title={Transition models for multivariate longitudinal binary data},
  author={Zeng, Leilei and Cook, Richard J},
  journal={Journal of the American Statistical Association},
  volume={102},
  number={477},
  pages={211--223},
  year={2007},
  publisher={Taylor \& Francis}
}

@article{liang1989class,
  title={A class of logistic regression models for multivariate binary time series},
  author={Liang, Kung-Yee and Zeger, Scott L},
  journal={Journal of the American Statistical Association},
  volume={84},
  number={406},
  pages={447--451},
  year={1989},
  publisher={Taylor \& Francis}
}

@article{lambert2002copula,
  title={A copula-based model for multivariate non-normal longitudinal data: analysis of a dose titration safety study on a new antidepressant},
  author={Lambert, Philippe and Vandenhende, Francois},
  journal={Statistics in Medicine},
  volume={21},
  number={21},
  pages={3197--3217},
  year={2002},
  publisher={Wiley Online Library}
}

@article{smith2010modeling,
  title={Modeling longitudinal data using a pair-copula decomposition of serial dependence},
  author={Smith, Michael and Min, Aleksey and Almeida, Carlos and Czado, Claudia},
  journal={Journal of the American Statistical Association},
  volume={105},
  number={492},
  pages={1467--1479},
  year={2010},
  publisher={Taylor \& Francis}
}

@article{oort2001three,
  title={Three-mode models for multivariate longitudinal data},
  author={Oort, Frans J},
  journal={British Journal of Mathematical and Statistical Psychology},
  volume={54},
  number={1},
  pages={49--78},
  year={2001},
  publisher={Wiley Online Library}
}

@article{panagiotelis2012pair,
  title={Pair copula constructions for multivariate discrete data},
  author={Panagiotelis, Anastasios and Czado, Claudia and Joe, Harry},
  journal={Journal of the American Statistical Association},
  volume={107},
  number={499},
  pages={1063--1072},
  year={2012},
  publisher={Taylor \& Francis}
}

@article{proust2013analysis,
  title={Analysis of multivariate mixed longitudinal data: a flexible latent process approach},
  author={Proust-Lima, C{\'e}cile and Amieva, H{\'e}l{\`e}ne and Jacqmin-Gadda, H{\'e}l{\`e}ne},
  journal={British Journal of Mathematical and Statistical Psychology},
  volume={66},
  number={3},
  pages={470--487},
  year={2013},
  publisher={Wiley Online Library}
}

@article{Benjamini_Yejutieli-2001-Aos,
  title={The control of the false discovery rate in multiple testing under dependency},
  author={Benjamini, Yoav and Yekutieli, Daniel},
  journal={Annals of statistics},
  pages={1165--1188},
  year={2001},
  publisher={JSTOR}
}

@article{sorensen2023longitudinal,
  title={Longitudinal modeling of age-dependent latent traits with generalized additive latent and mixed models},
  author={S{\o}rensen, {\O}ystein and Fjell, Anders M and Walhovd, Kristine B},
  journal={Psychometrika},
  volume={88},
  number={2},
  pages={456--486},
  year={2023},
  publisher={Springer}
}

@article{ounajim2023mixture,
  title={Mixture of longitudinal factor analyzers and their application to the assessment of chronic pain},
  author={Ounajim, Amine and Slaoui, Yousri and Louis, Pierre-Yves and Billot, Maxime and Frasca, Denis and Rigoard, Philippe},
  journal={Statistics in Medicine},
  volume={42},
  number={18},
  pages={3259--3282},
  year={2023},
  publisher={Wiley Online Library}
}

@article{liu2006mixed,
  title={A mixed-effects regression model for longitudinal multivariate ordinal data},
  author={Liu, Li C and Hedeker, Donald},
  journal={Biometrics},
  volume={62},
  number={1},
  pages={261--268},
  year={2006},
  publisher={Oxford University Press}
}

@article{wang2016second,
  title={A second-order longitudinal model for binary outcomes: Item response theory versus structural equation modeling},
  author={Wang, Chun and Kohli, Nidhi and Henn, Lisa},
  journal={Structural Equation Modeling: A Multidisciplinary Journal},
  volume={23},
  number={3},
  pages={455--465},
  year={2016},
  publisher={Taylor \& Francis}
}

\end{document}